\documentclass[11pt,a4paper]{article}
\pdfoutput=1
\usepackage{jheppub}
\usepackage{afterpage}
\usepackage{array}
\usepackage{MnSymbol}
\usepackage{dsfont}
\usepackage{slashed}
\usepackage{mathtools}
\usepackage{mathrsfs}
\usepackage[squaren,Gray]{SIunits}
\usepackage{booktabs}
\usepackage{dcolumn}
\usepackage{multirow}
\newcommand{\overbar}[1]{\mkern 1.5mu\overline{\mkern-1.5mu#1\mkern-1.5mu}\mkern 1.5mu}
\renewcommand{\underbar}[1]{\mkern 1.5mu\underline{\mkern-1.5mu#1\mkern-1.mu}\mkern 1.mu}
\newcommand{\MSbar}{\overbar{\text{MS}}}
\newcommand{\RI}{\text{RI}^\prime\mkern-3mu\raisebox{1pt}{$/$}\text{SMOM}}
\newcommand{\SU}[1]{\ensuremath{\text{SU}(#1)}}
\renewcommand{\H}[1]{\ensuremath{\overbar{\text{H}(#1)}}}
\newcommand{\2}{\phantom{2}}
\newcommand{\tabcolsepstd}{0.2cm}
\renewcommand{\tabcolsep}{\tabcolsepstd}
\newcommand{\mcemd}{\multicolumn{1}{c}{---}}
\newlength\uud
\newcommand{\flaket}[2]{\settowidth\uud{\ensuremath{#1uud}}\ensuremath{\lvert\kern.5\uud\mathclap{#2}\kern.5\uud\rangle}}
\newcommand{\fl}[1]{{\mathpalette\flaket{#1}}}
\newlength\up
\newcommand{\fla}[2]{\settowidth\up{\ensuremath{#1u}}\ensuremath{\kern.5\up\mathclap{#2}\kern.5\up}}
\newcommand{\f}[1]{{\mathpalette\fla{#1}}}
\newcommand{\goodarrow}[2]{\raisebox{\depth}{\resizebox*{!}{\height}{\ensuremath{#1#2}}}}
\newcommand{\gooduparrow}{{\mathpalette\goodarrow\uparrow}}
\newcommand{\gooddownarrow}{{\mathpalette\goodarrow\downarrow}}
\title{Light-cone distribution amplitudes of the baryon octet}
\author[a,b]{Gunnar~S.~Bali,}
\author[a]{Vladimir~M.~Braun,}
\author[a]{Meinulf~G{\"o}ckeler,}
\author[a]{Michael~Gruber,}
\author[a]{Fabian~Hutzler,}
\author[a]{Andreas~Sch{\"a}fer,}
\author[a]{Rainer~W.~Schiel,}
\author[a]{Jakob~Simeth,}
\author[a]{Wolfgang~S{\"o}ldner,}
\author[c]{Andre~Sternbeck}
\author[a]{and Philipp~Wein}
\emailAdd{michael1.gruber@physik.uni-regensburg.de}
\emailAdd{fabian.hutzler@physik.uni-regensburg.de}
\emailAdd{philipp.wein@physik.uni-regensburg.de}
\affiliation[a]{Institut f{\"u}r Theoretische Physik, Universit{\"a}t Regensburg,\\%
Universit{\"a}tsstra{\ss}e 31, D-93040 Regensburg, Germany}
\affiliation[b]{Department of Theoretical Physics, Tata Institute of Fundamental Research,\\%
Homi Bhabha Road, Mumbai 400005, India}
\affiliation[c]{Theoretisch-Physikalisches Institut, Friedrich-Schiller-Universit{\"a}t Jena,\\%
Max-Wien-Platz 1, D-07743 Jena, Germany}
\abstract{We present results of the first ab initio lattice QCD calculation of the normalization constants and first moments of the leading twist distribution amplitudes of the full baryon octet, corresponding to the small transverse distance limit of the associated $S$-wave light-cone wave functions. The $P$-wave (higher twist) normalization constants are evaluated as well. The calculation is done using $N_f=2+1$ flavors of dynamical (clover) fermions on lattices of different volumes and pion masses down to \unit{222}{\mega\electronvolt}. Significant \SU3 flavor symmetry violation effects in the shape of the distribution amplitudes are observed.}%
\arxivnumber{1512.02050}
\begin{document}
\maketitle%
\flushbottom%
\section{Introduction}%
Understanding the structure of matter in terms of quark and gluon degrees of freedom is the ultimate goal of the physics of strong interactions. For this purpose, the intuitive quantum-mechanical representation of a hadron as a superposition of Fock states with different numbers of partons in the infinite momentum frame (or using light-cone quantization) is very useful in order to develop the underlying physics picture. It also provides a good basis for theoretical modeling. Although a priori there is no reason to expect that, e.g., the nucleon wave function components with, say, 100 partons (quarks and gluons) are suppressed relative to those with only three valence quarks, the phenomenological success of quark models suggests that in many cases only the first few Fock components are important. Also the analysis of hard exclusive reactions involving large momentum transfer from the initial to the final state baryon within QCD perturbation theory~\cite{Lepage:1980fj,Efremov:1979qk,Chernyak:1983ej} suggests that such processes are dominated by the overlap of the valence light-cone wave functions at small transverse separations, usually referred to as light-cone distribution amplitudes (DAs).\par%
The DAs can be viewed as light-cone wave functions integrated over the quark transverse momenta~\cite{Lepage:1980fj}. They are fundamental nonperturbative functions that are complementary to conventional parton distributions, but are more elusive because their relation to experimental observables is less direct compared to quark parton densities. DAs are scale-dependent and for asymptotically large scales they are given by simple expressions, the so-called asymptotic DAs~\cite{Lepage:1980fj,Chernyak:1983ej}. There are many indications, however, that these asymptotic expressions poorly approximate the real DAs for the range of momentum transfers accessible in present experiments.\par%
The theoretical description of DAs is based on the relation of their moments, i.e., integrals over DAs weighted by powers of momentum fractions, to matrix elements of local operators. Such matrix elements can be estimated using nonperturbative techniques and the DAs can be reconstructed as an expansion in a suitable basis of polynomials in the momentum fractions.\par%
First estimates of the first and the second moments of the baryon DAs have been obtained more than 30 years ago using QCD sum rules (QCDSR)~\cite{Chernyak:1984bm,King:1986wi,Chernyak:1987nu,Chernyak:1987nv}. These early calculations suggested large deviations from the corresponding asymptotic values and were used extensively for model building of DAs~\cite{Chernyak:1984bm,King:1986wi,Chernyak:1987nu,Chernyak:1987nv,Stefanis:1992nw} that allowed for a reasonable description of the experimental data available at that time within a purely perturbative framework, see the review~\cite{Chernyak:1983ej}.\par%
Despite its phenomenological success, this approach has remained controversial over the years. In particular the QCD sum rules used to calculate the moments have been criticized as unreliable, see, e.g., ref.~\cite{Mikhailov:1986be}. Also, nowadays it is commonly accepted that perturbative contributions to hard exclusive reactions at accessible energy scales must be complemented by the so-called soft or end-point corrections. Estimates of the soft contributions using QCD sum rules~\cite{Radyushkin:1990te}, quark models~\cite{Bolz:1996sw} and light-cone sum rules~\cite{Braun:2001tj,Braun:2006hz,Anikin:2013aka} favor nucleon DAs that deviate only mildly from the asymptotic expressions.\par%
It has now become possible to calculate moments of the DAs from first principles using lattice QCD. The first quantitative results for the nucleon have been obtained by the QCDSF collaboration~\cite{Gockeler:2008xv,Braun:2008ur} using two flavors of dynamical (clover) fermions, followed by~\cite{Braun:2014wpa}, where a much larger set of lattices was used including ensembles at smaller pion masses, close to the physical point. The latter paper also contained an exploratory study of the DAs of negative parity nucleon resonances, see also ref.~\cite{Braun:2009jy}.\par%
In this work we extend the analysis of ref.~\cite{Braun:2014wpa} to the full $J^P = \frac12^+$ baryon octet. In addition to theoretical completeness, our study is motivated by applications to weak decays of heavy baryons such as $\Lambda_b$ and $\Lambda_c$. Such baryons are produced copiously at the LHC. As more data are collected, studies of rare $b$-baryon decays involving flavor-changing neutral currents offer interesting insights into the quark mixing matrix and, potentially, may reveal new physics beyond the Standard Model. In particular the $\Lambda_b \to \Lambda \mu^+\mu^-$ decays are receiving a lot of attention, see, e.g., refs.~\cite{Feldmann:2011xf,Wang:2015ndk} and references therein. $B$-meson decays into baryon-antibaryon pairs are also interesting. The pattern of \SU3 flavor symmetry breaking in weak decays is known in general to be nontrivial. In particular the large asymmetry observed in the decay $\Sigma^+\to p\gamma$ has been fueling a lot of discussions over many years and remains poorly understood at the parton level (see, e.g., ref.~\cite{Lach:1995we}). Another motivation comes from the emerging possibilities to study the transition form factors for the electroexcitation of nucleon resonances at large photon virtualities, planned for the JLAB \unit{12}{\giga\electronvolt} upgrade~\cite{Aznauryan:2012ba}, with the hope that similar transition form factors for hyperon production, e.g., in large-angle $\pi N$ scattering, will also become accessible in the future. This perspective already stimulated several theory studies, see, e.g., refs.~\cite{Liu:2010bh,Aliev:2014nla}.\par%
In this first study we will mainly address the development of the necessary formalism and methodical issues. Studies of hyperon DAs have a long history~\cite{Chernyak:1987nu}, however, we found that the definitions existing in the literature are not very convenient to study effects of \SU3 breaking and that the standard notation is, in part, contradictory. Therefore, we explain our notation and provide the necessary definitions in the introductory section~\ref{sect_bdas}. The physical interpretation of the DAs in terms of light-cone wave functions is considered as well. The related appendix~\ref{app_octet} explains the phase conventions for the flavor wave functions used in this work.\par%
Section~\ref{sect_lattice_formulation} is devoted to the lattice formulation of the problem at hand, the definition of correlation functions used in our analysis of the couplings and the first moments of baryon octet DAs, and the strategy to approach the physical limit of small pion mass. Calculations in this work are performed on a set of ensembles provided by the coordinated lattice simulations (CLS) effort~\cite{Bruno:2014jqa}. These are obtained using the tree-level Symanzik improved gauge action and $2+1$ dynamical Wilson (clover) quark flavors. In our calculation we start at the flavor symmetric point, where all quark masses are equal, and approach the real world in such a way that $u$ and $d$~quark masses decrease and simultaneously the $s$~quark mass increases so that the average mass is kept (approximately) constant~\cite{Bietenholz:2011qq,Bruno:2014jqa}.\par%
Section~\ref{sect_renormalization}, complemented by appendices~\ref{app_operator_relations} and~\ref{appendix_renormalization}, explains our renormalization procedure. We employ a nonperturbative method based on the well-known $\RI$ scheme, combined with matching factors calculated in continuum perturbation theory to convert our results to the $\MSbar$ scheme. The renormalization of flavor-octet operators turns out to be more complicated than the nucleon case and is discussed in some detail.\par%
Section~\ref{sect_chpt} contains a discussion of chiral extrapolation and \SU3 symmetry breaking in the framework of three-flavor baryon chiral perturbation theory (BChPT). Our presentation is based on the recent analysis in ref.~\cite{Wein:2015oqa}. One result is a simple relation between the DAs of the $\Sigma$ and $\Xi$ hyperons which has the same theory status as the famous Gell-Mann--Okubo sum rule for baryon masses and is satisfied to high accuracy for our lattice data.\par%
In section~\ref{sect_results} our final results are presented and compared with the existing lattice (for the nucleon) and QCD sum rule calculations. We find that deviations of the baryon DAs from their asymptotic form at hadronic scales are small, up to an order of magnitude smaller than in old QCD sum rule calculations. The \SU3 breaking in the corresponding shape parameters is, on the contrary, much larger than anticipated. Section~\ref{sect_conclusion} is reserved for a summary and conclusions.\par%
It has to be said that while our calculation provides the first qualitative insight into \SU3 breaking of octet baryon DAs from lattice QCD, it has not yet reached a quantitatively mature state. This is mainly due to the lack of a continuum extrapolation, which can have a significant impact on DAs (cf.\ ref.~\cite{Braun:2014wpa}). Actually, the whole CLS strategy to simulate with open boundary conditions is motivated by the fact that, at presently used lattice constants, discretization errors are significant. Moreover, on coarse lattices the lattice spacing will depend on the observable employed for scale setting. As our study is an exploratory one with significant systematic uncertainties, this fact is irrelevant in the present context. We use ensembles with the lattice spacing $a=\unit{0.0857(15)}{\femto\meter}$, which is determined from the Wilson flow method as described in ref.~\cite{Bruno:2014jqa}. The dimensionless flow time $t_0/a^2$ was extrapolated to the physical point and the lattice spacing was then assigned by matching to the continuum limit value $\sqrt{t_0}=\unit{0.1465(21)(13)}{\femto\meter}$ determined in ref.~\cite{Borsanyi:2012zs}. In the future we intend to include finer lattices, which are currently being generated within the CLS effort. This will then allow us to take the continuum limit and also eliminate scale setting ambiguities related to the nonzero lattice spacing.%
\section{Baryon distribution amplitudes\label{sect_bdas}}%
Baryon DAs~\cite{Lepage:1980fj,Efremov:1979qk,Chernyak:1983ej} are defined as matrix elements of renormalized three-quark operators at light-like separations:%
\begin{align}\label{eq_calB}%
 B_{\alpha\beta\gamma}^{fgh}(a_1,a_2,a_3;\mu) &= \langle 0 | \bigl[f_\alpha(a_1 n) g_\beta(a_2 n) h_\gamma(a_3 n)\bigr]^{\MSbar} | B(p,\lambda) \rangle\,,
\end{align}%
where $| B(p,\lambda) \rangle$ is the baryon state with momentum $p$ and helicity $\lambda$, while $\alpha,\beta,\gamma$ are Dirac indices, $n$ is a light-cone vector ($n^2=0$), the $a_i$ are real numbers, $\mu$ is the renormalization scale and $f,g,h$ are quark fields of the given flavor, chosen to match the valence quark content of the baryon $B$. The Wilson lines, which are needed for gauge invariance, as well as the color antisymmetrization, which is needed to form a color singlet, are not written out explicitly but always implied. Renormalization of three-quark operators using dimensional regularization and minimal subtraction requires some care; we will be using the renormalization scheme proposed in ref.~\cite{Kraenkl:2011qb}.\par%
Restricting ourselves to the analysis of the lowest $\frac12^+$ multiplet, neglecting electromagnetic interactions and assuming exact isospin symmetry ($m_l\equiv m_u=m_d$), it is sufficient to consider four cases:%
\begin{align}%
 B \, \in \, \{N\equiv p, \Sigma \equiv \Sigma^-, \Xi\equiv\Xi^0, \Lambda\} \,.
\end{align}%
For definiteness we choose the following flavor ordering:%
\begin{subequations}\label{eq_baryon_flavors}%
\begin{alignat}{2}%
 p:& &\quad\ (f,g,h)&=(\f{u},\f{u},\f{d})\,, \\*
 \Sigma^-:& & (f,g,h)&=(\f{d},\f{d},\f{s})\,, \\*
 \Xi^0:& & (f,g,h)&=(\f{s},\f{s},\f{u})\,, \\*
 \Lambda:& & (f,g,h)&=(\f{u},\f{d},\f{s})\,,
\end{alignat}%
\end{subequations}%
respectively. This choice is always implied so that in what follows we do not show flavor indices. Matrix elements for other baryons and/or with different flavor ordering can be obtained in a straightforward manner using isospin transformations.\par%
The general Lorentz decomposition of the matrix element~\eqref{eq_calB} consists of $24$ terms~\cite{Braun:2000kw} that are usually written in the form%
\begin{align}%
\label{eq_general_decomposition}
 B_{\alpha\beta\gamma}(a_1,a_2,a_3;\mu) &=
\sum_{\text{DA}} \bigl(\Gamma^{\text{DA}}\bigr)_{\alpha\beta} \bigl(\tilde \Gamma^{\text{DA}} u^B(p,\lambda)\bigr)_\gamma \int \! [dx] \; e^{-ip \cdot n \sum_i a_i x_i} \; \text{DA}^B(x_1,x_2,x_3;\mu)\,.
\end{align}%
Here $\Gamma^{\text{DA}}$ and $\tilde \Gamma^{\text{DA}}$ are the Dirac structures corresponding to the distribution amplitude $\text{DA}^B(x_i)$, see eq.~(2.9) of ref.~\cite{Braun:2000kw}, and $u^B(p,\lambda)$ is the Dirac spinor with on-shell momentum $p$ ($p^2=m_B^2$) and helicity $\lambda$. This decomposition can be organized in such a way that all DAs have definite collinear twist. The scale dependence will be suppressed from now on, unless it is explicitly needed. The variables $x_1,x_2,x_3$ are the momentum fractions carried by the quarks $f,g,h$, respectively, and the integration measure is defined as%
\begin{align}\label{eq_integrationmeasure}
 \int \! [dx] &= \int \limits_0^1 \!\!\!\!\! \int \limits_0^1 \!\!\!\!\! \int \limits_0^1 \!\! dx_1 dx_2 dx_3 \; \delta (1-x_1-x_2-x_3)\,.
\end{align}
The factor $\delta (1-x_1-x_2-x_3)$ enforces momentum conservation.%
\subsection{Leading twist distribution amplitudes}%
In this work we will mainly be concerned with the DAs of leading twist three. To this accuracy the general decomposition in eq.~\eqref{eq_general_decomposition} is simplified to three terms~\cite{Chernyak:1983ej}:%
\begin{align}\label{DAproton}%
\begin{split}
 \MoveEqLeft[4] 4 B_{\alpha\beta\gamma}(a_1,a_2,a_3) = \int \! [dx] \; e^{-ip \cdot n \sum_i a_i x_i} \\&\times\Bigl( v^B_{\alpha\beta;\gamma} V^B(x_1,x_2,x_3) + a^B_{\alpha\beta;\gamma} A^B(x_1,x_2,x_3) + t^B_{\alpha\beta;\gamma} T^B(x_1,x_2,x_3)+\ldots \Bigr) \,.
\end{split}
\end{align}%
Here%
\begin{subequations}%
\begin{align}%
 v^B_{\alpha\beta;\gamma} &= (\tilde{\vphantom{n}\smash{\slashed{n}}}C)_{\alpha\beta}(\gamma_5 u^B_+(p,\lambda))_\gamma\,, \\
 a^B_{\alpha\beta;\gamma} &= (\tilde{\vphantom{n}\smash{\slashed{n}}}\gamma_5C)_{\alpha\beta} (u^B_+(p,\lambda))_\gamma\,, \\
 t^B_{\alpha\beta;\gamma} &= (i\sigma_{\perp \tilde n} C)_{\alpha\beta}(\gamma^\perp\gamma_5 u^B_+(p,\lambda) )_\gamma\,,
\end{align}
\end{subequations}%
with the charge conjugation matrix $C$ and the notation%
\begin{subequations}%
\begin{align}%
 \tilde n_\mu &= p_\mu - \frac12 \frac{m^2_B}{p\cdot n} n_\mu \,, &
 u^B_+(p,\lambda) &= \frac12\frac{\tilde{\vphantom{n}\smash{\slashed{n}}}\slashed{n}}{\tilde n\cdot n} u^B(p,\lambda) \,, \\*
 \sigma_{\perp \tilde n}\otimes\gamma^\perp &= \sigma^{\mu\rho} \tilde n_\rho g^\perp_{\mu\nu} \otimes \gamma^\nu\,, &
 g_{\mu\nu}^\perp &= g_{\mu\nu} - \frac{\tilde n_\mu n_\nu+ \tilde n_\nu n_\mu}{\tilde n\cdot n} \,.
\end{align}%
\end{subequations}%
Our DAs $V^N$, $A^N$ and $T^N$ correspond to $V_1$, $A_1$ and $T_1$ in ref.~\cite{Braun:2000kw}.\par%
The equivalent definition in terms of the right- and left-handed components of the quark fields, $q^{\gooduparrow (\gooddownarrow)}=\frac12(\mathds 1 \pm \gamma_5) q$, is sometimes more convenient:%
\begin{subequations}%
\begin{align}%
\begin{split}\MoveEqLeft[4]
 \langle0|\bigl(f^{\gooduparrow T}(a_1n)C\slashed{n}g^\gooddownarrow(a_2n)\bigr)
\slashed{n}h^\gooduparrow(a_3n)|B(p,\lambda)\rangle =
\\&= -\tfrac12(p\cdot n)\slashed{n}u^{B\gooduparrow}(p,\lambda) \int \! [dx] \; e^{-ip \cdot n \sum_i a_i x_i} \; [V-A]^B(x_1,x_2,x_3) \, ,
\end{split}
\\
\begin{split}\MoveEqLeft[4]
 \langle0|\bigl(f^{\gooduparrow T}(a_1n)C\gamma^\mu\slashed{n}g^\gooduparrow(a_2n)\bigr)\gamma_\mu\slashed{n}h^\gooddownarrow(a_3n)|B(p,\lambda)\rangle =
\\&= 2(p\cdot n)\slashed{n}u^{B\gooduparrow}(p,\lambda) \int \! [dx] \; e^{-ip \cdot n \sum_i a_i x_i} \; T^B(x_1,x_2,x_3) \,,
\end{split}
\end{align}%
\end{subequations}%
where $u^{B\gooduparrow}(p,\lambda) = \frac12(\mathds 1 + \gamma_5) u^{B}(p,\lambda)$. In the nucleon case the combination $[V-A]^N$ appearing in the first of these equations is traditionally referred to as the leading twist nucleon DA $\Phi^N$. For the full octet we define%
\begin{subequations}%
\begin{align}%
 \Phi^{B\neq\Lambda}(x_1,x_2,x_3) &= [V-A]^B(x_1,x_2,x_3)\, , \\
 \Phi^{\Lambda}(x_1,x_2,x_3) &= -\sqrt{\tfrac23}\bigl\{[V-A]^\Lambda(x_1,x_2,x_3) -2 [V-A]^\Lambda(x_3,x_2,x_1) \bigr\}\, .
\end{align}%
\end{subequations}%
If $\Phi^B$ is given, the $V^B$ and $A^B$ components can be reconstructed due to their different symmetry properties under the exchange of the first and the second quark:%
\begin{subequations}\label{eq_symmetry_VAT}%
\begin{align}%
 V^{B\neq\Lambda}(x_2,x_1,x_3) &= + V^{B}(x_1,x_2,x_3) \,, & V^{\Lambda}(x_2,x_1,x_3) &= - V^{\Lambda}(x_1,x_2,x_3) \,, \\
 A^{B\neq\Lambda}(x_2,x_1,x_3) &= - A^{B}(x_1,x_2,x_3) \,, & A^{\Lambda}(x_2,x_1,x_3) &= + A^{\Lambda}(x_1,x_2,x_3) \,, \\
 T^{B\neq\Lambda}(x_2,x_1,x_3) &= + T^{B}(x_1,x_2,x_3) \,, & T^{\Lambda}(x_2,x_1,x_3) &= - T^{\Lambda}(x_1,x_2,x_3) \,.
\end{align}%
\end{subequations}%
Using isospin symmetry one can further show for the nucleon%
\begin{align}\label{eq_isospin_relation_nucleon}%
 T^N (x_1,x_3,x_2) &= \frac12\bigl[\Phi^N(x_1,x_2,x_3)+ \Phi^N(x_3,x_2,x_1)\bigr]\,,
\end{align}%
so that, to leading twist accuracy, $\Phi^N$ contains all necessary information. For other baryons this relation does not hold, so that the functions $T^B$ are independent of $[V-A]^B$.\par%
To fully exploit the benefits of \SU3 flavor symmetry it proves convenient to define the following set of DAs:%
\begin{subequations} \label{eq_convenient_DAs}%
\begin{align}%
 \Phi_{\pm}^{B\neq\Lambda}(x_1,x_2,x_3)&=\tfrac{1}{2} \bigl\{[ V-A ]^B(x_1,x_2,x_3) \pm [ V-A ]^B(x_3,x_2,x_1) \bigr\} \,,
\\*
 \Pi^{B\neq\Lambda}(x_1,x_2,x_3)&= T^B (x_1,x_3,x_2) \,,
\\*
 \Phi_{+}^{\Lambda}(x_1,x_2,x_3)&=\sqrt{\tfrac{1}{6}} \bigl\{[ V-A ]^\Lambda(x_1,x_2,x_3) + [ V-A ]^\Lambda(x_3,x_2,x_1) \bigr\} \,,
\\*
 \Phi_{-}^{\Lambda}(x_1,x_2,x_3)&=-\sqrt{\tfrac{3}{2}} \bigl\{[ V-A ]^\Lambda(x_1,x_2,x_3) - [ V-A ]^\Lambda(x_3,x_2,x_1) \bigr\} \,,
\\*
 \Pi^{\Lambda}(x_1,x_2,x_3)&=\sqrt{6} \; T^\Lambda (x_1,x_3,x_2) \,,
\end{align}%
\end{subequations}%
where for the nucleon $\Pi^N=\Phi_+^N$ up to isospin breaking effects. In the limit of \SU3 flavor symmetry, where $m_u=m_d=m_s$ (and in particular at the flavor symmetric point with physical average quark mass indicated by $\star$), the following relations hold:\footnote{Our phase conventions for the baryon states and the corresponding flavor wave functions are detailed in appendix~\ref{app_octet}.}%
\begin{subequations} \label{eq_su3_relations}%
\begin{align}%
 \Phi_{+}^\star &\equiv \Phi_{+}^{N\star} = \Phi_{+}^{\Sigma\star} = \Phi_{+}^{\Xi\star} = \Phi_{+}^{\Lambda\star} = \Pi^{N\star} = \Pi^{\Sigma\star} = \Pi^{\Xi\star} \,,
\\*
 \Phi_{-}^\star &\equiv \Phi_{-}^{N\star} = \Phi_{-}^{\Sigma\star} = \Phi_{-}^{\Xi\star} = \Phi_{-}^{\Lambda\star} = \Pi^{\Lambda\star}\, .
\end{align}%
\end{subequations}%
Therefore, the amplitudes $\Pi^B$ (or $T^B$) only need to be considered when flavor symmetry is broken. In the flavor symmetric limit $\Phi_{+}^\star$ and $\Phi_{-}^\star$ can again be combined to form a single distribution amplitude $\Phi^\star = \Phi_{+}^\star + \Phi_{-}^\star$. One can show (see the detailed discussion in section~\ref{sect_chpt}) that to first order in the symmetry breaking parameter, $m^2_K-m^2_\pi \propto m_s-m_l$, the following relation holds:%
\begin{align}%
 \Phi_+^\Sigma - \Pi^\Sigma &= \Pi^\Xi - \Phi_+^\Xi\,.
\end{align}\par%
To understand the physical meaning of the DAs it is instructive to work out their relation to light-front wave functions. The leading twist approximation corresponds to taking into account $S$-wave contributions in which case the helicities of the quarks sum up to the helicity of the baryon (cf. refs.~\cite{Ji:2002xn,Belitsky:2005qn}). Suppressing the transverse momentum dependence one finds%
\begin{align}\label{eq_wavefunction_not_lambda}%
\begin{split}
  | (B \neq \Lambda)^\gooduparrow \rangle &= \int \! \frac{[dx]}{8\sqrt{6x_1x_2x_3}} |f g h \rangle \otimes \bigl\{ \!\begin{aligned}[t]&[V+A]^B(x_1,x_2,x_3) \lvert \downarrow \uparrow \uparrow \rangle + [V-A]^B(x_1,x_2,x_3) \lvert \uparrow \downarrow \uparrow \rangle\\ &\mathord- 2 T^B(x_1,x_2,x_3) \lvert \uparrow \uparrow \downarrow \rangle \bigr\} \end{aligned} \\
&= \int \! \frac{[dx]}{8\sqrt{3x_1x_2x_3}} \lvert\uparrow \uparrow \downarrow \rangle \otimes \bigl\{ \!\begin{aligned}[t]&\mathord- \sqrt3\Phi_+^B(x_1,x_3,x_2) \bigl( | \text{MS},B \rangle - \sqrt{2}| \text{S},B \rangle \bigr)/3 \\ &\mathord-\sqrt3\Pi^B(x_1,x_3,x_2) \bigl( 2 | \text{MS},B \rangle + \sqrt2| \text{S},B \rangle \bigr)/3 \\&\mathord+ \Phi_-^B(x_1,x_3,x_2) | \text{MA},B \rangle \bigr\} \,,\end{aligned}
\end{split}\raisetag{2.02cm}%
\end{align}%
and%
\begin{align}\label{eq_wavefunction_lambda}%
\begin{split}
 | \Lambda^\gooduparrow \rangle &=
\int \! \frac{[dx]}{4\sqrt{6x_1x_2x_3}} |u d s \rangle \otimes
\bigl\{ \!\begin{aligned}[t]&[V+A]^\Lambda(x_1,x_2,x_3) \lvert \downarrow \uparrow \uparrow \rangle + [V-A]^\Lambda(x_1,x_2,x_3) \lvert \uparrow \downarrow \uparrow \rangle\\ &\mathord- 2 T^\Lambda(x_1,x_2,x_3) \lvert \uparrow \uparrow \downarrow \rangle \bigr\} \end{aligned} \\
  &= \int \! \frac{[dx]}{8\sqrt{3x_1x_2x_3}} \lvert\uparrow \uparrow \downarrow \rangle \otimes \bigl\{ \!\begin{aligned}[t] &\mathord-\sqrt3 \Phi_+^\Lambda(x_1,x_3,x_2)| \text{MS},\Lambda \rangle \\ &\mathord+ \Pi^\Lambda(x_1,x_3,x_2) \bigl( 2 |\text{MA},\Lambda\rangle + \sqrt2 | \text{A}, \Lambda \rangle \bigr)/3 \\&\mathord+ \Phi_-^\Lambda(x_1,x_3,x_2) \bigl( |\text{MA},\Lambda\rangle - \sqrt2 | \text{A}, \Lambda \rangle \bigr)/3 \bigr\} \,,\end{aligned}
\end{split}
\end{align}%
where $\lvert \uparrow \downarrow \uparrow \rangle$ etc.\ show quark helicities and $|f g h \rangle$ stands for the flavor ordering as specified in eq.~\eqref{eq_baryon_flavors}. $|\text{MS},B \rangle$ and $|\text{MA},B \rangle$ are the usual mixed-symmetric and mixed-antisymmetric octet flavor wave functions, respectively (see tables~\ref{table_flavor_octet_1} and~\ref{table_flavor_octet_2} of appendix~\ref{app_octet}). $|\text{A},\Lambda \rangle$ and $|\text{S},B\neq\Lambda \rangle$ are totally antisymmetric and symmetric flavor wave functions (see tables~\ref{table_flavor_singlet} and~\ref{table_flavor_decuplet}), which only occur in the octet if \SU3 symmetry is broken. From this representation it becomes obvious that $V^B$, $A^B$ and $T^B$ are convenient DAs if one sorts the quarks with respect to their flavor, while $\Phi^B_+$, $\Phi^B_-$ and $\Pi^B$ correspond to three distinct flavor structures in a helicity-ordered wave function. At the flavor symmetric point $\Phi_+^\star$ and $\Phi_-^\star$ isolate the mixed-symmetric and mixed-antisymmetric flavor wave functions:%
\begin{align}%
  | B^\gooduparrow \rangle^\star &= \int \! \frac{[dx]}{8\sqrt{3x_1x_2x_3}} \lvert\uparrow \uparrow \downarrow \rangle \otimes \bigl\{ - \sqrt{3} \Phi_+^\star(x_1,x_3,x_2) |\text{MS},B \rangle + \Phi_-^\star(x_1,x_3,x_2) |\text{MA},B \rangle \bigr\} \,.
\end{align}\par%
DAs can be expanded in a set of orthogonal polynomials (conformal partial wave expansion) in such a way that the coefficients have autonomous scale dependence at one loop. The first few polynomials are (see, e.g., ref.~\cite{Braun:2008ia})%
\begin{subequations}%
\begin{align}%
\mathcal{P}_{00} &= 1 \,, &
\mathcal{P}_{20} &= \tfrac{63}{10}[3 (x_1-x_3)^2 -3x_2(x_1+x_3)+ 2x_2^2]\,, \\
\mathcal{P}_{10} &= 21(x_1-x_3)\,, &
\mathcal{P}_{21} & = \tfrac{63}{2}(x_1-3x_2 + x_3)(x_1-x_3)\,, \\
\mathcal{P}_{11} &= 7(x_1-2x_2 + x_3)\,, &
\mathcal{P}_{22} & = \tfrac{9}{5} [x_1^2+9 x_2(x_1+x_3)-12x_1x_3-6x_2^2+x_3^2] \,.
\end{align}%
\end{subequations}%
Note that all $\mathcal{P}_{nk}$ have definite symmetry (being symmetric or antisymmetric) under the exchange of $x_1$ and $x_3$. Taking into account the corresponding symmetry of the DAs, defined in eq.~\eqref{eq_convenient_DAs}, a generic expansion reads%
\begin{subequations}\label{eq_da_parametrization}%
\begin{align}%
 \Phi_{+}^{B} &= 120 x_1 x_2 x_3 \bigl( \varphi^B_{00} \mathcal P_{00} + \varphi^B_{11} \mathcal P_{11} + \varphi^B_{20} \mathcal P_{20} + \varphi^B_{22} \mathcal P_{22} + \dots \bigr) \, , \\
 \Phi_{-}^{B} &= 120 x_1 x_2 x_3 \bigl( \varphi^B_{10} \mathcal P_{10} + \varphi^B_{21} \mathcal P_{21} + \dots \bigr) \,, \\
 \Pi^{B\neq\Lambda} &= 120 x_1 x_2 x_3 \bigl( \pi^B_{00} \mathcal P_{00} + \pi^B_{11} \mathcal P_{11} + \pi^B_{20} \mathcal P_{20} + \pi^B_{22} \mathcal P_{22} + \dots \bigr) \, , \\
 \Pi^{\Lambda} &= 120 x_1 x_2 x_3 \bigl( \pi^{\Lambda}_{10} \mathcal P_{10} + \pi^{\Lambda}_{21} \mathcal P_{21} + \dots \bigr) \,.
\label{eq_da_parametrization_pi_lambda}
\end{align}%
\end{subequations}%
In this way all nonperturbative information is encoded in the set of (scale-dependent) coefficients $\varphi^B_{nk}$, $\pi^B_{nk}$, which can be related to matrix elements of local operators. In each DA only polynomials of one type, either symmetric or antisymmetric under exchange of $x_1$ and $x_3$, appear.\par%
The leading contributions $120 x_1 x_2 x_3 \varphi^B_{00}$ and $120 x_1 x_2 x_3 \pi^{B\neq\Lambda}_{00}$ are usually referred to as the asymptotic DAs. The corresponding normalization coefficients $\varphi^B_{00}$ and $\pi^{B\neq\Lambda}_{00}$ can be thought of as the wave functions at the origin (in position space). In what follows we will use the notation%
\begin{align}%
  f^B &= \varphi^B_{00}\,, & f^{B\neq\Lambda}_T &= \pi^B_{00}\,.
\end{align}%
Note that for the nucleon the two couplings coincide, $f^N_T = f^N$. For the $\Lambda$ baryon the zeroth moment of $T^\Lambda$ vanishes. The higher-order coefficients are usually referred to as shape parameters. Note that, in contrast to ref.~\cite{Braun:2014wpa}, we do not separate the couplings $f^B$ and $f^{B\neq\Lambda}_T$ as overall normalization factors, so that our $\varphi^N_{nk}$ correspond to $f_N \varphi^N_{nk}$ of~\cite{Braun:2014wpa}. The one-loop scale evolution of the couplings and shape parameters is given by%
\begin{align}\label{eq_oneloop_evolution}%
 \varphi^B_{nk}(\mu)&=\varphi^B_{nk}(\mu_0) \biggl( \frac{\alpha_s(\mu)}{\alpha_s(\mu_0)} \biggr)^{\gamma_{nk}/\beta_0} \,, &
 \pi^B_{nk}(\mu)&=\pi^B_{nk}(\mu_0) \biggl( \frac{\alpha_s(\mu)}{\alpha_s(\mu_0)} \biggr)^{\gamma_{nk}/\beta_0} \,,
\end{align}%
where $\beta_0=11-2 N_f/3$ is the first coefficient of the QCD $\beta$-function. In this work we restrict ourselves to the contributions of first order polynomials $\mathcal P_{10}$, $\mathcal P_{11}$ and omit all higher terms. The relevant one-loop anomalous dimensions are%
\begin{align}%
\gamma_{00} &= \frac{2}{3}  \,, &
\gamma_{11} &= \frac{10}{3} \,, &
\gamma_{10} &= \frac{26}{9} \,.
\end{align}%
The scale dependence of $f^B$ and $f^{B\neq\Lambda}_T$ is identical and is known up to three-loop order, see refs.~\cite{Kraenkl:2011qb,Gracey:2012gx}.%
\subsection{Higher twist contributions}%
The general decomposition~\eqref{eq_general_decomposition} contains $21$ DAs of higher twist, which altogether involve only up to three new normalization constants (just two for $N$, $\Sigma$ and $\Xi$), for details see refs.~\cite{Braun:2000kw,Wein:2015oqa}. They can be defined as matrix elements of local three-quark twist four operators without derivatives. These twist four couplings are also interesting in a broader context, e.g., in studies of baryon decays in generic GUT models~\cite{Claudson:1981gh}, and as input parameters for QCD sum rule calculations, see, e.g., refs.~\cite{Belyaev:1982cd,Ioffe:1983ty,Aliev:2014nla}.\par%
We use the following definitions:%
\begin{subequations}\label{eq_higher_twist_normalization_constants_definition_non_singlet}%
\begin{align}%
 \langle 0 | \bigl(f^{\gooduparrow T}(0) C \gamma^\mu g^\gooddownarrow(0)\bigr) \gamma_\mu h^\gooduparrow(0) | (B \neq \Lambda) (p,\lambda) \rangle
&=
- \tfrac 12 \lambda_1^B m_B u^{B\gooddownarrow}(p,\lambda) \,,
\\
\langle 0 | \bigl(f^{\gooduparrow T}(0) C \sigma^{\mu\nu} g^\gooduparrow(0)\bigr) \sigma_{\mu\nu} h^\gooduparrow(0) | (B \neq \Lambda) (p,\lambda) \rangle
&= \lambda_2^B m_B u^{B\gooduparrow}(p,\lambda) \,,
\end{align}%
\end{subequations}%
for the isospin-nonsinglet baryons ($N$, $\Sigma$, $\Xi$) and%
\begin{subequations}\label{eq_higher_twist_normalization_constants_definition_singlet}%
\begin{align}%
 \langle 0 | \bigl(u^{\gooduparrow T}(0) C \gamma^\mu d^\gooddownarrow(0)\bigr) \gamma_\mu s^\gooduparrow(0) | \Lambda (p,\lambda) \rangle
&=
 \tfrac{1}{2\sqrt{6}} \lambda_1^\Lambda m_\Lambda u^{\Lambda\gooddownarrow}(p,\lambda) \,,
\\
 \langle 0 | \bigl(u^{\gooduparrow T}(0) C d^\gooduparrow(0)\bigr) s^\gooddownarrow(0) | \Lambda (p,\lambda) \rangle
&=
\tfrac{1}{2\sqrt{6}} \lambda_T^\Lambda m_\Lambda u^{\Lambda\gooddownarrow}(p,\lambda) \,,
\\
 \langle 0 | \bigl(u^{\gooduparrow T}(0) C d^\gooduparrow(0)\bigr) s^\gooduparrow(0) | \Lambda (p,\lambda) \rangle
&= \tfrac{-1}{4\sqrt{6}} \lambda_2^\Lambda m_\Lambda u^{\Lambda\gooduparrow}(p,\lambda) \,,
\end{align}%
\end{subequations}%
for the $\Lambda$ baryon. The definitions are chosen such that at the flavor symmetric point%
\begin{subequations}%
\begin{align}%
\lambda_1^\star &\equiv \lambda_1^{N\star} = \lambda_1^{\Sigma\star} = \lambda_1^{\Xi\star} = \lambda_1^{\Lambda\star} = \lambda_T^{\Lambda\star} \, , \\*
\lambda_2^\star &\equiv \lambda_2^{N\star} = \lambda_2^{\Sigma\star} = \lambda_2^{\Xi\star} = \lambda_2^{\Lambda\star} \, ,
\end{align}%
\end{subequations}%
cf.\ ref.~\cite{Wein:2015oqa}. For the nucleon the definitions in terms of chiral fields in eq.~\eqref{eq_higher_twist_normalization_constants_definition_non_singlet} are equivalent to the traditional definitions of $\lambda_1^N$ and $\lambda_2^N$ not involving chiral projections, as used in ref.~\cite{Braun:2014wpa}. Analogous definitions can also be given for the $\Lambda$ baryon:%
\begin{subequations}%
\begin{align}%
 \langle 0 | \bigl(u^T(0) C \gamma^\mu\gamma_5 d(0)\bigr) \gamma_\mu s(0) | \Lambda (p,\lambda) \rangle
&= \tfrac{-1}{\sqrt{6}} \lambda_1^\Lambda m_\Lambda u^\Lambda(p,\lambda) \,,
\\
  \langle 0 | \bigl(u^T(0) C d(0)\bigr) \gamma_5 s(0) | \Lambda (p,\lambda) \rangle
&= \tfrac{-1}{4\sqrt{6}} (\lambda_2^\Lambda+2\lambda_T^\Lambda) m_\Lambda u^\Lambda(p,\lambda) \,,
\label{eq_lambdaLambda2}\\
 \langle 0 | \bigl(u^T(0) C \gamma_5 d(0)\bigr) s(0) | \Lambda (p,\lambda) \rangle
&= \tfrac{-1}{4\sqrt{6}} (\lambda_2^\Lambda-2\lambda_T^\Lambda) m_\Lambda u^\Lambda(p,\lambda) \,.
\end{align}%
\end{subequations}\par%
The one-loop evolution for all twist four normalization constants is the same:%
\begin{align}%
 \lambda^B_{1,2,T}(\mu)&=\lambda^B_{1,2,T}(\mu_0) \biggl( \frac{\alpha_s(\mu)}{\alpha_s(\mu_0)} \biggr)^{-2/\beta_0} \, .
\end{align}%
The corresponding anomalous dimensions are known up to three-loop accuracy~\cite{Kraenkl:2011qb,Gracey:2012gx}. The scale dependence of the couplings $\lambda^B_1$ and $\lambda^\Lambda_T$ is the same to all orders, whereas for $\lambda^B_2$ it differs starting from the second loop.%
\section{Lattice formulation\label{sect_lattice_formulation}}%
In Euclidean spacetime a direct calculation of DAs is not possible, since this would require quark fields at light-like separations. However, lattice QCD allows us to access moments of the DAs, e.g.,%
\begin{align}%
 V^B_{lmn} &= \int \! [dx] \; x_1^l x_2^m x_3^n V^B(x_1,x_2,x_3) \,,
\end{align}%
and similarly for the other functions. They are related to matrix elements of local three-quark operators, whose general form reads%
\begin{align}\label{eq_local_lattice_operators}%
 \mathcal X_{\bar r \bar l \bar m \bar n}^{B,lmn} &= \epsilon^{ijk} \Bigl( \bigl[ i^l D_{\bar l} f^T\!(0) \bigr]^i C \Gamma_{\bar r}^{\mathcal X_{\bar r}} \bigl[ i^m D_{\bar m} g(0) \bigr]^j \Bigr) \tilde \Gamma^{\mathcal X_{\bar r}} \bigl[ i^n D_{\bar n} h(0) \bigr]^k \,.
\end{align}%
Here we use a multi-index notation for the covariant derivatives, $D_{\bar l} \equiv D_{\lambda_1} \cdots D_{\lambda_l}$. The Dirac structures that we consider, $\Gamma_{\bar r}^{\mathcal X_{\bar r}}$ and $\tilde \Gamma^{\mathcal X_{\bar r}}$, are listed in table~\ref{table_operators}.\footnote{Starting from this section all equations refer to Euclidean spacetime; we use the gamma matrix convention of~\cite{Braun:2008ur}.} %
\begin{table}[t]%
\centering%
\caption{\label{table_operators}Definition of the Dirac matrix structures that appear in the local operators which are used in the lattice calculation, see eq.~\eqref{eq_local_lattice_operators}. Lorentz indices appearing in both $\Gamma_{\bar r}^{\mathcal X_{\bar r}}$ and $\tilde \Gamma^{\mathcal X_{\bar r}}$ are summed over implicitly.}%
\begin{tabular}{lc@{}c@{}c@{}c@{}c@{}c@{}c@{}c}%
  \toprule
  $\mathcal X_{\bar r}$ & $\hspace{1.6em}\mathclap{\mathcal S}\hspace{1.6em}$ & $\hspace{1.6em}\mathclap{\mathcal P}\hspace{1.6em}$ & $\hspace{1.6em}\mathclap{\mathcal V}\hspace{1.6em}$ & $\hspace{1.6em}\mathclap{\mathcal A}\hspace{1.6em}$ & $\hspace{1.6em}\mathclap{\mathcal T}\hspace{1.6em}$ & $\hspace{1.6em}\mathclap{\mathcal V_{\rho}}\hspace{1.6em}$ & $\hspace{1.6em}\mathclap{\mathcal A_{\rho}}\hspace{1.6em}$ & $\hspace{1.6em}\mathclap{\mathcal T_{\rho}}\hspace{1.6em}$ \\
  \midrule
  $\Gamma_{\bar r}^{\mathcal X_{\bar r}}$ & $\mathds 1$ & $\gamma_5$& $ \gamma_\eta$ & $ \gamma_\eta\gamma_5$ & $ \sigma_{\eta_1\eta_2}$ & $ \gamma_{\rho}$& $ \gamma_{\rho} \gamma_5$ & $ i \sigma_{\rho\eta}$ \\
  $\tilde \Gamma^{\mathcal X_{\bar r}}$ & $\gamma_5$& $\mathds 1$ & $\gamma_\eta \gamma_5$ & $\gamma_\eta$ & $ \sigma_{\eta_1\eta_2} \gamma_5$ & $\gamma_5$ & $\mathds 1$ & $\gamma_\eta \gamma_5$ \\
  \bottomrule
\end{tabular}%
\end{table}%
As sources for the baryon fields we have used the interpolating currents%
\begin{subequations}%
\begin{align}%
\mathcal N^N &= \bigl( u^T C \gamma_5 d \bigr) u \,, \\
\mathcal N^{\Sigma} &= \bigl( d^T C \gamma_5 s \bigr) d \,, \\
\mathcal N^{\Xi} &= \bigl( s^T C \gamma_5 u \bigr) s \,, \\
\mathcal N^{\Lambda} &= \tfrac{1}{\sqrt{6}} \bigl( 2 \bigl( u^T C \gamma_5 d \bigr) s + \bigl( u^T C \gamma_5 s \bigr) d + \bigl( s^T C \gamma_5 d \bigr) u \bigr) \,,
\end{align}%
\end{subequations}%
with an optimized number of smearing steps in the quark sources to suppress excited state contributions. The other baryons can then be obtained by means of isospin symmetry.%
\subsection{Correlation functions\label{sect_correlation_functions}}%
Moments of baryon DAs can be extracted from the ground state contribution to the two-point correlation functions. Neglecting the exponentially suppressed excited states the correlation functions can be written as%
\begin{align}%
 \langle \mathcal O_\tau(t,\mathbf p) \bar{\mathcal N}^B_{\tau^\prime}(0,\mathbf p) \rangle &= \frac{\sqrt{Z_B}}{2 E_B} \sum_\lambda \langle 0 | \mathcal O_\tau(0) | B(\mathbf p,\lambda) \rangle \; \bar u^B_{\tau^\prime}(\mathbf p,\lambda) e^{- E_B t} \,,
\end{align}%
with the energy $E_B=E_B(\mathbf p)=\sqrt{\smash{m_B^2}+\mathbf p^2}$, where we assume the continuum dispersion relation. The momentum-dependent coupling $Z_B=Z_B(\mathbf p)$ describes the overlap between the smeared source operator and the physical baryon ground state and can be obtained from the correlator%
\begin{align}\label{eq_smeared_smeard_correlation_function}%
 \langle \mathcal N^B_\tau(t,\mathbf p) \bar{\mathcal N}^B_{\tau^\prime}(0,\mathbf p) (\gamma _+)_{\tau^\prime \tau} \rangle &= Z_B \frac{m_B+k E_B}{E_B} e^{- E_B t} \,,
\end{align}%
where $\gamma_+=(\mathds 1+k \gamma_4)/2$ with $k=m_{B^*}/E_{B^*}$ suppresses the negative parity contribution~\cite{Leinweber:2004it,Braun:2014wpa}.\footnote{$B^*$ denotes the negative parity partner of the baryon $B$.}%
\subsubsection{Leading twist -- zeroth moments}%
In order to extract the leading twist normalization constants, the following linear combinations of operators are constructed such that their matrix elements do not contain any contributions of higher twist:%
\begin{subequations}\label{eq_operators_zeroth_moments}%
\begin{align}%
\mathcal O^{B,000}_{\mathcal X, \mathfrak{A}} &= - \gamma_1 \mathcal X^{B,000}_1 + \gamma_2 \mathcal X^{B,000}_2 \,, \\
\mathcal O^{B,000}_{\mathcal X, \mathfrak{B}} &= - \gamma_3 \mathcal X^{B,000}_3 + \gamma_4 \mathcal X^{B,000}_4 \,, \\
\mathcal O^{B,000}_{\mathcal X, \mathfrak{C}} &= - \gamma_1 \mathcal X^{B,000}_1 - \gamma_2 \mathcal X^{B,000}_2 + \gamma_3 \mathcal X^{B,000}_3 + \gamma_4 \mathcal X^{B,000}_4 \,,
\end{align}%
\end{subequations}%
where $\mathcal X$ can be $\mathcal V$, $\mathcal A$ or $\mathcal T$. The leading twist baryon couplings can be determined from the following correlation functions:%
\begin{subequations}\label{eq_correlators_zeroth_moments}%
\begin{align}%
\begin{split}
 C_{\mathcal X,\mathfrak{A}}^{B,000} &=\langle \bigl( \gamma _4 \mathcal O_{\mathcal X, \mathfrak{A}}^{B,000} (t,\mathbf p)\bigr)_\tau \bar{\mathcal N}^B_{\tau^\prime} (0,\mathbf p) (\gamma _+)_{\tau^\prime \tau} \rangle \\
 &= c_X X^B_{000} \sqrt{Z_B} \frac{k (p_1^2 - p_2^2)}{E_B} e^{-E_Bt} \,,
\end{split}\\
\begin{split}
 C_{\mathcal X,\mathfrak{B}}^{B,000} &=\langle \bigl( \gamma _4 \mathcal O_{\mathcal X, \mathfrak{B}}^{B,000} (t,\mathbf p)\bigr)_\tau \bar{\mathcal N}^B_{\tau^\prime} (0,\mathbf p) (\gamma _+)_{\tau^\prime \tau} \rangle \\
 &= c_X X^B_{000} \sqrt{Z_B} \frac{E_B(m_B+ k E_B) + k p_3^2}{E_B} e^{-E_Bt} \,,
\end{split}\\
\begin{split}
 C_{\mathcal X, \mathfrak{C}}^{B,000} &=\langle \bigl( \gamma _4 \mathcal O_{ \mathcal X,\mathfrak{C}}^{B,000} (t,\mathbf p)\bigr)_\tau \bar{\mathcal N}^B_{\tau^\prime} (0,\mathbf p) (\gamma _+)_{\tau^\prime \tau} \rangle \\
 &= c_X X^B_{000} \sqrt{Z_B} \frac{E_B(m_B+ k E_B) + k (p_1^2+p_2^2-p_3^2)}{E_B} e^{-E_Bt} \,,
\end{split}
\end{align}%
\end{subequations}%
where $c_V=c_A=1$ and $c_T=-2$. Again, $\mathcal X$ can be $\mathcal V$, $\mathcal A$ or $\mathcal T$. In practice we only consider the zero momentum correlators $C_{\mathcal X,\mathfrak{B}}^{B,000}$ and $C_{\mathcal X,\mathfrak{C}}^{B,000}$ as they are less noisy and, therefore, can be measured with higher accuracy. The couplings of interest are related to the calculated zeroth moments as follows:%
\begin{align}%
 f^{B\neq\Lambda} &\equiv \varphi^B_{00}=V^{B}_{000} \,, &
 f^\Lambda &\equiv \varphi^\Lambda_{00} = -\sqrt{\tfrac{2}{3}} A^{\Lambda}_{000} \,, &
 f_T^{B\neq\Lambda} &\equiv \pi^B_{00}=T^{B}_{000} \,,
\end{align}%
where $f^N_T = f^N$ due to isospin symmetry. The remaining zeroth moments of the leading twist DAs $V^B$, $A^B$ and $T^B$ vanish:%
\begin{align}%
 V^{\Lambda}_{000}&=A^{B\neq\Lambda}_{000}=T^{\Lambda}_{000}=0 \,.
\end{align}%
\subsubsection{Leading twist -- first moments}%
First moments of DAs can be calculated utilizing operators containing one covariant derivative. For $l+m+n=1$ we define the leading twist combinations%
\begin{subequations}\label{eq_operators_first_moments}%
\begin{align}%
 \mathcal{O}^{B,lmn}_{\mathcal X,\mathfrak{A}} &=
 + \gamma_1 \gamma_3 \mathcal{X}^{B,lmn}_{\{13\}} + \gamma_1 \gamma_4 \mathcal{X}^{B,lmn}_{\{14\}}
 - \gamma_2 \gamma_3 \mathcal{X}^{B,lmn}_{\{23\}} - \gamma_2 \gamma_4 \mathcal{X}^{B,lmn}_{\{24\}} - 2 \gamma_1 \gamma_2 \mathcal{X}^{B,lmn}_{\{12\}} \,, \\
 \mathcal{O}^{B,lmn}_{\mathcal X,\mathfrak{B}} &=
 + \gamma_1 \gamma_3 \mathcal{X}^{B,lmn}_{\{13\}} - \gamma_1 \gamma_4 \mathcal{X}^{B,lmn}_{\{14\}}
 + \gamma_2 \gamma_3 \mathcal{X}^{B,lmn}_{\{23\}} - \gamma_2 \gamma_4 \mathcal{X}^{B,lmn}_{\{24\}} + 2 \gamma_3 \gamma_4 \mathcal{X}^{B,lmn}_{\{34\}} \,, \\
 \mathcal{O}^{B,lmn}_{\mathcal X,\mathfrak{C}} &=
 - \gamma_1 \gamma_3 \mathcal{X}^{B,lmn}_{\{13\}} + \gamma_1 \gamma_4 \mathcal{X}^{B,lmn}_{\{14\}}
 + \gamma_2 \gamma_3 \mathcal{X}^{B,lmn}_{\{23\}} - \gamma_2 \gamma_4 \mathcal{X}^{B,lmn}_{\{24\}} \,,
\end{align}%
\end{subequations}%
where the braces indicate symmetrization. For the calculation of the first moments of the leading twist DAs one can use the correlation functions ($l+m+n=1$)%
\begin{subequations}%
\begin{align}%
\begin{split}
 C_{\mathcal X,\mathfrak{A},1}^{B,lmn} &=\langle \bigl( \gamma _4 \gamma_1 \mathcal O_{\mathcal X,\mathfrak{A}}^{B,lmn} (t,\mathbf p)\bigr)_\tau \bar{\mathcal N}^B_{\tau^\prime} (0,\mathbf p) (\gamma_+)_{\tau^\prime \tau} \rangle \\
 &= - c_X X^B_{lmn} \sqrt{Z_B} p_1 \frac{E_B(m_B+ k E_B) + k (2 p_2^2 - p_3^2)}{E_B} e^{-E_Bt} \,,
\end{split}\\
\begin{split}
 C_{\mathcal X,\mathfrak{A},2}^{B,lmn} &=\langle \bigl( \gamma _4 \gamma_2 \mathcal O_{ \mathcal X,\mathfrak{A}}^{B,lmn} (t,\mathbf p)\bigr)_\tau \bar{\mathcal N}^B_{\tau^\prime} (0,\mathbf p) (\gamma_+)_{\tau^\prime \tau} \rangle \\
 &= + c_X X^B_{lmn} \sqrt{Z_B} p_2 \frac{E_B(m_B+ k E_B) + k (2 p_1^2 - p_3^2)}{E_B} e^{-E_Bt} \,,
\end{split}\\
\begin{split}
 C_{\mathcal X,\mathfrak{A},3}^{B,lmn} &=\langle \bigl( \gamma _4 \gamma_3 \mathcal O_{\mathcal X,\mathfrak{A}}^{B,lmn} (t,\mathbf p)\bigr)_\tau \bar{\mathcal N}^B_{\tau^\prime} (0,\mathbf p) (\gamma_+)_{\tau^\prime \tau} \rangle \\
 &= - c_X X^B_{lmn} \sqrt{Z_B} p_3 \frac{ k ( p_1^2 - p_2^2)}{E_B} e^{-E_Bt} \,,
\end{split}\\[\baselineskip]
\begin{split}
 C_{\mathcal X,\mathfrak{B},1}^{B,lmn} &=\langle \bigl( \gamma _4 \gamma_1 \mathcal O_{\mathcal X,\mathfrak{B}}^{B,lmn} (t,\mathbf p)\bigr)_\tau \bar{\mathcal N}^B_{\tau^\prime} (0,\mathbf p) (\gamma_+)_{\tau^\prime \tau} \rangle \\
 &= + c_X X^B_{lmn} \sqrt{Z_B} p_1 \frac{E_B(m_B+ k E_B) + k p_3^2}{E_B} e^{-E_Bt} \,,
\end{split}\\
\begin{split}
 C_{\mathcal X,\mathfrak{B},2}^{B,lmn} &=\langle \bigl( \gamma _4 \gamma_2 \mathcal O_{\mathcal X,\mathfrak{B}}^{B,lmn} (t,\mathbf p)\bigr)_\tau \bar{\mathcal N}^B_{\tau^\prime} (0,\mathbf p) (\gamma_+)_{\tau^\prime \tau} \rangle \\
 &= + c_X X^B_{lmn} \sqrt{Z_B} p_2 \frac{E_B(m_B+ k E_B) + k p_3^2}{E_B} e^{-E_Bt} \,,
\end{split}\\
\begin{split}
 C_{\mathcal X,\mathfrak{B},3}^{B,lmn} &=\langle \bigl( \gamma _4 \gamma_3 \mathcal O_{ \mathcal X,\mathfrak{B}}^{B,lmn} (t,\mathbf p)\bigr)_\tau \bar{\mathcal N}^B_{\tau^\prime} (0,\mathbf p) (\gamma_+)_{\tau^\prime \tau} \rangle \\
 &= - c_X X^B_{lmn} \sqrt{Z_B} p_3 \frac{2 E_B(m_B+ k E_B) + k (p_1^2+p_2^2)}{E_B} e^{-E_Bt} \,,
\end{split} \\[\baselineskip]
\begin{split}
 C_{\mathcal X,\mathfrak{C},1}^{B,lmn} &=\langle \bigl( \gamma _4 \gamma_1 \mathcal O_{\mathcal X,\mathfrak{C}}^{B,lmn} (t,\mathbf p)\bigr)_\tau \bar{\mathcal N}^B_{\tau^\prime} (0,\mathbf p) (\gamma_+)_{\tau^\prime \tau} \rangle \\
 &= - c_X X^B_{lmn} \sqrt{Z_B} p_1 \frac{E_B(m_B+ k E_B) + k p_3^2}{E_B} e^{-E_Bt} \,,
\end{split}\\
\begin{split}
 C_{\mathcal X,\mathfrak{C},2}^{B,lmn} &=\langle \bigl( \gamma _4 \gamma_2 \mathcal O_{\mathcal X,\mathfrak{C}}^{B,lmn} (t,\mathbf p)\bigr)_\tau \bar{\mathcal N}^B_{\tau^\prime} (0,\mathbf p) (\gamma_+)_{\tau^\prime \tau} \rangle \\
 &= + c_X X^B_{lmn} \sqrt{Z_B} p_2 \frac{E_B(m_B+ k E_B) + k p_3^2}{E_B} e^{-E_Bt} \,,
\end{split}\\
\begin{split}
 C_{\mathcal X,\mathfrak{C},3}^{B,lmn} &=\langle \bigl( \gamma _4 \gamma_3 \mathcal O_{ \mathcal X,\mathfrak{C}}^{B,lmn} (t,\mathbf p)\bigr)_\tau \bar{\mathcal N}^B_{\tau^\prime} (0,\mathbf p) (\gamma_+)_{\tau^\prime \tau} \rangle \\
 &= + c_X X^B_{lmn} \sqrt{Z_B} p_3 \frac{k ( p_1^2 - p_2^2)}{E_B} e^{-E_Bt} \,.
\end{split}
\end{align}%
\end{subequations}%
One immediately notices that at least one nonzero component of spatial momentum is required to extract the first moments. We evaluate $ C_{\mathcal X,\mathfrak{A},1}^{B,lmn}$, $C_{\mathcal X,\mathfrak{B},1}^{B,lmn}$ and $C_{\mathcal X,\mathfrak{C},1}^{B,lmn}$ with momentum in $x$ direction ($\mathbf{p}=(\pm 1 ,0,0)$),\footnote{All momentum components are given as multiples of $2 \pi/L$ ($L$ being the spatial extent of the lattice).} and $C_{\mathcal X,\mathfrak{A},2}^{B,lmn}$, $C_{\mathcal X,\mathfrak{B},2}^{B,lmn}$ and $C_{\mathcal X,\mathfrak{C},2}^{B,lmn}$ with momentum in $y$ direction ($\mathbf{p}=(0 ,\pm 1,0)$). For momentum in $z$ direction ($\mathbf{p}=(0 ,0,\pm 1)$) only the correlator $C_{\mathcal X,\mathfrak{B},3}^{B,lmn}$ can be used. We do not consider the remaining two correlators as they require a higher number of nonvanishing momentum components, which would lead to larger statistical uncertainties.\par%
The shape parameters defined in eq.~\eqref{eq_da_parametrization} can be expressed as linear combinations of $V^B_{lmn}$, $A^B_{lmn}$ and $T^B_{lmn}$ via eq.~\eqref{eq_convenient_DAs}. For the $N$, $\Sigma$ and $\Xi$ baryons,%
\begin{subequations}%
\begin{align}%
 \varphi_{11}^{B\neq\Lambda} &= \tfrac{1}{2}\bigl( [V-A]^B_{100} -2 [V-A]^B_{010} + [V-A]^B_{001} \bigr) \,, \\
 \varphi_{10}^{B\neq\Lambda} &= \tfrac{1}{2}\bigl( [V-A]^B_{100} - [V-A]^B_{001} \bigr) \,, \\
 \pi_{11}^{B\neq\Lambda} &= \tfrac{1}{2}\bigl( T^B_{100} + T^B_{010} - 2 T^B_{001} \bigr) \,,
\end{align}%
\end{subequations}%
where $\pi^N_{11} = \varphi^N_{11}$ due to isospin symmetry. For the $\Lambda$ baryon,%
\begin{subequations}%
\begin{align}%
 \varphi_{11}^{\Lambda} &= \tfrac{1}{\sqrt{6}}\bigl( [V-A]^\Lambda_{100} -2 [V-A]^\Lambda_{010} + [V-A]^\Lambda_{001} \bigr) \,, \\
 \varphi_{10}^{\Lambda} &= -\sqrt{\tfrac{3}{2}}\bigl( [V-A]^\Lambda_{100} - [V-A]^\Lambda_{001} \bigr) \,, \\
 \pi_{10}^{\Lambda} &= \sqrt{\tfrac{3}{2}}\bigl( T^\Lambda_{100} -  T^\Lambda_{010} \bigr) \,.
\end{align}%
\end{subequations}%
In addition we define combinations corresponding to the sum of contributions with the derivative acting on each of the three quarks:%
\begin{subequations}\label{eq_momentum-sums}%
\begin{align}%
 \varphi_{00,(1)}^{B\neq\Lambda} &= [V-A]^B_{100} + [V-A]^B_{010} + [V-A]^B_{001} \,, \\
 \pi_{00,(1)}^{B\neq\Lambda} &= T^B_{100} +  T^B_{010} + T^B_{001} \,, \\
 \varphi_{00,(1)}^{\Lambda} &= \sqrt{\tfrac{2}{3}}\bigl( [V-A]^\Lambda_{100} + [V-A]^\Lambda_{010} + [V-A]^\Lambda_{001} \bigr) \,,
\end{align}%
\end{subequations}%
where $\pi^N_{00,(1)} = \varphi^N_{00,(1)}$ due to isospin symmetry. Thanks to the Leibniz product rule for derivatives, this sum can be written as a total derivative acting on a local three-quark operator without derivatives so that in the continuum%
\begin{align}\label{eq_sumrule}%
 \varphi_{00,(1)}^B &= \varphi_{00}^B \,, &
 \pi_{00,(1)}^{B\neq\Lambda} &= \pi_{00}^B \,,
\end{align}%
corresponding to the momentum conservation condition $x_1+x_2+x_3=1$, see eq.~\eqref{eq_integrationmeasure}. However, the Leibniz rule is violated by lattice discretization and this relation can only be expected to hold after continuum extrapolation in a renormalization scheme which respects the Lorentz symmetry. Note that under renormalization $\varphi_{00,(1)}^B$ and $\pi_{00,(1)}^{B\neq\Lambda}$ mix with the other first moments, see section~\ref{sect_renormalization}. It turns out that for the bare lattice values the equalities~\eqref{eq_sumrule} are violated significantly. After renormalization and conversion to the $\MSbar$ scheme we find that the sum rules~\eqref{eq_sumrule} are fulfilled to an accuracy between $\approx96\%$ and $\approx98\%$ for our value of the lattice spacing $a\approx\unit{0.0857}{\femto\meter}$, see tables~\ref{table_extrapolated} and~\ref{table_extrapolatedUC}. A violation of the momentum sum rule of this size is in perfect agreement with the results in ref.~\cite{Braun:2014wpa}, where similar discretization effects have been observed.%
\subsubsection{Higher twist}%
Higher twist normalization constants can be calculated from the correlation functions%
\begin{align}\label{eq_correlators_higher_twist}%
 \langle \mathcal X^{B,000}_\tau(t,\mathbf p) \bar{\mathcal N}^B_{\tau^\prime}(0,\mathbf p) (\gamma _+)_{\tau^\prime \tau} \rangle &= \kappa^B_{\mathcal X} m_B \sqrt{Z_B} \frac{m_B+k E_B}{E_B}  e^{- E_B t} \,,
\end{align}%
where $\mathcal X$ can be $\mathcal S$, $\mathcal P$, $\mathcal V$, $\mathcal A$ or $\mathcal T$, cf.\ eq.~\eqref{eq_local_lattice_operators} and table~\ref{table_operators}. The twist four couplings of interest defined in eqs.~\eqref{eq_higher_twist_normalization_constants_definition_non_singlet} and~\eqref{eq_higher_twist_normalization_constants_definition_singlet} are given by%
\begin{subequations}%
\begin{align}%
 \lambda_1^{B\neq\Lambda} &= -\kappa_{\mathcal V}^{B} \,, & \lambda_2^{B\neq\Lambda} &= \kappa_{\mathcal T}^B \,, \\
 \lambda_1^\Lambda &= - \sqrt{6} \kappa_{\mathcal A}^{\Lambda} \,, & \lambda_2^{\Lambda} &= - 2 \sqrt{6} \bigl( \kappa_{\mathcal S}^{\Lambda} + \kappa_{\mathcal P}^{\Lambda} \bigr) \,, & \lambda_T^{\Lambda} &= - \sqrt{6} \bigl( \kappa_{\mathcal S}^{\Lambda} - \kappa_{\mathcal P}^{\Lambda} \bigr) \,.
\end{align}%
\end{subequations}%
Due to symmetry properties of the associated operators it follows that%
\begin{align}%
\kappa_{\mathcal S}^{B\neq\Lambda}&=\kappa_{\mathcal P}^{B\neq\Lambda}=\kappa_{\mathcal V}^{\Lambda}=\kappa_{\mathcal A}^{B\neq\Lambda}=\kappa_{\mathcal T}^{\Lambda}=0 \,,
\end{align}%
and the corresponding correlators vanish.%
\subsection{Details and strategy of the lattice simulation}%
In this analysis we use lattice ensembles generated within the coordinated lattice simulations (CLS) effort. These $N_f=2+1$ simulations employ the nonperturbatively order $a$ improved Wilson (clover) quark action and the tree-level Symanzik improved gauge action. We used a modified version of the Chroma software system~\cite{Edwards:2004sx,Arts:2015jia}, the LibHadronAnalysis library and efficient inverters~\cite{Heybrock:2014iga,Frommer:2013fsa,Luscher:2012av}. To enhance the ground state overlap the source interpolators are Wuppertal-smeared~\cite{Gusken:1989qx}, employing spatially APE-smeared~\cite{Falcioni:1984ei} transporters. A special feature of CLS configurations is the use of open boundary conditions in time direction~\cite{Luscher:2012av,Luscher:2011kk}. This will eventually allow for simulations at very fine lattices without topological freezing. We achieve an efficient and stable hybrid Monte Carlo (HMC) sampling by applying twisted-mass determinant reweighting~\cite{Luscher:2012av}, which avoids near-zero modes of the Wilson Dirac operator.\par
\begin{figure}[t]%
\centering%
\includegraphics[width=.75\textwidth]{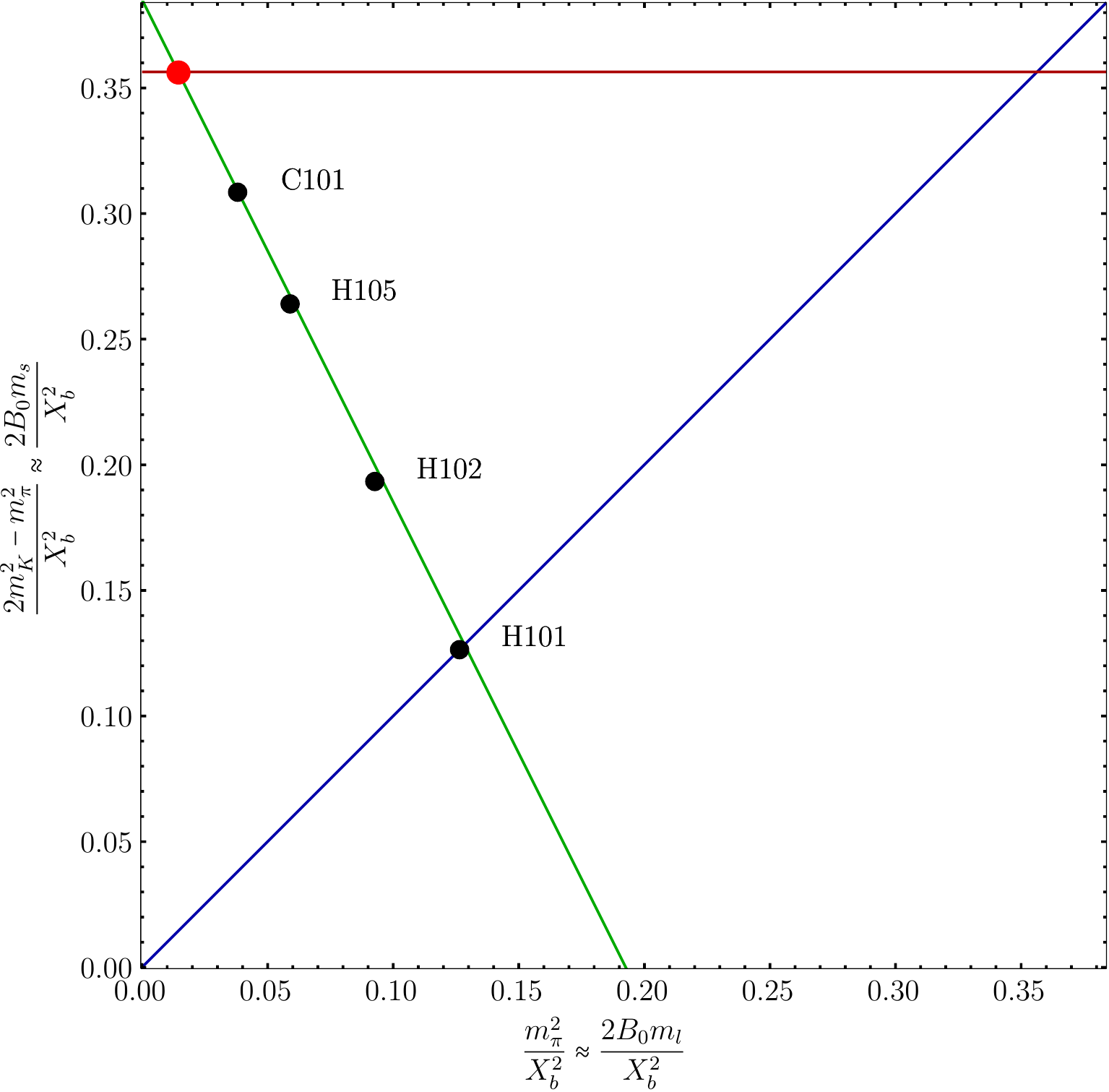}%
\caption{\label{figure_ensembles}Plot showing the meson masses of the lattice ensembles used in this study. All quantities are made dimensionless using the average octet baryon mass $X_b$, cf.\ section~\ref{sect_chpt}. Along the flavor symmetric line (blue) all pseudoscalar mesons have equal mass ($m_K^2=m_\pi^2$), which is equivalent to equal quark masses ($m_l=m_s$). The (green) line of physical normalized average quadratic meson mass ($(2 m_K^2+m_\pi^2)/X_b^2=\text{phys.}$) corresponds to an approximately physical mean quark mass ($2m_l+m_s\approx\text{phys.}$). The red line is defined by $(2 m_K^2-m_\pi^2)/X_b^2=\text{phys.}$ and indicates an approximately physical strange quark mass ($m_s\approx\text{phys.}$). The red dot marks the physical point.}%
\end{figure}

%
A list of the CLS ensembles used in this work is given in table~\ref{table_ensembles}. As schematically represented in figure~\ref{figure_ensembles}, these ensembles are tuned such that the average quark mass reproduces (approximately) the physical value. They have rather large spatial volumes ($L>\unit{2.7}{\femto\meter}$, with $m_\pi L \gtrsim 4$) and high statistics. Consecutive gauge field configurations are separated by four molecular dynamics units.\par%
\begin{table}[t]%
\centering%
\caption{\label{table_ensembles}List of the ensembles used in this work, labeled by their CLS identifier. The pion and kaon masses have been obtained from two-point functions. $\beta=3.4$ corresponds to the lattice spacing $a\approx \unit{0.0857}{\femto\metre}$. An in-depth description of these lattices can be found in ref.~\cite{Bruno:2014jqa}.}%
\renewcommand{\tabcolsep}{0.15cm}%
\begin{tabular}{ccccD{.}{.}{1.8}D{.}{.}{1.9}cccc}%
  \toprule
  id &   $\beta$ &  $N_\text{s}$  &  $N_\text{t}$  &  \multicolumn{1}{c}{$\kappa_l$} & \multicolumn{1}{c}{$\kappa_s$} & $m_\pi \, [\mega\electronvolt]$ &   $m_K \, [\mega\electronvolt]$ &  $m_\pi L$ & $\#\text{conf.}$\\
  \midrule
  H101 & $3.40$ & $32$ & $96$ & 0.13675962 & 0.13675962  & $420$ & $420$ & $5.8$ & $2000$\\
  H102 & $3.40$ & $32$ & $96$ & 0.136865   & 0.136549339 & $355$ & $440$ & $4.9$ & $1997$\\
  H105 & $3.40$ & $32$ & $96$ & 0.136970   & 0.136340790 & $280$ & $465$ & $3.9$ & $2833$\\
  C101 & $3.40$ & $48$ & $96$ & 0.137030   & 0.136222041 & $222$ & $474$ & $4.6$ & $1552$\\
  \bottomrule
\end{tabular}%
\renewcommand{\tabcolsep}{\tabcolsepstd}%
\end{table}%
Lattice calculations with the average quark mass fixed at the physical value have already been carried out for hadron masses and some form factors~\cite{Bietenholz:2011qq,Gockeler:2011ze,Cooke:2013qqa}. At the flavor symmetric point hadrons form \SU3 multiplets and their properties are related by symmetry. For example the masses have to be equal for all octet baryons. The real world is then approached in such a way that $u$ and $d$~quark masses decrease and simultaneously the $s$~quark mass increases so that their average is kept constant.\par%
For each configuration we have carried out all measurements with $3$ different source positions $t_{\text{src}}=30a$, $47a$ and $65a$. Taking the average of correlators from all these different sources is not advisable as the open boundary conditions break translational invariance in time. Instead, we average suitable forward and backward propagating states, i.e., the forward direction from $t_{\text{src}}=30a$ and the backward direction from $t_{\text{src}}=65a$ as well as the forward and the backward running state from $t_{\text{src}}=47a$. The two remaining ones (backward from $t_{\text{src}}=30a$ and forward from $t_{\text{src}}=65a$) are not considered in this analysis, as sink positions closer than $\sim20$ time slices to the boundary can show significant boundary effects due to the open boundary conditions.\par%
The second step of the data analysis is conducted by averaging over appropriate two-point functions and momenta as outlined in section~\ref{sect_correlation_functions}. For the statistical analysis we then generate $1000$ bootstrap samples per ensemble using a binsize of $8$ to eliminate autocorrelations. For each sample we use a $\chi^2$-measure to simultaneously fit the two correlation functions resulting from the forward-backward averaging procedure described above.\par%
\begin{figure}[t]%
\centering%
\includegraphics[width=0.4915\textwidth,trim=0.355cm 0.4125cm 0.355cm 0.365cm,clip]{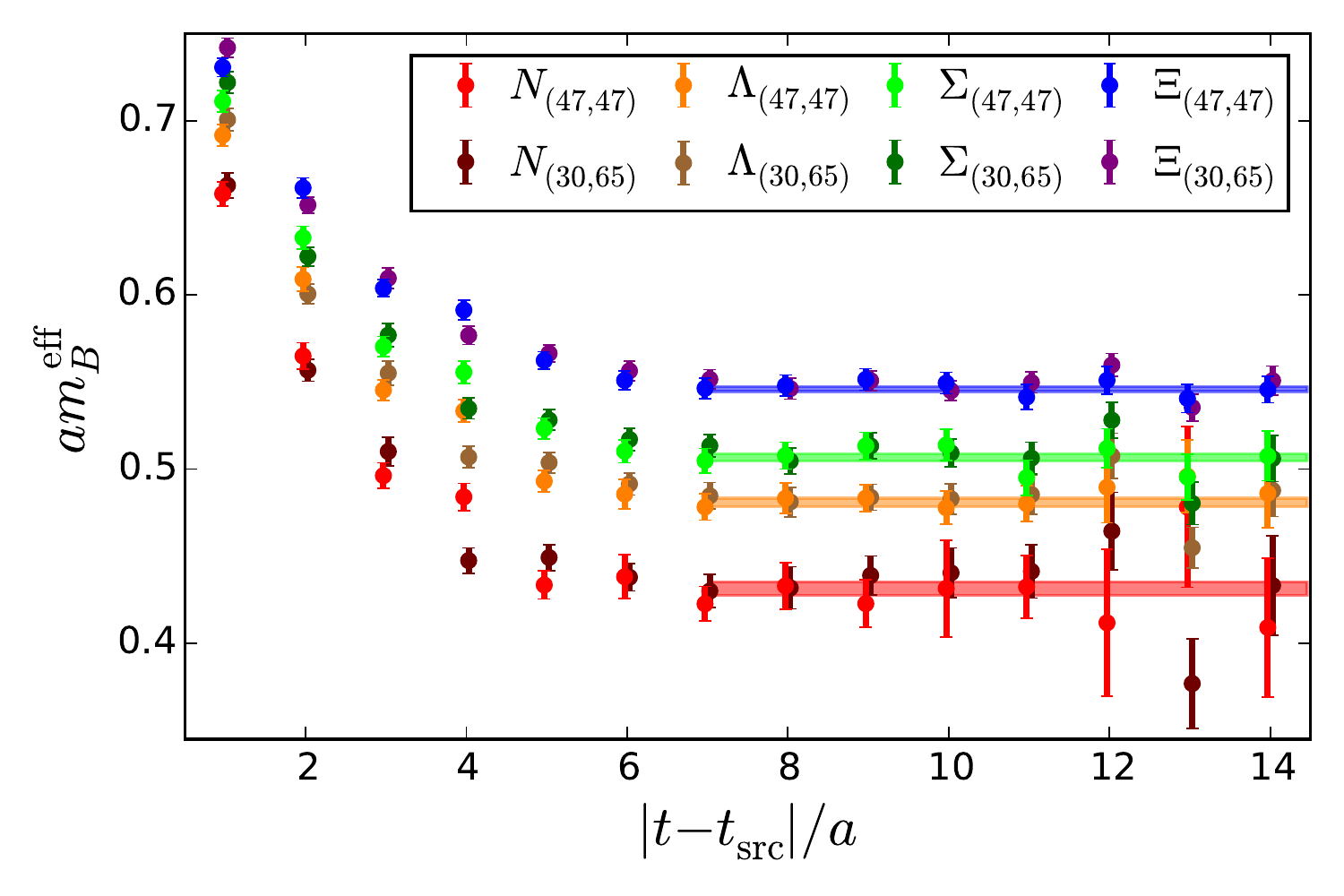}\hspace{0.017\textwidth}\includegraphics[width=0.4915\textwidth,trim=0.355cm 0.4125cm 0.355cm 0.365cm,clip]{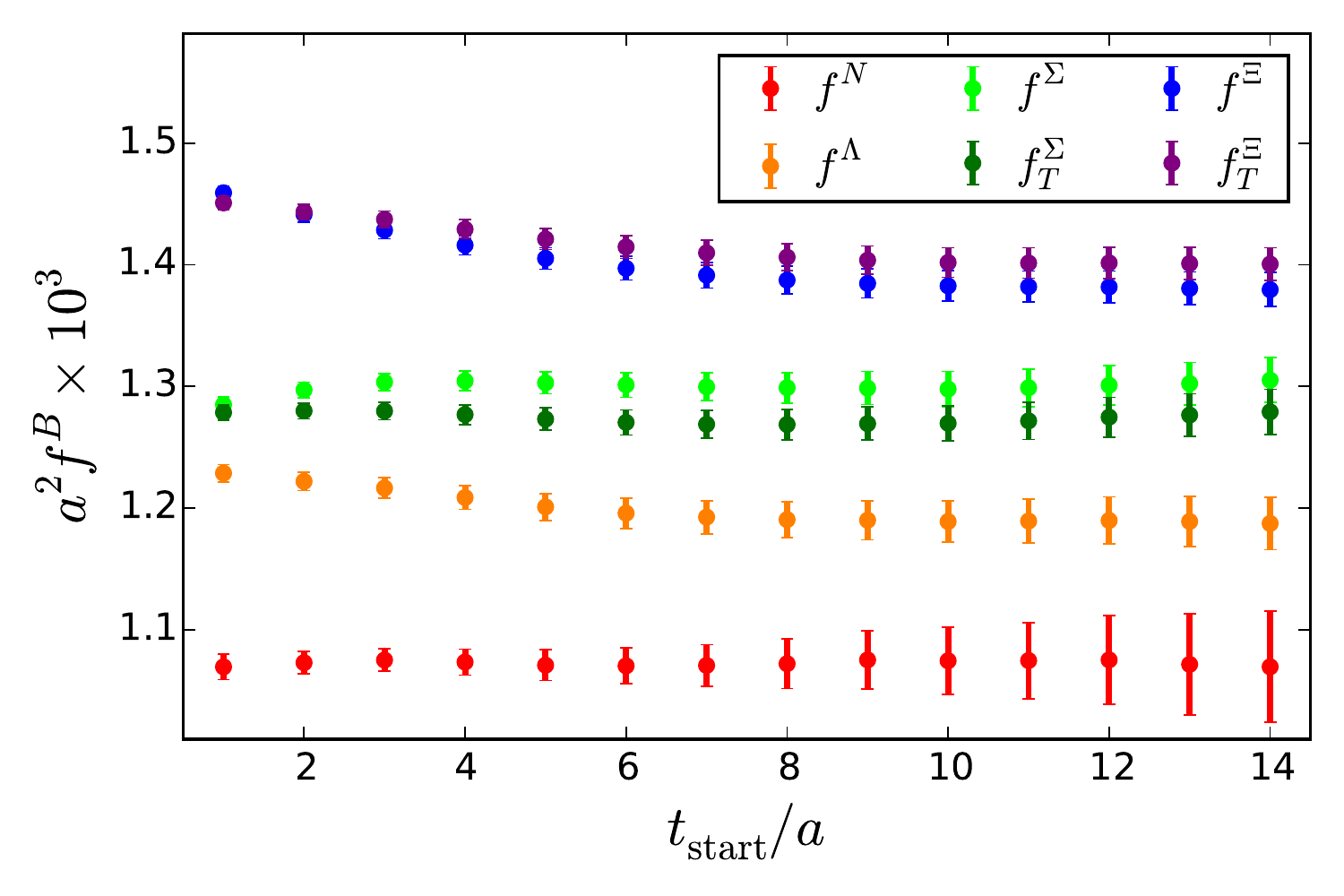}\newline%
\begin{minipage}[t]{0.4915\textwidth}%
\caption{\label{figure_meff} The data points in this plot show the effective baryon masses obtained from the two forward-backward averaged smeared-smeared correlation functions (as described in the main text) calculated on the C101 ensemble with zero three-momentum. The plateaus start at $|t-t_{\text{src}}|=7a$, where excited states are sufficiently suppressed. For each baryon the horizontal line represents the result of a combined fit to both correlators in the range $7a \leq |t-t_{\text{src}}| \leq 20a$.}%
\end{minipage}\hspace{0.017\textwidth}\begin{minipage}[t]{0.4915\textwidth}
\caption{\label{figure_start} Typical plot (from the C101 ensemble) used for the determination of the fit ranges by varying the value of the minimal source-sink distance $t_{\text{start}}$. It shows the leading twist normalization constants obtained from the correlators given in eq.~\eqref{eq_correlators_zeroth_moments}. A conservative choice is $t_{\text{start}}=9a$, where the results have fully saturated for all leading twist couplings. A variation of the maximal source-sink separation within reasonable bounds did not have any significant impact on the result. Here it is always set to $t_{\text{end}}=20a$.}%
\end{minipage}%
\end{figure}%
In order to exclude contributions from excited states the choice for the lower bound of the fit range is crucial. Figure~\ref{figure_meff} demonstrates, that, with increasing source-sink distance, the excited states decay and clear plateaus in the effective masses emerge. To determine the optimum minimal source-sink distance $t_{\text{start}}$ we perform multiple fits with varying fit ranges for all observables. $t_{\text{start}}$ is chosen in such a way that fits with even larger starting times no longer show any systematic trend in the fit results. As an example, figure~\ref{figure_start} shows the fitted leading twist coupling constants as a function of $t_{\text{start}}$.%
\section{Renormalization\label{sect_renormalization}}%
\begin{table}[t]%
\centering%
\caption[]{\label{table_multiplets}List of three-quark operator multiplets transforming irreducibly under \H4, sorted by operator dimension and representation. For zero derivatives all operators are listed. For the operators with derivatives only the leading twist multiplets are shown. The dots indicate the position of the remaining higher twist operators. The nomenclature follows ref.~\cite{Kaltenbrunner:2008pb}.}%
\renewcommand{\tabcolsep}{0.15cm}%
\begin{tabular}{>{$}l<{$}>{$}c<{$}>{$}c<{$}>{$}c<{$}}%
    \toprule
                          & \text{no derivatives} & \text{1 derivative}   & \text{2 derivatives}\\
                          & \text{dimension 9/2}  & \text{dimension 11/2} & \text{dimension 13/2}\\
    \midrule%
    \tau^{\underbar{4}}_1 & \mathcal{O}_1, \mathcal{O}_2, \mathcal{O}_3, \mathcal{O}_4, \mathcal{O}_5 & ... & \mathcal{O}_{DD1}, \mathcal{O}_{DD2}, \mathcal{O}_{DD3}, ...\\[2pt]
    \tau^{\underbar{4}}_2 & & & \mathcal{O}_{DD4}, \mathcal{O}_{DD5}, \mathcal{O}_{DD6}, ...\\[2pt]
    \tau^{\underbar{8}}_{\phantom{1}}    & \mathcal{O}_6 & \mathcal{O}_{D1}, ... & \mathcal{O}_{DD7}, \mathcal{O}_{DD8}, \mathcal{O}_{DD9}, ...\\[2pt]
    \tau^{\underbar{12}}_1 & \mathcal{O}_7, \mathcal{O}_8, \mathcal{O}_9& \mathcal{O}_{D2}, \mathcal{O}_{D3}, \mathcal{O}_{D4}, ... & \mathcal{O}_{DD10}, \mathcal{O}_{DD11}, \mathcal{O}_{DD12}, \mathcal{O}_{DD13}, ...\\[2pt]
    \tau^{\underbar{12}}_2 & & \mathcal{O}_{D5}, \mathcal{O}_{D6},\mathcal{O}_{D7}, \mathcal{O}_{D8} & \mathcal{O}_{DD14}, \mathcal{O}_{DD15}, \mathcal{O}_{DD16}, \mathcal{O}_{DD17}, \mathcal{O}_{DD18}, ...\\
    \bottomrule
\end{tabular}%
\renewcommand{\tabcolsep}{\tabcolsepstd}%
\end{table}%
Bare lattice results have to be renormalized. The preferred scheme in phenomenological applications is based on dimensional regularization where, for baryons, there are subtleties due to contributions of evanescent operators that have to be taken into account, see refs.~\cite{Kraenkl:2011qb,Gracey:2012gx}. For simplicity, we refer to the prescription suggested in~\cite{Kraenkl:2011qb} as the $\MSbar$ scheme. In principle, lattice perturbation theory could be used to compute the required renormalization coefficients, however, such calculations are nontrivial and often poorly convergent. Therefore, we employ a nonperturbative method combined with matching factors calculated in continuum perturbation theory to convert our results to the $\MSbar$ scheme in the end. The details of our renormalization procedure are described in appendix~\ref{appendix_renormalization}.\par%
Studying the renormalization of our three-quark operators, we face the problem of the reduced symmetry of the four-dimensional lattice relative to the continuum. The lattice symmetry group for fermions is known as the spinorial hypercubic group \H4, which has five irreducible spinorial representations: $\tau^{\underbar{4}}_1$, $\tau^{\underbar{4}}_2$, $\tau^{\underbar{8}}_{\phantom{1}}$, $\tau^{\underbar{12}}_1$ and $\tau^{\underbar{12}}_2$. (The superscripts indicate the dimension of these representations.) Multiplets of three-quark operators which transform according to these representations have been given in ref.~\cite{Kaltenbrunner:2008pb}. The resulting classification is summarized in table~\ref{table_multiplets}. In appendix~\ref{app_operator_relations_lattice} the operators relevant for the leading twist moments (defined in eqs.~\eqref{eq_operators_zeroth_moments} and~\eqref{eq_operators_first_moments}) as well as the operators for the higher twist normalization constants (see eq.~\eqref{eq_correlators_higher_twist}) are expressed in terms of the operators constructed in ref.~\cite{Kaltenbrunner:2008pb}. The leading twist normalization constants $f^B$ and $f_T^{B\neq\Lambda}$ are related to the three multiplets $\mathcal{O}_{7-9}$ in the representation $\tau^{\underbar{12}}_1$. The higher twist normalization constants $\lambda_1^B$, $\lambda_T^\Lambda$ ($\mathcal{O}_{3-5}$) and $\lambda_2^B$ ($\mathcal{O}_{1-2}$) are obtained from operators belonging to the same representation, $\tau^{\underbar{4}}_1$, so that they can mix under renormalization.\par%
Operators of higher dimension (e.g., with derivatives) can in general mix with operators of lower dimension transforming according to the same representation. This mixing is highly undesirable as the admixture of the lower dimensional operators is proportional to negative powers of the lattice spacing $a$ and leads to severe problems when taking the continuum limit. It can be avoided if one chooses operators from a \H4 representation where no lower dimensional multiplets exist. For single-derivative operators this happens in the case of the representation $\tau^{\underbar{12}}_2$ (see table~\ref{table_multiplets}). Therefore, we use $\mathcal O_{D5-D7}$ (and do not use $\mathcal O_{D2-D4}$) to calculate the leading twist first moments. There exists a fourth multiplet, $\mathcal O_{D8}$, in this representation which can in principle mix with these operators. However, these operators have different chirality. Hence, an admixture is a power-suppressed $O(a)$ effect. Furthermore, in the continuum limit all octet-baryon-to-vacuum matrix elements of operators within $\mathcal O_{D8}$ vanish identically, even though the operators themselves are nonzero. We have verified this property numerically on the lattice. In summary, the admixture of $\mathcal O_{D8}$ to $\mathcal O_{D5-D7}$ is completely negligible and can safely be ignored in our analysis.\par%
The classification of three-quark operators according to irreducible representations of \H4 and their behavior under renormalization is independent of the flavor structure. Indeed, in ref.~\cite{Kaltenbrunner:2008pb} the irreducible multiplets have been given assuming three generic flavors. However, the mixing is further restricted by the behavior of the operators under permutations of the three quarks. Consider operator multiplets that transform irreducibly also with respect to the permutation group $\mathcal S_3$. Such operators are given in appendix~\ref{app_operator_relations_renormalization}. In a flavor symmetric world the three inequivalent irreducible representations of $\mathcal S_3$ would correspond to a decuplet of SU(3) (trivial representation of $\mathcal S_3$, labeled $\mathscr D$), an SU(3) singlet (totally antisymmetric representation, labeled $\mathscr S$) and a doublet of octets (two-dimensional representation of $\mathcal S_3$, labeled $\mathscr O$). Of course, in the real world flavor symmetry is broken, which can lead to the appearance of renormalization constants from different representations of $\mathcal S_3$ in a single matrix element.\par%
To be more specific, we obtain the renormalized leading twist couplings from operators belonging to the representation $\tau^{\underbar{12}}_1$ of \H4. These can be arranged into a doublet of octet multiplets (with a common renormalization factor $Z^{\mathscr{O}f}$) and a decuplet multiplet (with renormalization factor $Z^{\mathscr{D}f}$), so that we end up with%
\begin{subequations}%
\begin{align}%
 \begin{pmatrix}
  f^{B\neq\Lambda}\\f_T^{B\neq\Lambda}
 \end{pmatrix}^{\MSbar}
 &= \frac13\begin{pmatrix}
  Z^{\mathscr{O}f}+2Z^{\mathscr{D}f}  & 2Z^{\mathscr{O}f}-2Z^{\mathscr{D}f}\\
  Z^{\mathscr{O}f}-Z^{\mathscr{D}f}\2 & 2Z^{\mathscr{O}f}+Z^{\mathscr{D}f}\2
 \end{pmatrix}
 \begin{pmatrix}
  f^B\\f_T^B
 \end{pmatrix}^{\mathrlap{\text{lat}}} \,,
\\
 \bigl(f^\Lambda\bigr)^{\MSbar}
 &= Z^{\mathscr{O}f} \bigl(f^\Lambda\bigr)^\text{lat} \,.
\end{align}%
\end{subequations}%
If $f_T^{B\neq\Lambda}=f^B$ (as is the case for the nucleon and for the \SU3 symmetric limit), the first equation reduces to a multiplicative renormalization with one and the same factor $Z^{\mathscr{O}f}$.\par%
As detailed above, for the higher twist normalization constants we use the \H4 representation $\tau^{\underbar{4}}_1$. In this case one obtains a singlet multiplet (renormalization factor $Z^{\mathscr{S}\lambda}$) and two doublets of octet multiplets, which can mix under renormalization (with a $2 \times 2$ renormalization matrix $Z_{m m^\prime}^{\mathscr{O}\lambda}$). Thus, we have%
\begin{subequations}%
\begin{align}%
 \begin{pmatrix}
  \lambda_1^{B\neq\Lambda}\\\lambda_2^{B\neq\Lambda}
 \end{pmatrix}^{\MSbar}
 &= \begin{pmatrix}
  Z^{\mathscr{O}\lambda}_{11}         & \tfrac{1}{\sqrt6}Z^{\mathscr{O}\lambda}_{12}\\
  \sqrt{6}Z^{\mathscr{O}\lambda}_{21} & Z^{\mathscr{O}\lambda}_{22}
 \end{pmatrix}
 \begin{pmatrix}
  \lambda_1^{B}\\\lambda_2^{B}
 \end{pmatrix}^{\mathrlap{\text{lat}}} \,,
\\
 \begin{pmatrix}
  \lambda_1^\Lambda\\\lambda_T^\Lambda\\\lambda_2^\Lambda
 \end{pmatrix}^{\MSbar}
 &= \frac13\begin{pmatrix}
  Z^{\mathscr{O}\lambda}_{11}+2Z^{\mathscr{S}\lambda}  & 2Z^{\mathscr{O}\lambda}_{11}-2Z^{\mathscr{S}\lambda}  & \sqrt{\tfrac{3}{2}}Z^{\mathscr{O}\lambda}_{12}\\
  Z^{\mathscr{O}\lambda}_{11}-Z^{\mathscr{S}\lambda}\2 & 2Z^{\mathscr{O}\lambda}_{11}+Z^{\mathscr{S}\lambda}\2 & \sqrt{\tfrac{3}{2}}Z^{\mathscr{O}\lambda}_{12}\\
  \sqrt{6}Z^{\mathscr{O}\lambda}_{21}\2                & 2\sqrt{6}Z^{\mathscr{O}\lambda}_{21}                  & 3Z^{\mathscr{O}\lambda}_{22}
 \end{pmatrix}
 \begin{pmatrix}
  \lambda_1^\Lambda\\\lambda_T^\Lambda\\\lambda_2^\Lambda
 \end{pmatrix}^{\mathrlap{\text{lat}}} \,.
\end{align}
\end{subequations}%
At the flavor symmetric point, where $\lambda_T^\Lambda=\lambda_1^\Lambda$, the second equation reduces to the first one, which is then valid for all octet baryons.\par%
In the case of the first moments of the leading twist DAs we work with the \H4 representation $\tau^{\underbar{12}}_2$. Here all three representations of $\mathcal S_3$ appear: one singlet multiplet (renormalization factor $Z^{\mathscr{S}\varphi}$), four doublets of octet multiplets (renormalization matrix $Z_{m m^\prime}^{\mathscr{O}\varphi}$) and three decuplet multiplets (renormalization matrix $Z_{m m^\prime}^{\mathscr{D}\varphi}$). The resulting renormalization pattern is the following:%
\begin{subequations}%
\begin{align}%
\begin{pmatrix}\varphi_{00,(1)}^{B\neq\Lambda}\\ \pi_{00,(1)}^{B\neq\Lambda}\\ \sqrt{2} \varphi_{11}^{B\neq\Lambda}\\ \sqrt{2} \pi_{11}^{B\neq\Lambda} \\ \sqrt{2}
\varphi_{10}^{B\neq\Lambda}\end{pmatrix}^{\mathrlap{\MSbar}}&=\frac13
\begin{pmatrix}
Z^{\mathscr{O}\varphi}_{11} + 2 Z^{\mathscr{D}\varphi}_{11} & 2 Z^{\mathscr{O}\varphi}_{11} - 2Z^{\mathscr{D}\varphi}_{11} &
Z^{\mathscr{O}\varphi}_{12} + 2 Z^{\mathscr{D}\varphi}_{12} & 2 Z^{\mathscr{O}\varphi}_{12} - 2Z^{\mathscr{D}\varphi}_{12} &
3 Z^{\mathscr{O}\varphi}_{13} \\
Z^{\mathscr{O}\varphi}_{11} - Z^{\mathscr{D}\varphi}_{11} \2& 2 Z^{\mathscr{O}\varphi}_{11} + Z^{\mathscr{D}\varphi}_{11} \2&
Z^{\mathscr{O}\varphi}_{12} - Z^{\mathscr{D}\varphi}_{12} \2& 2 Z^{\mathscr{O}\varphi}_{12} + Z^{\mathscr{D}\varphi}_{12} \2&
3 Z^{\mathscr{O}\varphi}_{13} \\
Z^{\mathscr{O}\varphi}_{21} + 2 Z^{\mathscr{D}\varphi}_{21} & 2 Z^{\mathscr{O}\varphi}_{21} - 2Z^{\mathscr{D}\varphi}_{21} &
Z^{\mathscr{O}\varphi}_{22} + 2 Z^{\mathscr{D}\varphi}_{22} & 2 Z^{\mathscr{O}\varphi}_{22} - 2Z^{\mathscr{D}\varphi}_{22} &
3 Z^{\mathscr{O}\varphi}_{23} \\
Z^{\mathscr{O}\varphi}_{21} - Z^{\mathscr{D}\varphi}_{21} \2& 2 Z^{\mathscr{O}\varphi}_{21} + Z^{\mathscr{D}\varphi}_{21} \2&
Z^{\mathscr{O}\varphi}_{22} - Z^{\mathscr{D}\varphi}_{22} \2& 2 Z^{\mathscr{O}\varphi}_{22} + Z^{\mathscr{D}\varphi}_{22} \2&
3 Z^{\mathscr{O}\varphi}_{23} \\
Z^{\mathscr{O}\varphi}_{31}\2 & 2 Z^{\mathscr{O}\varphi}_{31} & Z^{\mathscr{O}\varphi}_{32}\2 & 2 Z^{\mathscr{O}\varphi}_{32} &
3 Z^{\mathscr{O}\varphi}_{33}\end{pmatrix}
\begin{pmatrix}\varphi_{00,(1)}^{B}\\ \pi_{00,(1)}^{B}\\ \sqrt{2} \varphi_{11}^{B}\\ \sqrt{2} \pi_{11}^{B} \\ \sqrt{2} \varphi_{10}^{B}\end{pmatrix}^{\mathrlap{{\mathrlap{\text{lat}}}}} \,, \\
\begin{pmatrix}\varphi_{00,(1)}^{\Lambda}\\ \sqrt{2} \varphi_{11}^{\Lambda} \\ \sqrt{2} \varphi_{10}^{\Lambda} \\ \sqrt{2} \pi_{10}^{\Lambda} \end{pmatrix}^{\mathrlap{\MSbar}}&=\frac13
\begin{pmatrix}
3Z^{\mathscr{O}\varphi}_{11} & 3Z^{\mathscr{O}\varphi}_{12} &
Z^{\mathscr{O}\varphi}_{13}\2 & 2Z^{\mathscr{O}\varphi}_{13} \\
3Z^{\mathscr{O}\varphi}_{21} & 3Z^{\mathscr{O}\varphi}_{22} &
Z^{\mathscr{O}\varphi}_{23}\2 & 2Z^{\mathscr{O}\varphi}_{23} \\
3Z^{\mathscr{O}\varphi}_{31} & 3Z^{\mathscr{O}\varphi}_{32} &
Z^{\mathscr{O}\varphi}_{33} + 2Z^{\mathscr{S}\varphi} & 2Z^{\mathscr{O}\varphi}_{33} - 2Z^{\mathscr{S}\varphi} \\
3Z^{\mathscr{O}\varphi}_{31} & 3Z^{\mathscr{O}\varphi}_{32} &
Z^{\mathscr{O}\varphi}_{33} -Z^{\mathscr{S}\varphi} \2& 2Z^{\mathscr{O}\varphi}_{33} + Z^{\mathscr{S}\varphi}\2
\end{pmatrix}
\begin{pmatrix}\varphi_{00,(1)}^{\Lambda}\\ \sqrt{2} \varphi_{11}^{\Lambda} \\ \sqrt{2} \varphi_{10}^{\Lambda} \\ \sqrt{2} \pi_{10}^{\Lambda} \end{pmatrix}^{\mathrlap{\text{lat}}} \,.
\end{align}%
\end{subequations}%
In the \SU3 symmetric limit $\pi_{00,(1)}^{B\neq\Lambda}=\varphi_{00,(1)}^{B}$, $\pi_{11}^{B\neq\Lambda}=\varphi_{11}^{B}$ and $\pi_{10}^{\Lambda}=\varphi_{10}^{\Lambda}$ such that, similar to the above, the two equations become equivalent.%
\section{\label{sect_chpt}Chiral extrapolation and SU(3) flavor breaking}%
For the chiral extrapolation we use the three-flavor baryon chiral perturbation theory expressions derived in ref.~\cite{Wein:2015oqa}. All data points used in the present study have approximately physical average quark mass. Therefore, we use the simplified version of the extrapolation formulas, where the mean quark mass is kept fixed and all quantities are expanded around the flavor symmetric point. This scenario corresponds to the green line of figure~\ref{figure_ensembles}. Using the average octet baryon mass $X_b\equiv(2m_N+3m_\Sigma+2m_\Xi+m_\Lambda)/8$, we define the dimensionless quantity%
\begin{align}\label{eq_delta_m_definition}%
 \delta m &= \frac{4 (m_K^2 - m_\pi^2)}{3 X_b^2} \propto (m_s - m_l) + \mathcal O((m_s-m_l)^2) \,,
\end{align}%
to parametrize this path in a natural way starting from the flavor symmetric point at $\delta m=0$ and hitting the physical point at $\delta m_\text{phys} \approx 0.228$. For the leading twist DAs, defined in eq.~\eqref{eq_convenient_DAs}, the extrapolation formulas read%
\begin{subequations}%
\begin{align}%
 \Phi^B_+ &= g^B_{\Phi+} (\delta m) \bigl( \Phi^\star_+ + \delta m \Delta \Phi^B_+ \bigr) \,, \\
 \Phi^B_- &= g^B_{\Phi-} (\delta m) \bigl( \Phi^\star_- + \delta m \Delta \Phi^B_- \bigr) \,, \\
 \Pi^{B\neq\Lambda} &= g^B_{\Pi} (\delta m) \bigl( \Phi^\star_+ + \delta m \Delta \Pi^B \bigr) \,, \\
 \Pi^{\Lambda} &= g^\Lambda_{\Pi} (\delta m) \bigl( \Phi^\star_- + \delta m \Delta \Pi^\Lambda \bigr) \,.
\end{align}%
\end{subequations}%
The formulas for the higher twist normalization constants are similar:\footnote{In eq.~\eqref{eq_gXi} the subscript $\Xi$ refers to the chiral even higher twist DAs $\Xi_{\pm,4/5}^B$, see ref.~\cite{Wein:2015oqa}.}%
\begin{subequations}%
\begin{align}%
 \lambda^B_1 &= g^B_{\Phi-} (\delta m) \bigl( \lambda^\star_1 + \delta m \Delta \lambda^B_1 \bigr) \,, \\*
 \lambda^\Lambda_T &= g^\Lambda_{\Pi} (\delta m) \bigl( \lambda^\star_1 + \delta m \Delta \lambda^\Lambda_T \bigr) \,, \\*
 \lambda^B_2 &= g^B_{\Xi} (\delta m) \bigl( \lambda^\star_2 + \delta m \Delta \lambda^B_2 \bigr) \,. \label{eq_gXi}
\end{align}%
\end{subequations}%
The functions $g^B_{\text{DA}}(\delta m)$ contain the nonanalytic contributions from the leading one-loop diagrams and of the wave function renormalization. These are normalized such that $g^B_{\text{DA}}(0)=1$, which means that $\Phi^\star_\pm$ are the distribution amplitudes at the flavor symmetric point given in eq.~\eqref{eq_su3_relations}. The functional form of $g^B_{\text{DA}}(\delta m)$ is known and can be extracted from eq.~(5.18) of ref.~\cite{Wein:2015oqa}.\footnote{In our calculation we use $F_\star=\unit{112}{\mega\electronvolt}$ (cf.\ ref.~\cite{Bruns:2012eh}), $D=0.72$ and $F=0.54$ as input values. The latter lie within the range of typical estimates used in the literature, see, e.g., refs.~\cite{Jenkins:1991es,Close:1993mv,Borasoy:1998pe,Zhu:2000zf,Lin:2007ap,WalkerLoud:2011ab}.} It is important that these nonanalytic terms entering as multiplicative factors do not depend on the quark momentum fractions. The remaining quantities $\Phi^\star_\pm$, $\Delta \Phi^B_\pm$, $\Delta \Pi^B$, $\lambda^\star_{1,2}$, $\Delta\lambda^B_{1,2}$ and $\Delta \lambda^\Lambda_T$ play the role of low energy constants, meaning that they are independent of $\delta m$. However, note that $\Phi^\star_\pm$, $\Delta \Phi^B_\pm$, $\Delta \Pi^B$ still depend on $x_1,x_2,x_3$ and their functional forms cannot be predicted by an effective low energy theory. Chiral perturbation theory~\cite{Wein:2015oqa} imposes, however, certain relations between the DAs $\Delta \Phi^B_\pm$ and $\Delta \Pi^B$ which parametrize the \SU3 breaking:%
\begin{subequations}\label{eq_symmetry_constraints}%
\begin{align}%
 \Delta \Phi^N_\pm + \Delta \Phi^\Sigma_\pm + \Delta \Phi^\Xi_\pm &= 0 \,,
\end{align}%
and%
\begin{align}%
 \Delta \Pi^N &=  \Delta \Phi^N_+ \,, & \Delta \Pi^\Sigma &=  - \frac{1}{2}\Delta \Phi^\Sigma_+ - \frac{3}{2}\Delta \Phi^\Lambda_+ \,, \\*
 \Delta \Pi^\Xi &=  \frac{1}{2}\Delta \Phi^\Sigma_+ + \frac{3}{2}\Delta \Phi^\Lambda_+  - \Delta \Phi^N_+ \,, & \Delta \Pi^\Lambda &=   - \frac{1}{2}\Delta \Phi^\Lambda_- - \frac{3}{2}\Delta \Phi^\Sigma_- \,.
\end{align}%
Analogously, the \SU3 breaking parameters of the higher twist couplings should satisfy the constraints%
\begin{align}%
 \Delta \lambda^N_{1,2} + \Delta \lambda^\Sigma_{1,2} + \Delta \lambda^\Xi_{1,2} &= 0 \,, & \Delta \lambda^\Sigma_{2} + \Delta \lambda^\Lambda_{2}&= 0 \,,
\end{align}%
and%
\begin{align}%
  \Delta \lambda^\Lambda_T &= - \frac{1}{2}\Delta \lambda^\Lambda_1 - \frac{3}{2}\Delta \lambda^\Sigma_1 \,.
\end{align}%
\end{subequations}%
In the following we will call fits to the lattice data \emph{constrained}, if the relations~\eqref{eq_symmetry_constraints} are imposed, and \emph{unconstrained} otherwise.\par%
Combining eqs.~\eqref{eq_symmetry_constraints} with the explicit form of $g^B_{\text{DA}}(\delta m)$ one can find specific linear combinations of DAs for which  all terms linear in $\delta m$ cancel so that the \SU3 breaking is minimized. Similar combinations exist for the baryon masses:%
\begin{subequations}%
\begin{align}%
 0+\mathcal{O}(\delta m^2) &= 2 m_N - m_\Sigma + 2 m_\Xi - 3 m_\Lambda \,, \label{eq_GMO_sum_rule} \\
 8m_\star+\mathcal{O}(\delta m^2) &= 2 m_N + 3 m_\Sigma + 2 m_\Xi + m_\Lambda \,.
\end{align}%
\end{subequations}%
The first relation is the famous Gell-Mann--Okubo (GMO) sum rule for baryon masses~\cite{GellMann:1961ky}, whose almost exact realization in nature is widely known. The second one cannot be checked for the physical masses since it depends on $m_\star$, the baryon mass at the flavor symmetric point, which is inherently inaccessible in experiment. The analogous expressions for the leading twist DAs read:%
\begin{subequations}\label{eq_GHW_DAs}%
\begin{align}%
0 + \mathcal O (\delta m^2) &= \Phi_+^\Sigma - \Pi^\Sigma + \Phi_+^\Xi - \Pi^\Xi \,, \\
 8 \cdot 3 \Phi_+^\star + \mathcal O (\delta m^2) &=  2 \cdot 3 \Phi_+^N + 3 \cdot ( \Phi_+^\Sigma + 2 \Pi^\Sigma ) + 2 \cdot ( \Phi_+^\Xi + 2 \Pi^\Xi ) + 1 \cdot 3 \Phi_+^\Lambda  \,, \\\label{eq_Phi-relation}
 8 \cdot 3 \Phi_-^\star + \mathcal O (\delta m^2) &=2 \cdot 3 \Phi_-^N + 3 \cdot 3 \Phi_-^\Sigma + 2 \cdot 3 \Phi_-^\Xi + 1 \cdot ( \Phi_-^\Lambda + 2 \Pi^\Lambda )  \, .
\end{align}%
\end{subequations}%
For appropriately defined higher twist DAs (see ref.~\cite{Wein:2015oqa}) similar relations hold. For the normalization constants one has%
\begin{subequations}%
\begin{align}%
0 + \mathcal O (\delta m^2) &= f^\Sigma - f_T^\Sigma + f^\Xi - f_T^\Xi \,, \\
 8 \cdot 3 f^\star + \mathcal O (\delta m^2) &=  2 \cdot 3 f^N + 3 \cdot ( f^\Sigma + 2 f_T^\Sigma ) + 2 \cdot ( f^\Xi + 2 f_T^\Xi ) + 1 \cdot 3 f^\Lambda \,,
\\\label{eq_lambda1-relation}
 8 \cdot 3 \lambda_1^\star + \mathcal O (\delta m^2) &=2 \cdot 3 \lambda_1^N + 3 \cdot 3 \lambda_1^\Sigma + 2 \cdot 3 \lambda_1^\Xi + 1 \cdot ( \lambda_1^\Lambda + 2 \lambda_T^\Lambda )  \,, \\
 8 \lambda_2^\star + \mathcal O (\delta m^2) &=2 \lambda_2^N + 3 \lambda_2^\Sigma + 2 \lambda_2^\Xi + \lambda_2^\Lambda   \,,
\end{align}%
\end{subequations}%
where the first and the second equation follow directly from eq.~\eqref{eq_GHW_DAs}, and the remaining ones result from the corresponding relations for higher twist DAs. The similarity between the relations~\eqref{eq_Phi-relation} and~\eqref{eq_lambda1-relation} is not accidental. A chiral perturbation theory analysis~\cite{Wein:2015oqa} reveals that the DAs of arbitrary twist can be categorized into classes with definite chiral behavior, and $\lambda_1^B$ is the normalization of several higher twist DAs within the same class as $\Phi_-^B$. Similar relations hold for the higher moments of the DAs. In particular, the relations for the first moments of the leading twist DAs are obtained from eq.~\eqref{eq_GHW_DAs} by replacing $\Phi_+ \mapsto \varphi_{11}$, $\Pi^{B\neq\Lambda} \mapsto \pi^B_{11}$,  $\Phi_- \mapsto \varphi_{10}$ and $\Pi^{\Lambda} \mapsto \pi^\Lambda_{10}$.\par%
To visualize the size of higher order \SU3 breaking terms it is convenient to form dimensionless expressions that vanish in the flavor symmetric limit ($\delta m \to 0$):%
\begin{subequations}\label{eq_GHW_deviations}%
\begin{align}%
 \delta_1 f &= 1 - \frac{f^\Sigma + f^\Xi}{f_T^\Sigma + f_T^\Xi} \,, \\*
 \delta_2 f &= 1 - \frac{1}{8 \cdot 3 f^\star} \Bigl(  2 \cdot 3 f^N + 3 \cdot ( f^\Sigma + 2 f_T^\Sigma ) + 2 \cdot ( f^\Xi + 2 f_T^\Xi ) + 1 \cdot 3 f^\Lambda \Bigr) \,, \\*
 \delta \lambda_1 &= 1 - \frac{1}{8 \cdot 3 \lambda_1^\star} \Bigl( 2 \cdot 3 \lambda_1^N + 3 \cdot 3 \lambda_1^\Sigma + 2 \cdot 3 \lambda_1^\Xi + 1 \cdot ( \lambda_1^\Lambda + 2 \lambda_T^\Lambda )   \Bigr) \,, \\*
 \delta \lambda_2 &= 1 - \frac{1}{8  \lambda_2^\star} \Bigl( 2 \lambda_2^N + 3 \lambda_2^\Sigma + 2 \lambda_2^\Xi + \lambda_2^\Lambda \Bigr) \,.
\end{align}%
\end{subequations}%
\begin{figure}[t]%
\centering%
\includegraphics[width=0.495\textwidth]{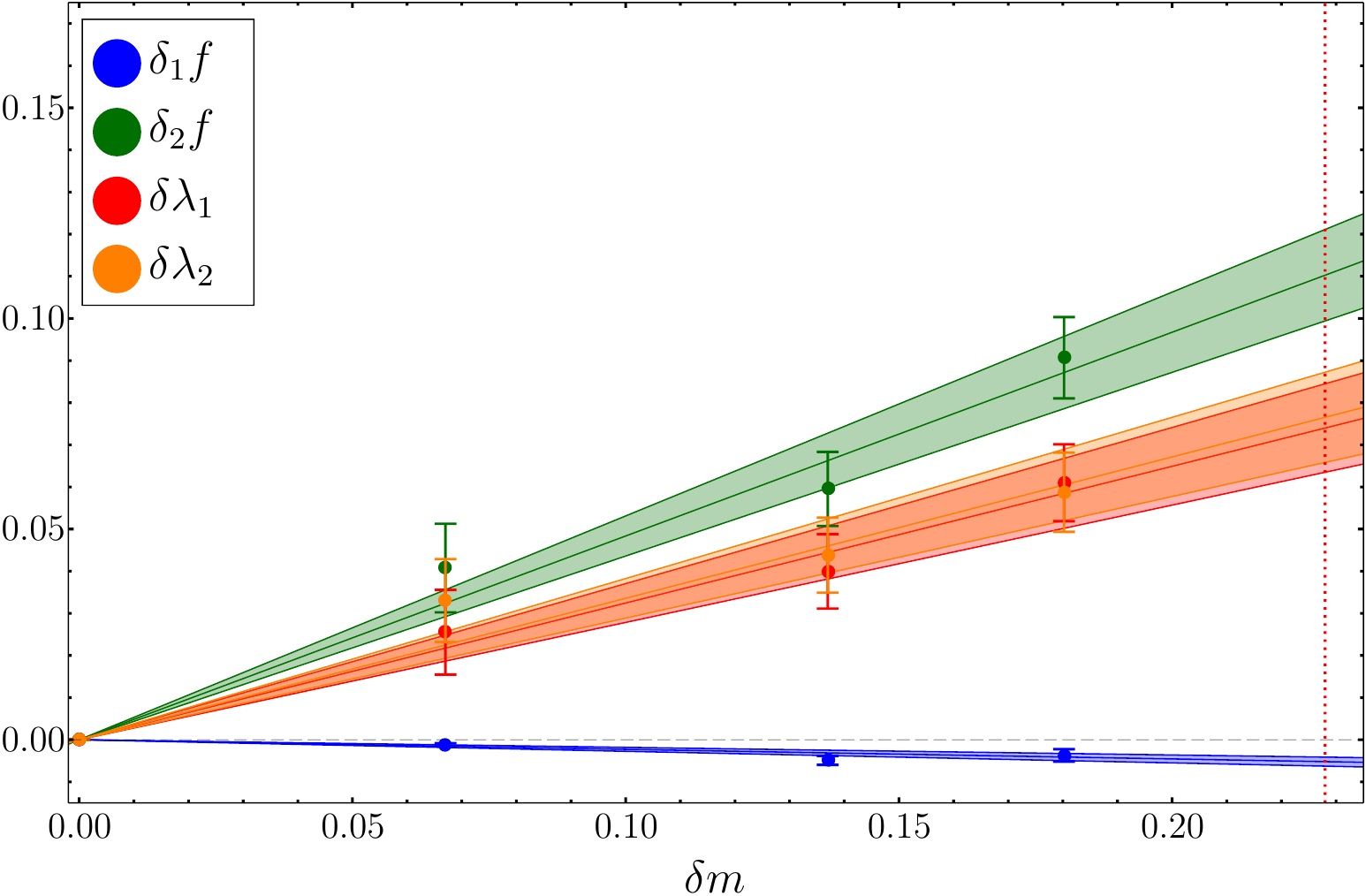}\hspace{0.01\textwidth}\includegraphics[width=0.495\textwidth]{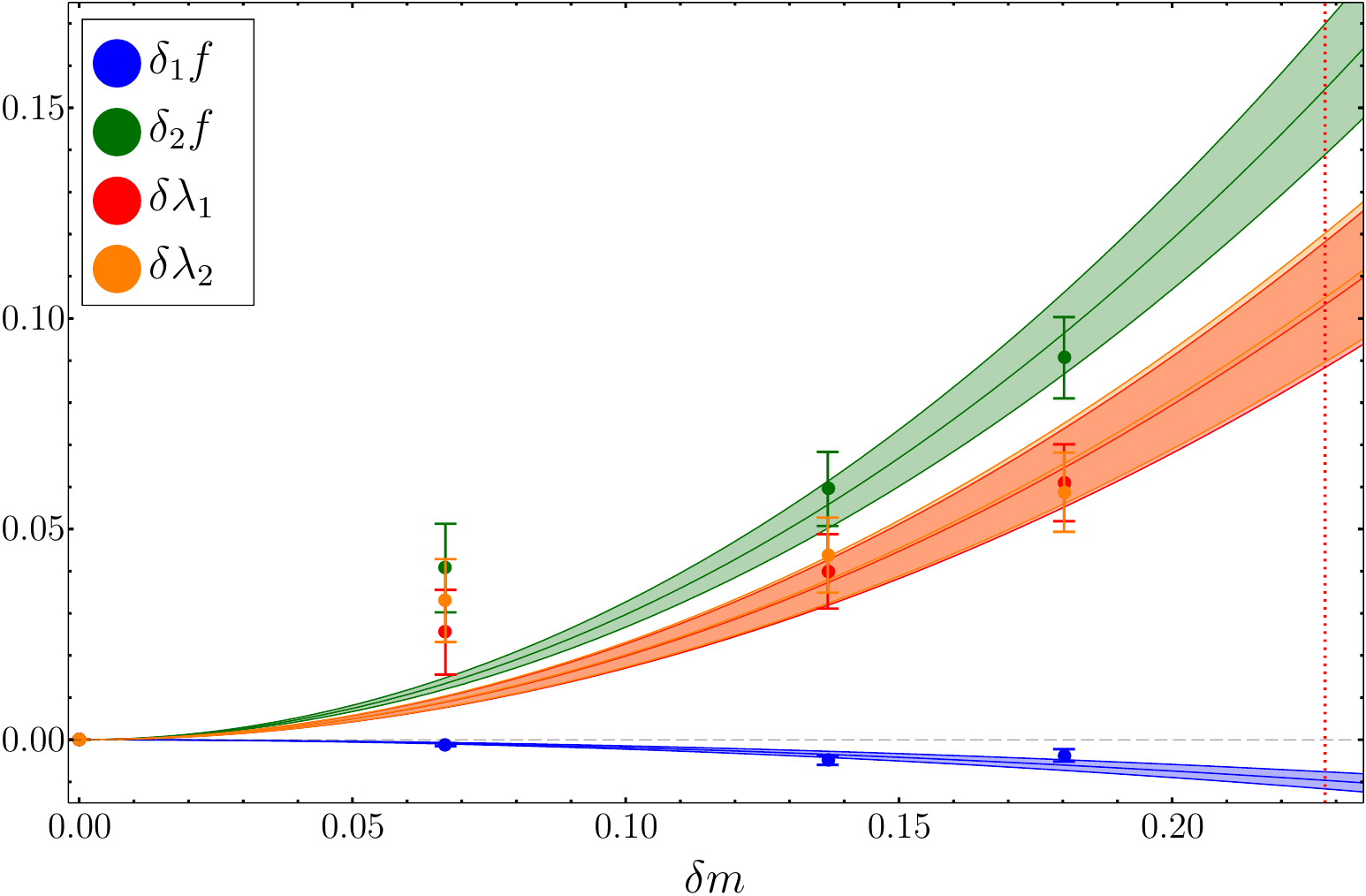}%
\caption{\label{figure_GHW}Results for the quantities defined in eq.~\eqref{eq_GHW_deviations} are shown, along with linear and quadratic fits. Note that the curves for $\delta \lambda_1$ and $\delta \lambda_2$ lie almost on top of each other.}%
\end{figure}%
In figure~\ref{figure_GHW} we show linear and quadratic fits to the data. Even though for all these combinations the expected $\delta m$ dependence is quadratic, we find that a linear dependence cannot be excluded. The largest deviation at the physical point is found for $\delta_2 f$ (up to $\approx 15\%$). Most remarkably, the deviation from the GMO-like relation for the leading twist DAs, $\delta_1 f$, is very small ($|\delta_1 f| \approx 1\%$ at the physical point). For comparison, the violation of the GMO sum rule~\eqref{eq_GMO_sum_rule} using the experimental values of baryon masses is%
\begin{align}%
 1-\frac{2m_N+2m_\Xi}{m_\Sigma+3m_\Lambda} \approx 0.57\% \,.
\end{align}\par
\begin{figure}[p]%
\hspace*{-0.075\textwidth}\begin{minipage}{1.15\textwidth}%
\includegraphics[width=0.495\textwidth]{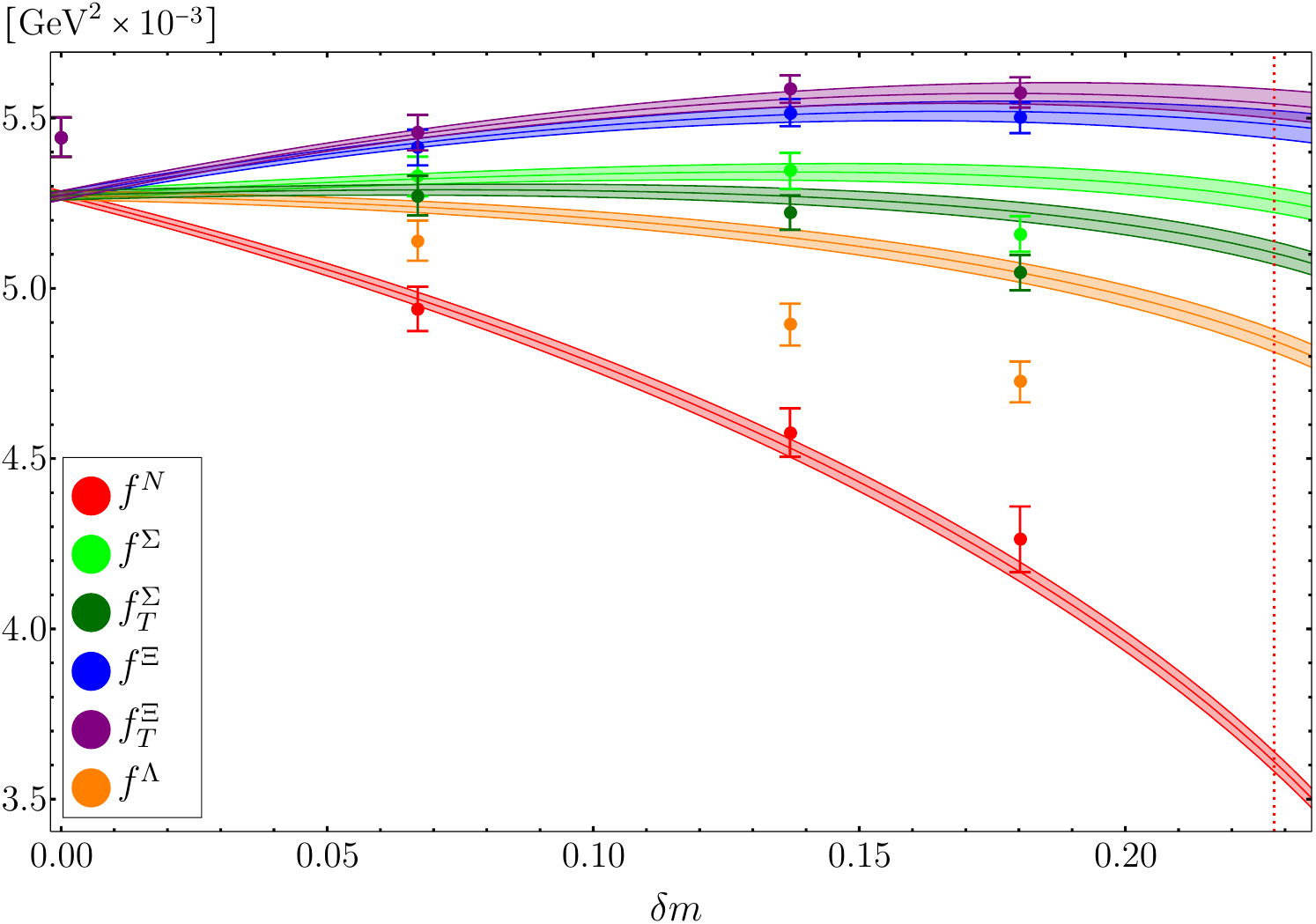}\hspace{0.01\textwidth}%
\includegraphics[width=0.495\textwidth]{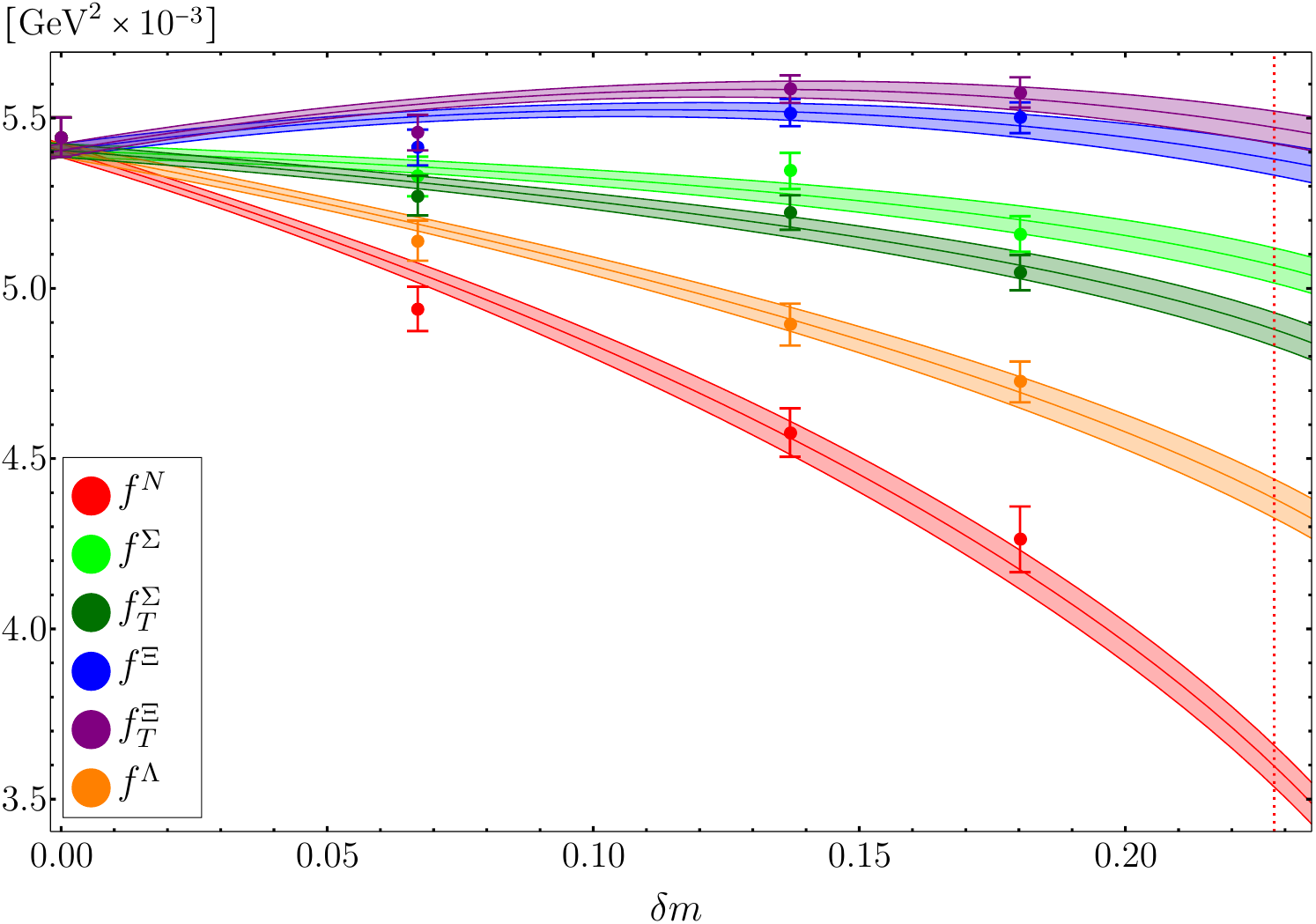}\vspace{-8pt}%
\caption{\label{figure_f}Constrained fit (left, $4$ parameters) and unconstrained fit (right, $7$ parameters) for the leading twist normalization constants $f^N$, $f^\Sigma$, $f^\Sigma_T$, $f^\Xi$, $f^\Xi_T$ and $f^\Lambda$.}%
\vspace{.9\baselineskip}%
\includegraphics[width=0.495\textwidth]{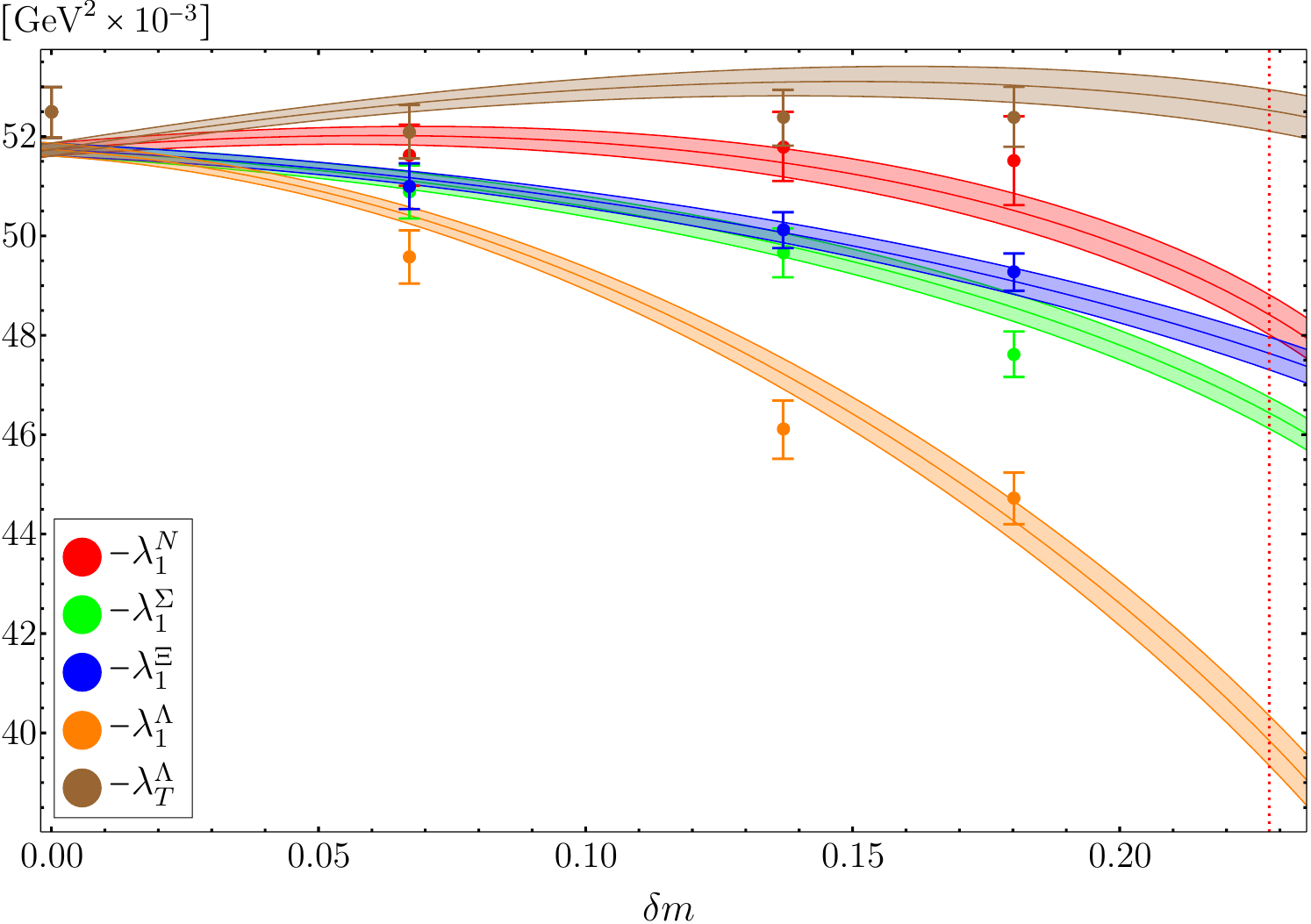}\hspace{0.01\textwidth}%
\includegraphics[width=0.495\textwidth]{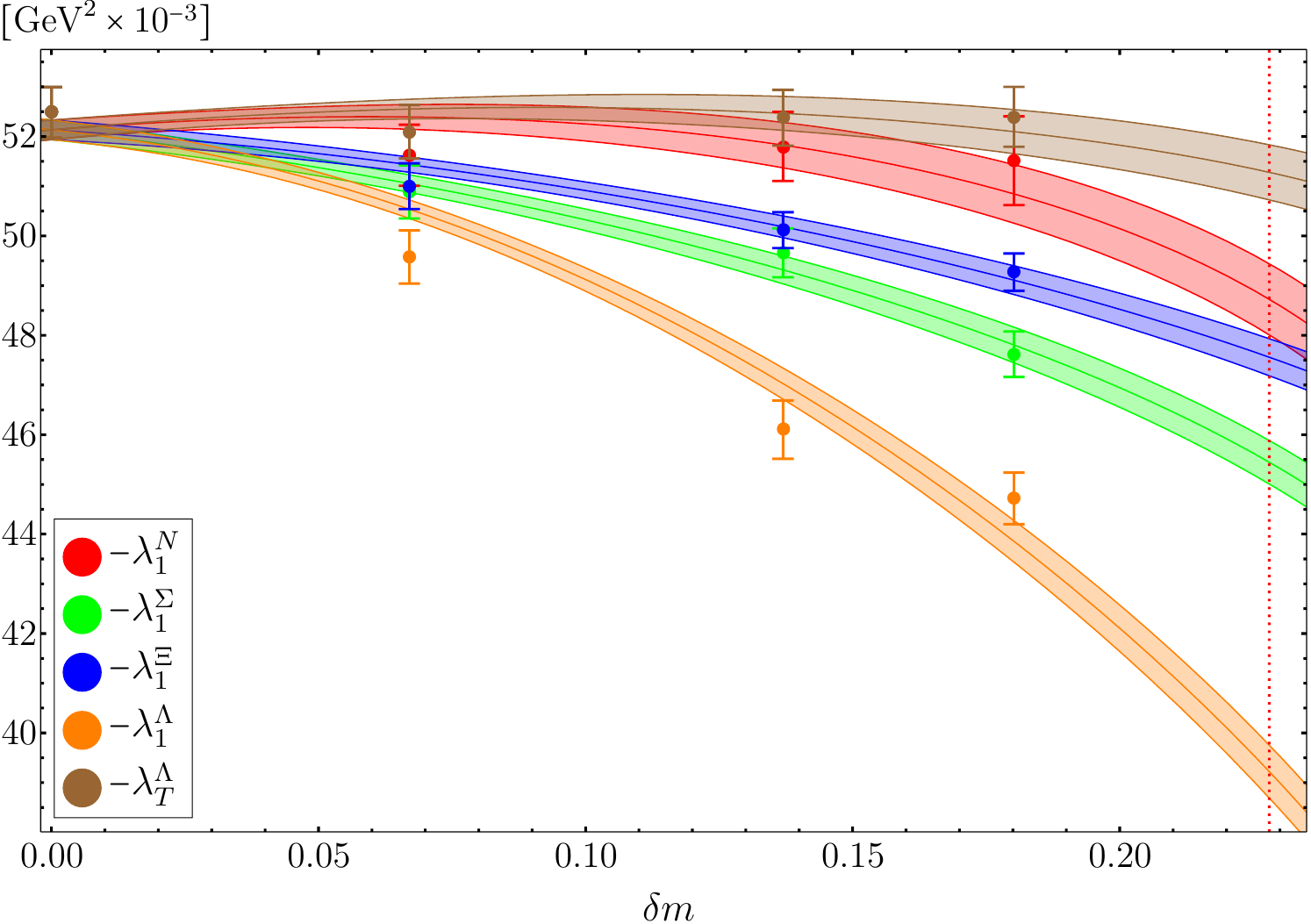}\vspace{-8pt}%
\caption{\label{figure_l1}Constrained fit (left, $4$ parameters) and unconstrained fit (right, $6$ parameters) for the chiral odd higher twist normalization constants $\lambda_1^N$, $\lambda_1^\Sigma$, $\lambda_1^\Xi$, $\lambda_1^\Lambda$ and $\lambda_T^\Lambda$.}%
\vspace{.9\baselineskip}%
\includegraphics[width=0.495\textwidth]{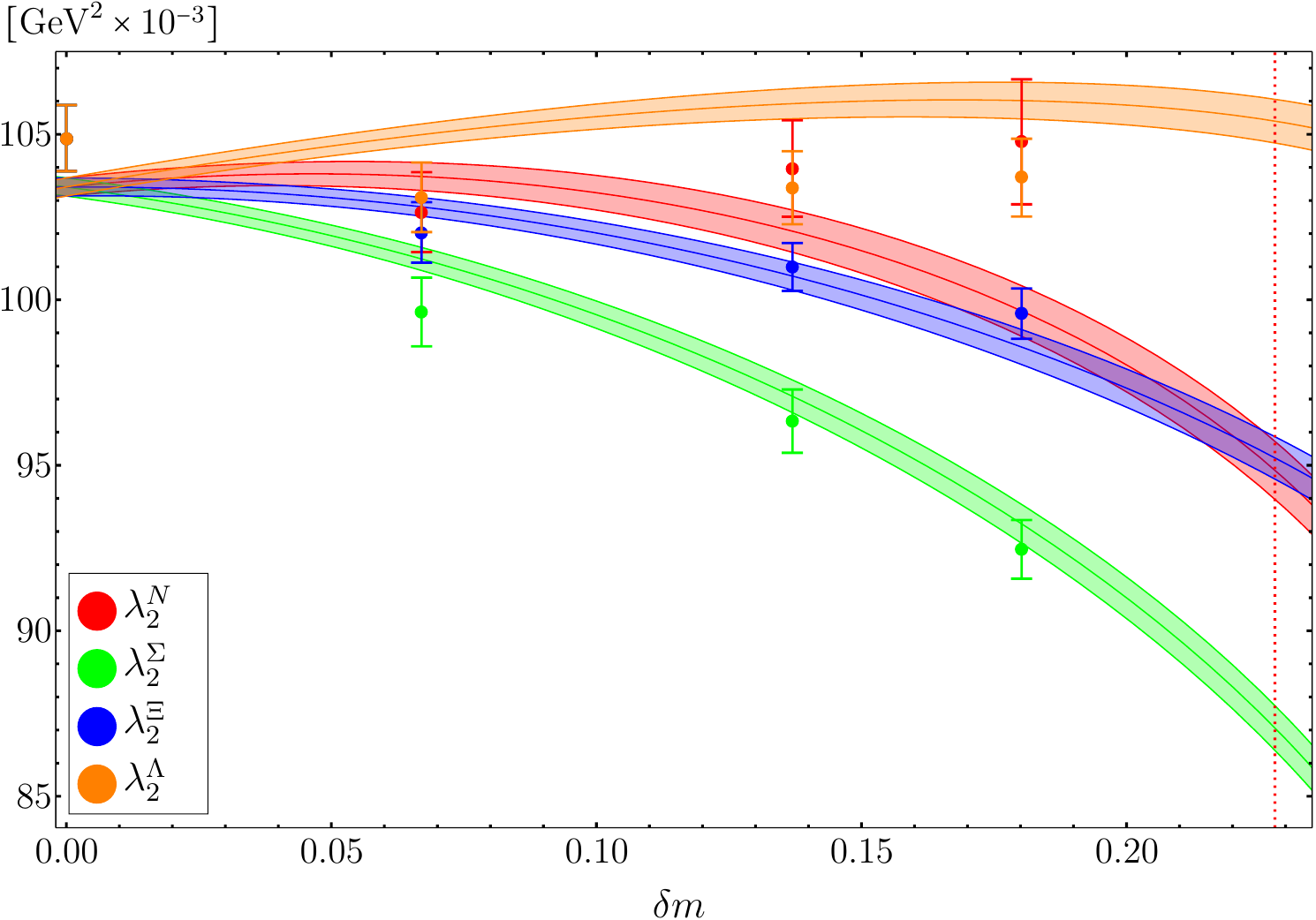}\hspace{0.01\textwidth}%
\includegraphics[width=0.495\textwidth]{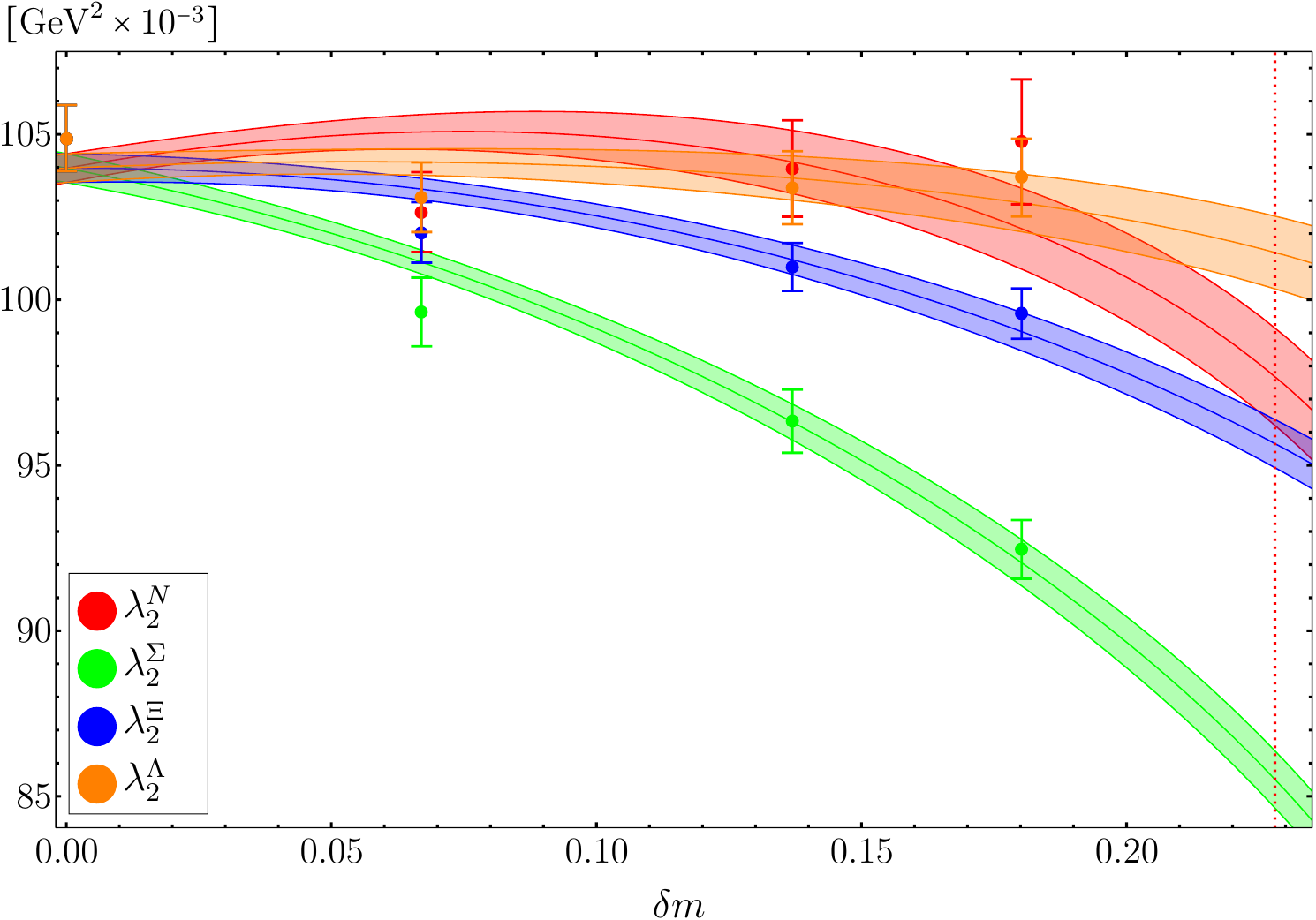}\vspace{-8pt}%
\caption{\label{figure_l2}Constrained fit (left, $3$ parameters) and unconstrained fit (right, $5$ parameters) for the chiral even higher twist normalization constants $\lambda_2^N$, $\lambda_2^\Sigma$, $\lambda_2^\Xi$ and $\lambda_2^\Lambda$.}%
\end{minipage}\hspace*{-0.075\textwidth}%
\end{figure}

\begin{figure}[p]%
\hspace*{-0.075\textwidth}\begin{minipage}{1.15\textwidth}%
\includegraphics[width=0.495\textwidth]{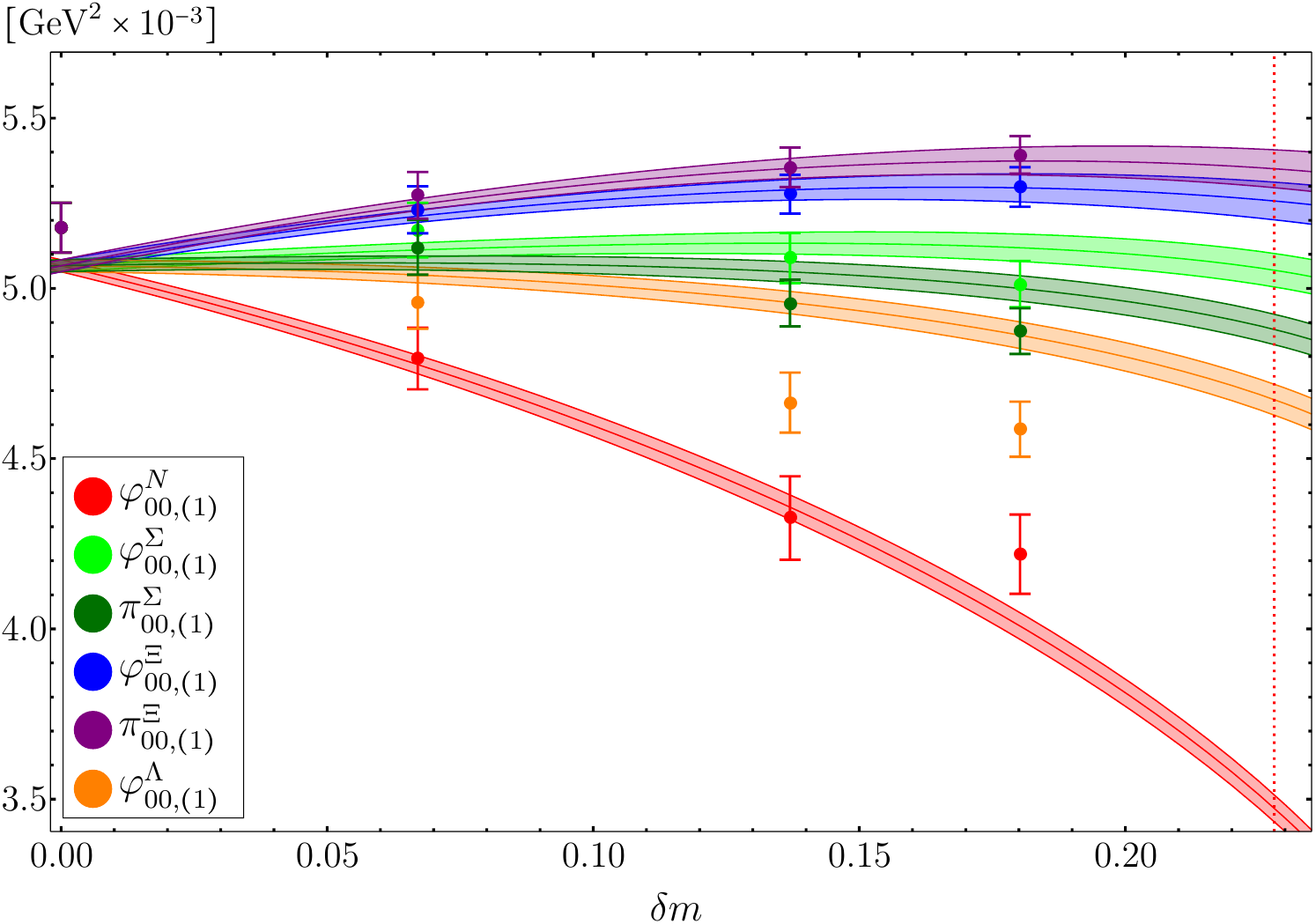}\hspace{0.01\textwidth}%
\includegraphics[width=0.495\textwidth]{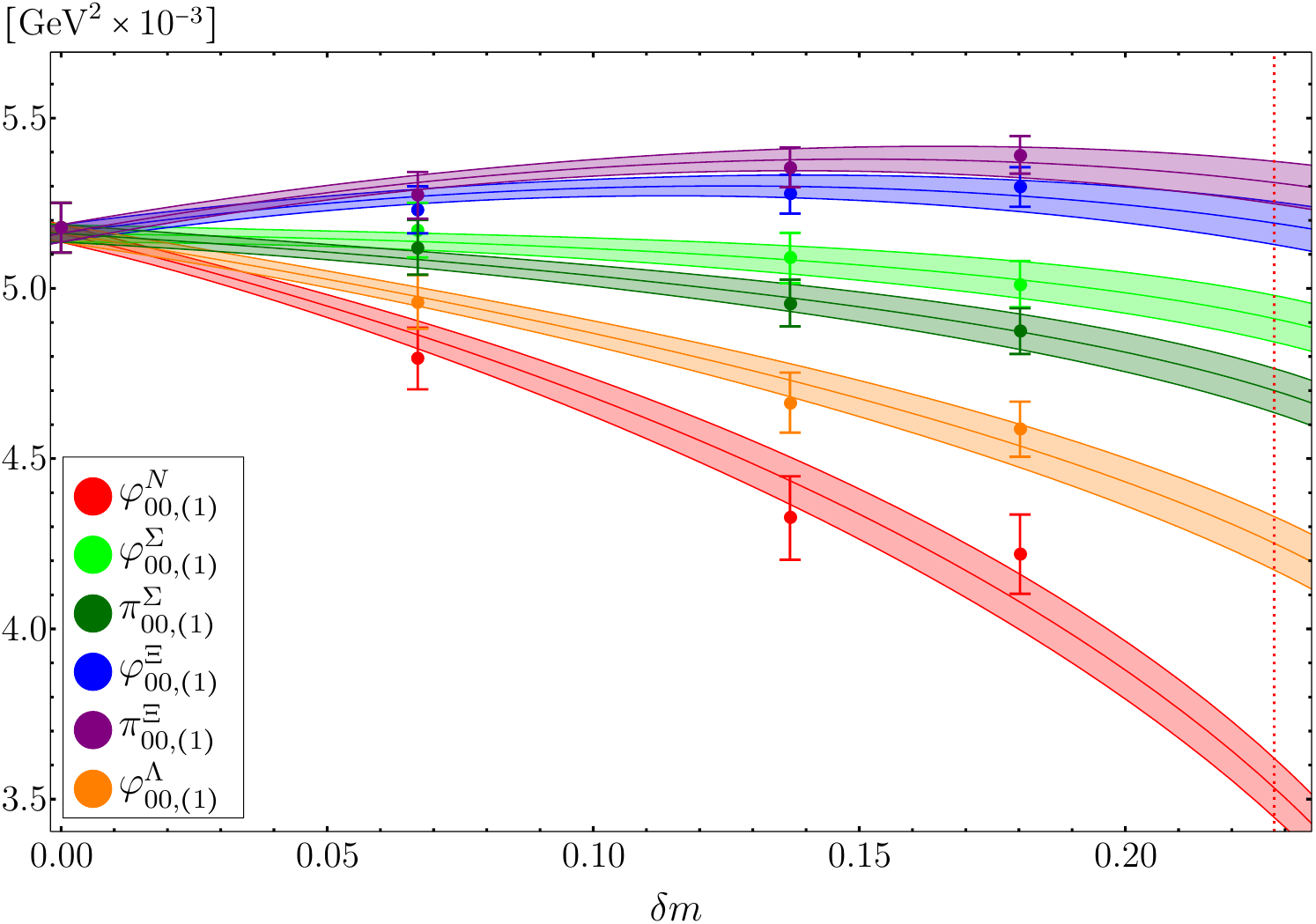}\vspace{-8pt}%
\caption{\label{figure_phi001}Constrained fit (left, $4$ parameters) and unconstrained fit (right, $7$ parameters) for $\varphi_{00,(1)}^N$, $\varphi_{00,(1)}^\Sigma$, $\pi_{00,(1)}^\Sigma$, $\varphi_{00,(1)}^\Xi$, $\pi_{00,(1)}^\Xi$ and $\varphi_{00,(1)}^\Lambda$ of the leading twist DAs $\Phi_+^B$ and $\Pi^{B\neq\Lambda}$. These moments should be equivalent to the leading twist normalization constants $f^B$ and $f^{B\neq\Lambda}_T$ in the continuum (cf.\ figure~\ref{figure_f}).}%
\vspace{.9\baselineskip}%
\includegraphics[width=0.495\textwidth]{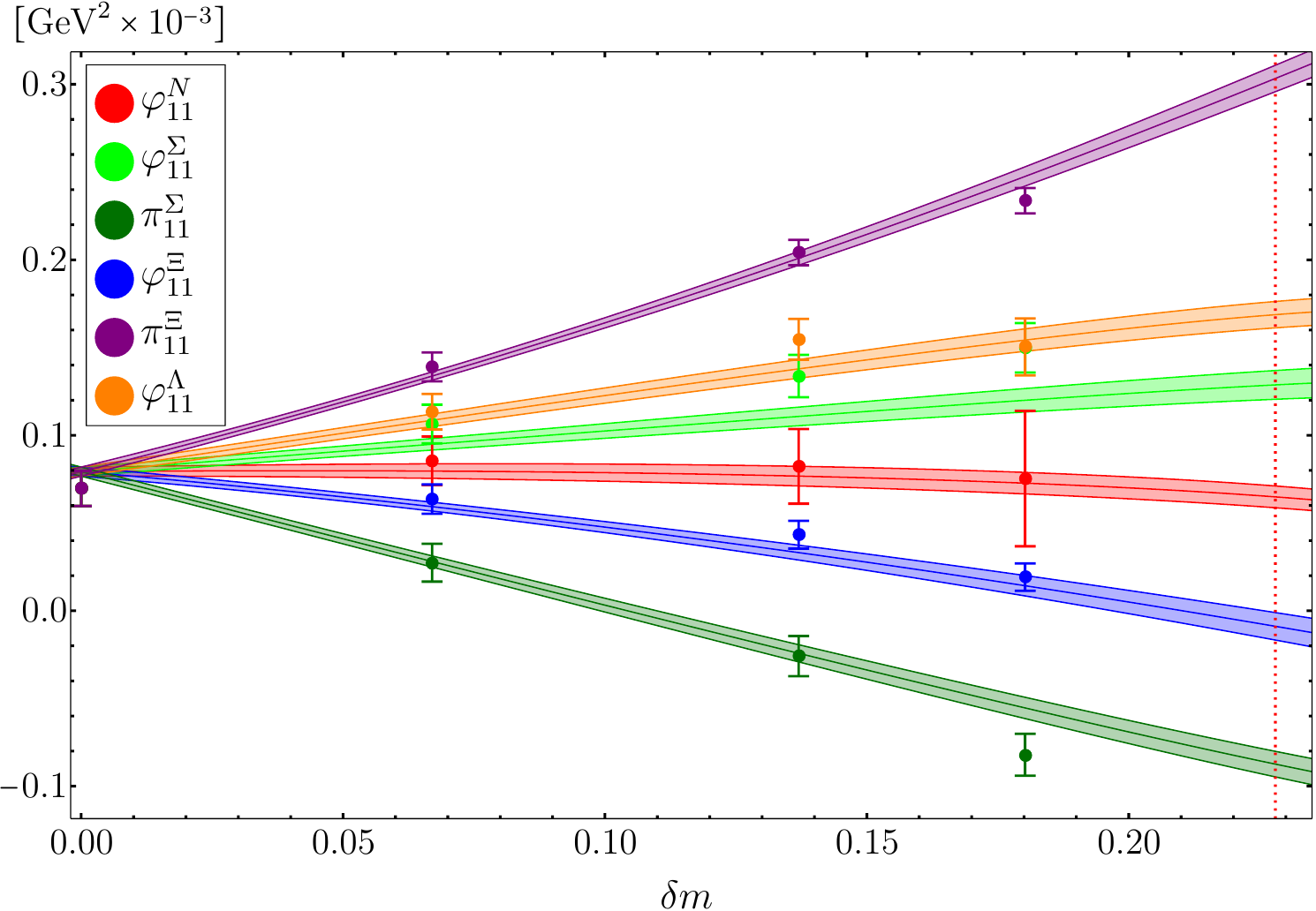}\hspace{0.01\textwidth}%
\includegraphics[width=0.495\textwidth]{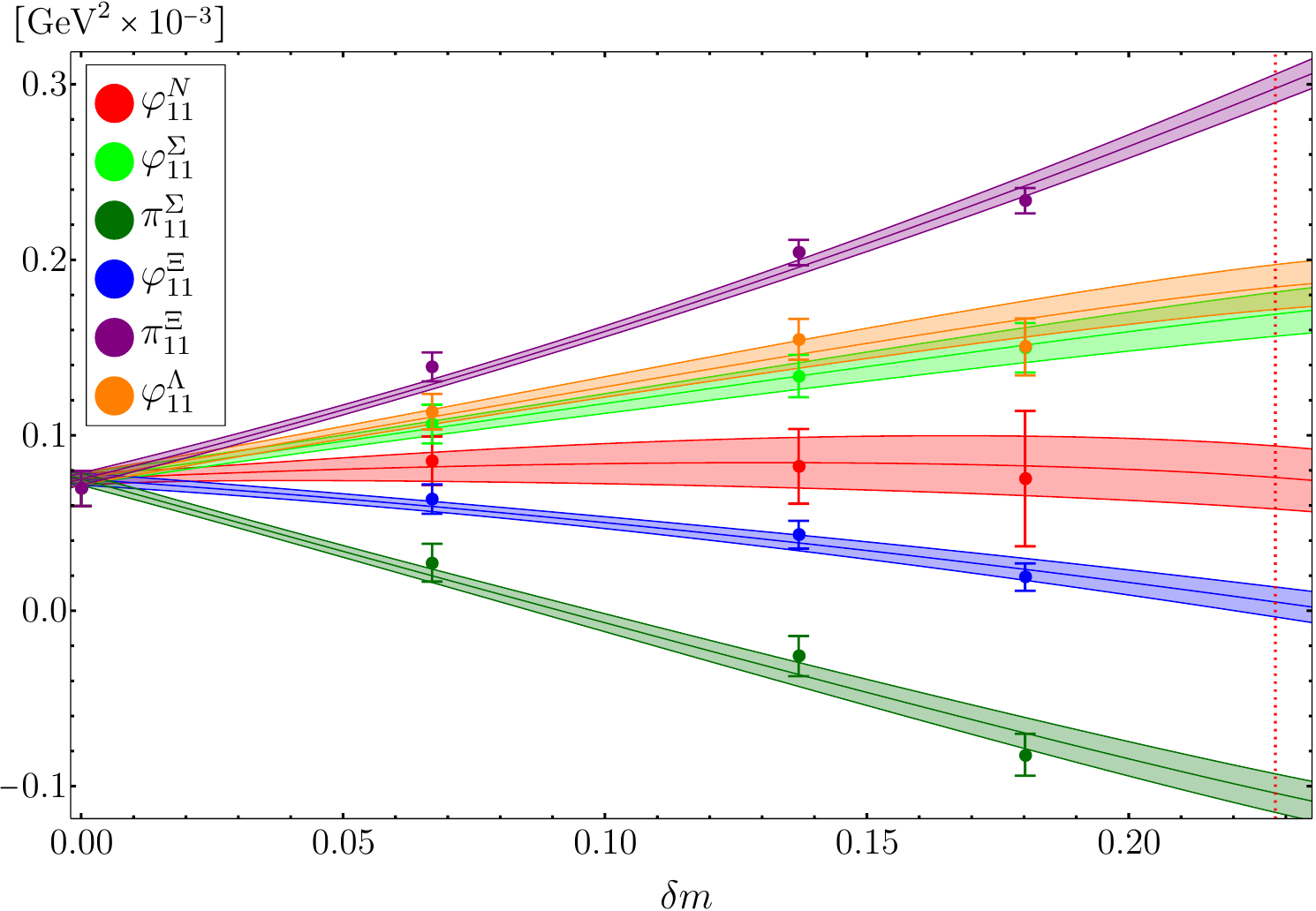}\vspace{-8pt}%
\caption{\label{figure_phi11}Constrained fit (left, $4$ parameters) and unconstrained fit (right, $7$ parameters) for the first moments $\varphi_{11}^N$, $\varphi_{11}^\Sigma$, $\pi_{11}^\Sigma$, $\varphi_{11}^\Xi$, $\pi_{11}^\Xi$ and $\varphi_{11}^\Lambda$ of the leading twist DAs $\Phi_+^B$ and $\Pi^{B\neq\Lambda}$.}%
\vspace{.9\baselineskip}%
\includegraphics[width=0.495\textwidth]{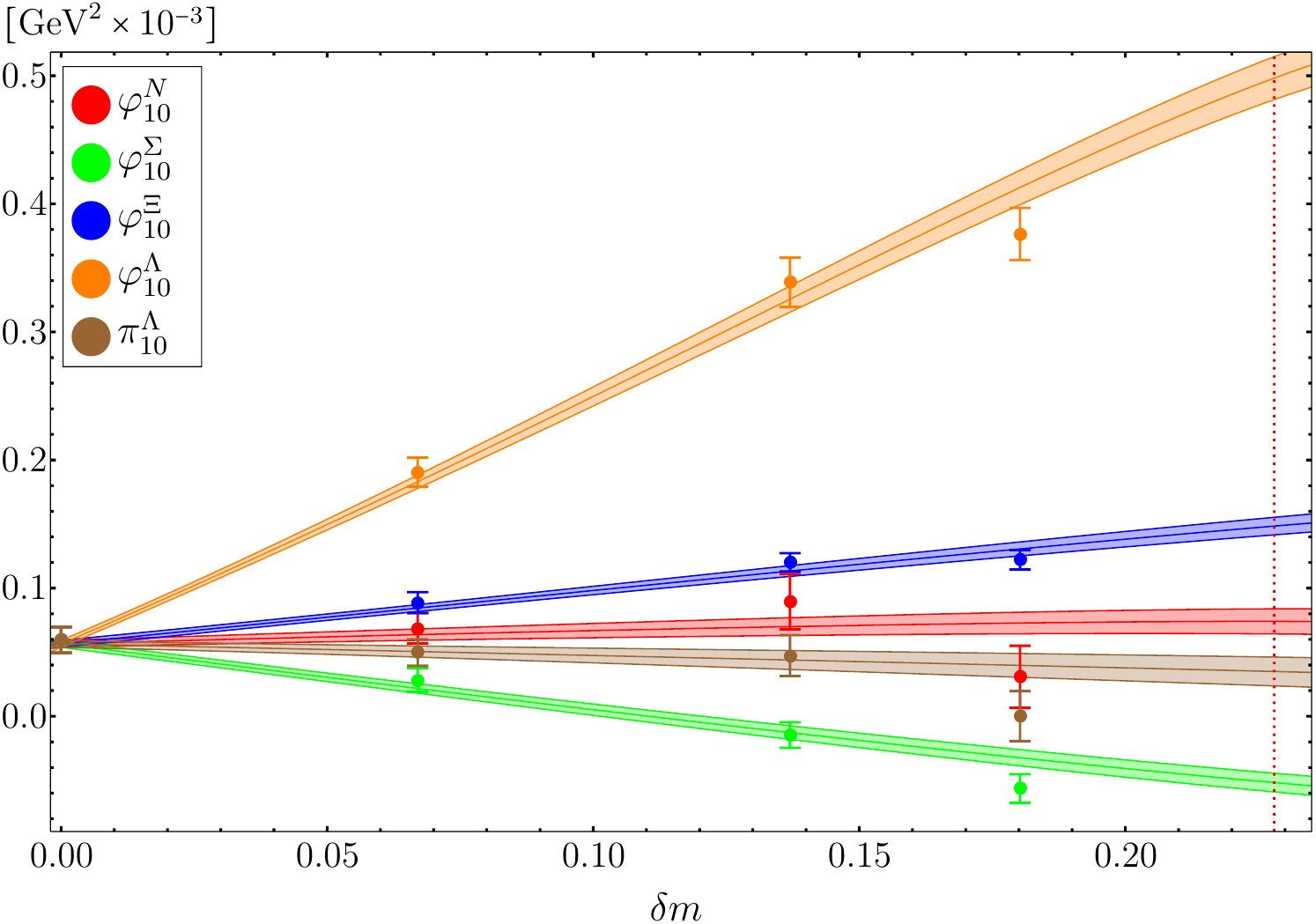}\hspace{0.01\textwidth}%
\includegraphics[width=0.495\textwidth]{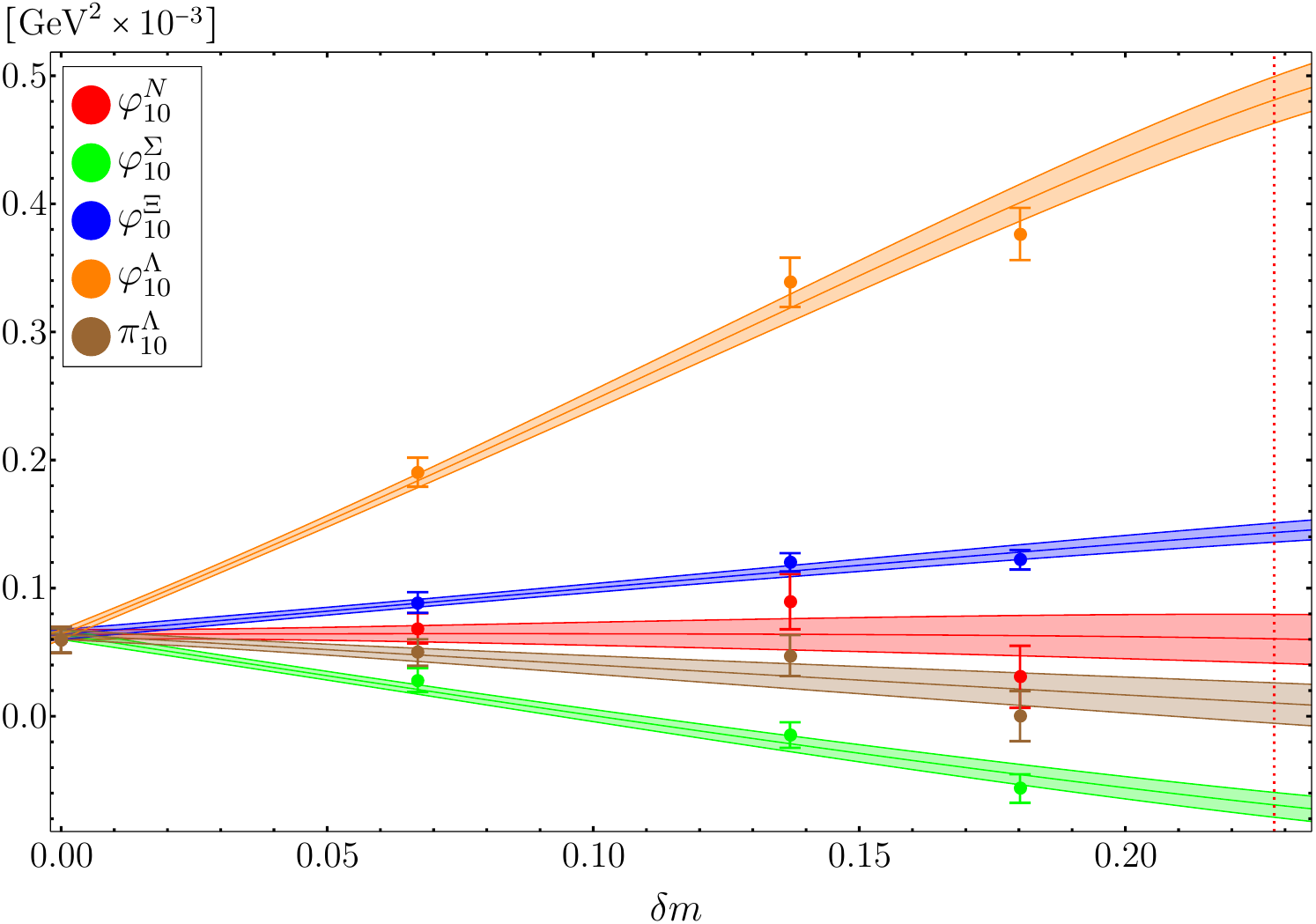}\vspace{-8pt}%
\caption{\label{figure_phi10}Constrained fit (left, $4$ parameters) and unconstrained fit (right, $6$ parameters) for the first moments $\varphi_{10}^N$, $\varphi_{10}^\Sigma$, $\varphi_{10}^\Xi$, $\varphi_{10}^\Lambda$ and $\pi_{10}^\Lambda$ of the leading twist DAs $\Phi_-^B$ and $\Pi^{\Lambda}$.}%
\vspace{-.9\baselineskip}\end{minipage}\hspace*{-0.075\textwidth}%
\end{figure}

%
In figures~\ref{figure_f}--\ref{figure_phi10} we show constrained (left) and unconstrained (right) combined fits to the lattice data. For most of the measured quantities we find that the constraints in eqs.~\eqref{eq_symmetry_constraints} are fulfilled reasonably well. This manifests itself in comparable values of $\chi^2$ per degree of freedom for both, the unconstrained fit, where the symmetry constraints are ignored, and the constrained fit, where the symmetry relations are enforced. Especially for $\lambda^B_1$ and $\lambda^\Lambda_T$, as well as for the first moments of $\Phi_-^B$ and $\Pi^\Lambda$ ($\varphi_{10}^B$ and $\pi_{10}^\Lambda$), which have the same chiral behavior as $\lambda^B_1$ and $\lambda^\Lambda_T$, one finds an extraordinarily good agreement with the lattice data (cf.\ figures~\ref{figure_l1} and~\ref{figure_phi10}). Also for the first moments $\varphi_{11}^B$ and $\pi_{11}^{B\neq\Lambda}$, which appear in $\Phi_+^B$ and $\Pi^{B\neq\Lambda}$, and are predicted to have the same chiral logarithms as the couplings $f^B$ and $f_T^{B\neq\Lambda}$, the constraints are fulfilled within errors (cf.\ figure~\ref{figure_phi11}). In contrast, for the leading twist normalization constants $f^B$ and $f^{B\neq\Lambda}_T$, as well as for $\varphi_{00,(1)}^B$ and $\pi_{00,(1)}^{B\neq\Lambda}$ (which have to coincide with $f^B$ and $f^{B\neq\Lambda}_T$ in the continuum), these relations seem to be broken rather badly (cf.\ figures~\ref{figure_f} and~\ref{figure_phi001}). Also for $\lambda_2^B$ the agreement is not really flawless (cf.\ figure~\ref{figure_l2}).\par%
\afterpage{\clearpage}%
We can summarize that leading one-loop BChPT can qualitatively describe our data, even though in some cases the observed \SU3 breaking cannot be reproduced by the constrained fit. This might indicate that for these quantities higher order BChPT effects are particularly large. However, the observed discrepancies could also be caused by systematic errors in the lattice data, for which finite volume and discretization effects are prominent candidates. In particular lattice spacing effects have already been identified as a major source of systematic uncertainty in the two-flavor calculation of ref.~\cite{Braun:2014wpa}, where it was also argued that for the leading twist normalization constants discretization effects are expected to be larger than for the higher twist couplings.\par%
A heuristic parametrization of the leading discretization effects can be constructed by introducing a multiplicative factor into the extrapolation formulas. The leading corrections are linear in the lattice spacing, since the operators we use are not $\mathcal{O}(a)$ improved. At fixed mean quark mass this would yield, for instance, for the leading twist couplings:%
\begin{subequations}%
\begin{align}%
f^B &= g^B_{\Phi+} (\delta m) \bigl( 1 + a C + a \delta m D^B  \bigr) \bigl( f^\star + \delta m \Delta f^B \bigr) \,, \\
f_T^{B\neq\Lambda} &= g^B_{\Pi} (\delta m) \bigl( 1 + a C  + a \delta m D_T^B  \bigr) \bigl( f^\star + \delta m \Delta f_T^B \bigr) \,.
\end{align}%
\end{subequations}%
The constant $C$ has to be equal for all baryons in the octet while the $D_{(T)}^B$ can be different and are not necessarily subject to the same constraints as $\Delta f_{(T)}^B$. One can easily convince oneself that, at nonzero lattice spacing, terms $\mathcal{O}(a\delta m)$ can override the effect of the constraints given in eq.~\eqref{eq_symmetry_constraints}.\par%
In this work we only use data at a single lattice spacing and cannot study discretization effects. Therefore, for the time being, the difference between chiral extrapolations using constrained and unconstrained fits has to be interpreted as evidence for systematic uncertainties.%
\section{Results\label{sect_results}}%
The results of the chiral extrapolations as shown in figures~\ref{figure_f}--\ref{figure_phi10} are summarized in table~\ref{table_extrapolated} (constrained fit) and table~\ref{table_extrapolatedUC} (unconstrained fit). For all quantities the first error refers to a combined statistical and extrapolation error, and the second error is an estimate of the uncertainty due to the renormalization procedure as described in appendix~\ref{appendix_renormalization}. We do not expect significant finite volume effects~\cite{Wein:2011ix,Braun:2014wpa,Schiel:2011av} since all our ensembles have values of $m_\pi L \gtrsim 4$ and at the same time $L>\unit{2.7}{\femto\meter}$, cf.\ table~\ref{table_ensembles}. As discussed above, for some quantities the difference between constrained and unconstrained chiral extrapolations is sizable and can be viewed as part of the systematic uncertainty. Since the overall quality of the unconstrained fit is better ($\chi^2$ per degree of freedom is smaller than $1.5$ for all unconstrained fits), we present the corresponding numbers as our final results for this lattice spacing (see table~\ref{table_extrapolatedUC}). All further tables and figures in this section are generated using these values.\par%
\begin{table}[t]%
\centering%
\caption{\label{table_extrapolated} Couplings and shape parameters obtained by the constrained fit method. All values are given in units of $\squaren{\giga\electronvolt}$ in the $\MSbar$ scheme at a scale $\mu^2=\unit{4}{\squaren{\giga\electronvolt}}$. The number in the first parentheses gives a combined statistical and chiral extrapolation error. The second one is an estimate of the error due to the renormalization procedure.}%
\begin{tabular}{lD{.}{.}{3.10}D{.}{.}{3.9}D{.}{.}{3.8}D{.}{.}{3.10}}%
  \toprule
  $B$ & \multicolumn{1}{c}{$N$} & \multicolumn{1}{c}{$\Sigma$} & \multicolumn{1}{c}{$\Xi$} & \multicolumn{1}{c}{$\Lambda$}\\
  \midrule
$f^B\times 10^3$ & 3.61(3)(1) & 5.26(4)(2) & 5.48(4)(2) & 4.85(3)(2)\\
$f^B_T\times 10^3$ & 3.61(3)(1) & 5.10(3)(2) & 5.54(4)(2) & \mcemd\\
$\varphi_{11}^B\times 10^3$ & 0.06(1)(1) & 0.13(1)(2) & -0.01(1)(3) & 0.17(1)(1)\\
$\pi_{11}^B\times 10^3$ & 0.06(1)(1) & -0.09(1)(3) & 0.30(1)(1) & \mcemd\\
$\varphi_{10}^B\times 10^3$ & 0.074(10)(4) & -0.052(7)(2) & 0.15(1)(1) & 0.50(2)(3)\\
$\pi_{10}^B\times 10^3$ & \mcemd & \mcemd & \mcemd & 0.035(11)(2)\\
\midrule
$\varphi_{00,(1)}^B\times 10^3$ & 3.47(4)(2) & 5.05(5)(2) & 5.26(6)(2) & 4.67(5)(2)\\
$\pi_{00,(1)}^B\times 10^3$ & 3.47(4)(2) & 4.88(4)(2) & 5.35(6)(2) & \mcemd\\
\midrule
$\lambda_1^B\times 10^3$ & -48.4(4)(23) & -46.4(3)(22) & -47.6(3)(23) & -40\mathrlap{(1)(2)}\\
$\lambda_T^B\times 10^3$ & \mcemd & \mcemd & \mcemd & -52.5(4)(25)\\
$\lambda_2^B\times 10^3$ & 95\mathrlap{(1)(5)} & 87\mathrlap{(1)(4)} & 95\mathrlap{(1)(5)} & 105\mathrlap{(1)(5)}\\
  \bottomrule
\end{tabular}%
\end{table}%
Our results for the nucleon normalization constants (at $a\approx \unit{0.0857}{\femto\metre}$  with $N_f=2+1$) are approximately $30\%$ larger for $f^N$ and about $20\%$ larger in the case of $\lambda_1^N$ and $\lambda_2^N$, in comparison to the $N_f=2$ lattice study~\cite{Braun:2014wpa}, where a continuum extrapolation was performed. As one can see from figure~$7$ therein,\footnote{We refer to the figure numbers of the journal version of ref.~\cite{Braun:2014wpa}.} the continuum extrapolation from lattices with  $a\approx \unit{0.06-0.08}{\femto\metre}$ resulted in a decrease of $f^N$ by $\approx 30\%$ and a somewhat smaller decrease for $\lambda_{1,2}^N$, so that our results are in fact very compatible. Given that we use a similar lattice action, we have to expect discretization effects of the same magnitude as in~\cite{Braun:2014wpa}, and therefore, a thorough continuum extrapolation will be of utmost importance and is a primary goal for future studies. Note, however, that our results for the momentum sums $\varphi_{00,(1)}^B$ and $\pi_{00,(1)}^B$ defined in eq.~\eqref{eq_momentum-sums} are within $5\%$ of the corresponding couplings, cf. eq.~\eqref{eq_sumrule}, indicating that discretization errors in the derivatives are under control, see also figure~$8$ in ref.~\cite{Braun:2014wpa}.\par%
\begin{table}[t]%
\centering%
\caption{\label{table_extrapolatedUC} Couplings and shape parameters obtained fron the unconstrained fits. All values are given in units of $\squaren{\giga\electronvolt}$ in the $\MSbar$ scheme at a scale $\mu^2=\unit{4}{\squaren{\giga\electronvolt}}$. The number in the first parentheses gives a combined statistical and chiral extrapolation error. The second one is an estimate of the error due to the renormalization procedure. The numbers from this table should be quoted as the final results at our lattice spacing.}%
\begin{tabular}{lD{.}{.}{3.10}D{.}{.}{3.10}D{.}{.}{3.10}D{.}{.}{3.10}}%
  \toprule
  $B$ & \multicolumn{1}{c}{$N$} & \multicolumn{1}{c}{$\Sigma$} & \multicolumn{1}{c}{$\Xi$} & \multicolumn{1}{c}{$\Lambda$}\\
  \midrule
$f^B\times 10^3$ & 3.60(6)(2) & 5.07(5)(2) & 5.38(5)(2) & 4.38(6)(2)\\
$f^B_T\times 10^3$ & 3.60(6)(2) & 4.88(5)(2) & 5.47(5)(2) & \mcemd\\
$\varphi_{11}^B\times 10^3$ & 0.08(2)(1) & 0.17(1)(2) & 0.01(1)(2) & 0.18(1)(1)\\
$\pi_{11}^B\times 10^3$ & 0.08(2)(1) & -0.10(1)(3) & 0.30(1)(1) & \mcemd\\
$\varphi_{10}^B\times 10^3$ & 0.060(19)(3) & -0.069(10)(3) & 0.14(1)(1) & 0.48(2)(3)\\
$\pi_{10}^B\times 10^3$ & \mcemd & \mcemd & \mcemd & 0.010(16)(1)\\
\midrule
$\varphi_{00,(1)}^B\times 10^3$ & 3.53(9)(2) & 4.91(7)(2) & 5.19(6)(2) & 4.25(8)(2)\\
$\pi_{00,(1)}^B\times 10^3$ & 3.53(9)(2) & 4.70(6)(2) & 5.31(6)(2) & \mcemd\\
\midrule
$\lambda_1^B\times 10^3$ & -49\mathrlap{(1)(2)} & -45.4(4)(21) & -47.6(4)(23) & -39\mathrlap{(1)(2)}\\
$\lambda_T^B\times 10^3$ & \mcemd & \mcemd & \mcemd & -51\mathrlap{(1)(2)}\\
$\lambda_2^B\times 10^3$ & 98\mathrlap{(1)(5)} & 86\mathrlap{(1)(4)} & 96\mathrlap{(1)(5)} & 101\mathrlap{(1)(5)}\\
  \bottomrule
\end{tabular}%
\end{table}%
Our results for the first order shape parameters of the leading twist DA of the nucleon, $\varphi_{11}^N=\pi_{11}^N$ and  $\varphi_{10}^N$, agree with the results of ref.~\cite{Braun:2014wpa} within errors,\footnote{In contrast to the normalization constants, the shape parameters have not been extrapolated to the continuum in ref.~\cite{Braun:2014wpa}.} and also with the parameters extracted from the study of the nucleon electromagnetic form factors in light-cone sum rules~\cite{Anikin:2013aka}. Note that our $\varphi^N_{nk}$ correspond to $f_N \varphi^N_{nk}$ in refs.~\cite{Anikin:2013aka,Braun:2014wpa}. We also confirm the approximate equality $\varphi_{10}^N\approx \varphi_{11}^N$ found in~ref.~\cite{Braun:2014wpa}. Our results for the shape parameters of hyperons are, however, up to an order of magnitude smaller than the values obtained using QCD sum rules~\cite{Chernyak:1987nu}, see table~\ref{table_cozparison}. In ref.~\cite{Braun:2014wpa} it has already been reported that, in general, modern lattice simulations and light-cone sum rule calculations yield estimates of the first moments of the nucleon DA that are one order of magnitude smaller than in ``old'' phenomenology, cf. refs.~\cite{Chernyak:1983ej,Chernyak:1987nu}. Our measurements confirm this observation also for the hyperons.\par%
\begin{table}[t]%
\centering%
\caption{\label{table_cozparison}Comparison of the central values of our $N_f=2+1$ results (unconstrained fit, see table~\ref{table_extrapolatedUC}) with the $N_f=2$ lattice study for the nucleon~\cite{Braun:2014wpa} and the Chernyak--Ogloblin--Zhitnitsky (COZ) model~\cite{Chernyak:1987nu}. All values are given in units of $\giga\electronvolt\squared$. All quantities have been converted to the conventions established in this work and rescaled to $\mu^2=\unit{4}{\giga\electronvolt\squared}$, using the three-loop evolution equation for the couplings with the anomalous dimensions calculated in ref.~\cite{Gracey:2012gx}, and the one-loop equation~\eqref{eq_oneloop_evolution} for the shape parameters. Note that $f^T_\Lambda$ in ref.~\cite{Chernyak:1987nu} is proportional to the first moment $\pi_{10}^\Lambda$ in our nomenclature.}%
\begin{tabular}{lccD{.}{.}{1.2}D{.}{.}{1.2}D{.}{.}{1.3}D{.}{.}{2.3}D{.}{.}{2.3}D{.}{.}{1.2}}%
\toprule
$B$ & \clap{work} & method & \multicolumn{1}{c}{$f^B\times 10^3$} & \multicolumn{1}{c}{$f_T^B\times 10^3$} & \multicolumn{1}{c}{$\varphi_{11}^B\times 10^3$} & \multicolumn{1}{c}{$\pi_{11}^B\times 10^3$} & \multicolumn{1}{c}{$\varphi_{10}^B\times 10^3$} & \multicolumn{1}{c}{$\pi_{10}^B\times 10^3$}\\
\midrule
\multirow{3}{*}{$N$}
&\text{ours}                   & $N_f=2+1$             & 3.60 & 3.60 & 0.08   & 0.08   & 0.06  & \mcemd\\
&\text{\cite{Braun:2014wpa}}   & $N_f=2\phantom{{}+1}$ & 2.84 & 2.84 & 0.085  & 0.085  & 0.082  & \mcemd\\
&\text{\cite{Chernyak:1987nu}} & COZ                   & 4.55 & 4.55 & 0.885  & 0.885  & 0.748  & \mcemd\\
\midrule
\multirow{2}{*}{$\Sigma$}
&\text{ours}                   & $N_f=2+1$             & 5.07 & 4.88 & 0.17  & - 0.10  & - 0.069 & \mcemd\\
&\text{\cite{Chernyak:1987nu}} & COZ                   & 4.65 & 4.46 & 1.11  &   0.511 &   0.523 & \mcemd\\
\midrule
\multirow{2}{*}{$\Xi$}
&\text{ours}                   & $N_f=2+1$             & 5.38 & 5.47 & 0.01  & 0.30  & 0.14  & \mcemd\\
&\text{\cite{Chernyak:1987nu}} & COZ                   & 4.83 & 4.92 & 0.685 & 1.10  & 0.883 & \mcemd\\
\midrule
\multirow{2}{*}{$\Lambda$}
&\text{ours}                   & $N_f=2+1$             & 4.38 & \mcemd & 0.18  & \mcemd & 0.48  & 0.01\\
&\text{\cite{Chernyak:1987nu}} & COZ                   & 4.69 & \mcemd & 1.05  & \mcemd & 1.39  & 1.32\\
\bottomrule
\end{tabular}%
\end{table}%
Interestingly, the \SU3 breaking in the shape parameters of the octet baryons turns out to be very large, e.g., $\pi_{11}^\Xi\gtrsim3\varphi_{11}^N$ and $\varphi_{10}^\Lambda\gtrsim7\varphi_{10}^N$. This effect is much stronger than in QCD sum rule calculations~\cite{Chernyak:1987nu}, even though the absolute values are much smaller. This large \SU3 breaking is somewhat surprising as shape parameters have autonomous scale dependence and should be viewed as independent nonperturbative parameters, and is in stark contrast to the situation for the normalization constants where the differences between octet baryons are at most $50\%$. As a consequence, \SU3 breaking in hard exclusive reactions that are sensitive to the deviations of the DAs from their asymptotic form can be enhanced.\par%
\begin{figure}[p]%
\centering%
\includegraphics[width=0.46\textwidth]{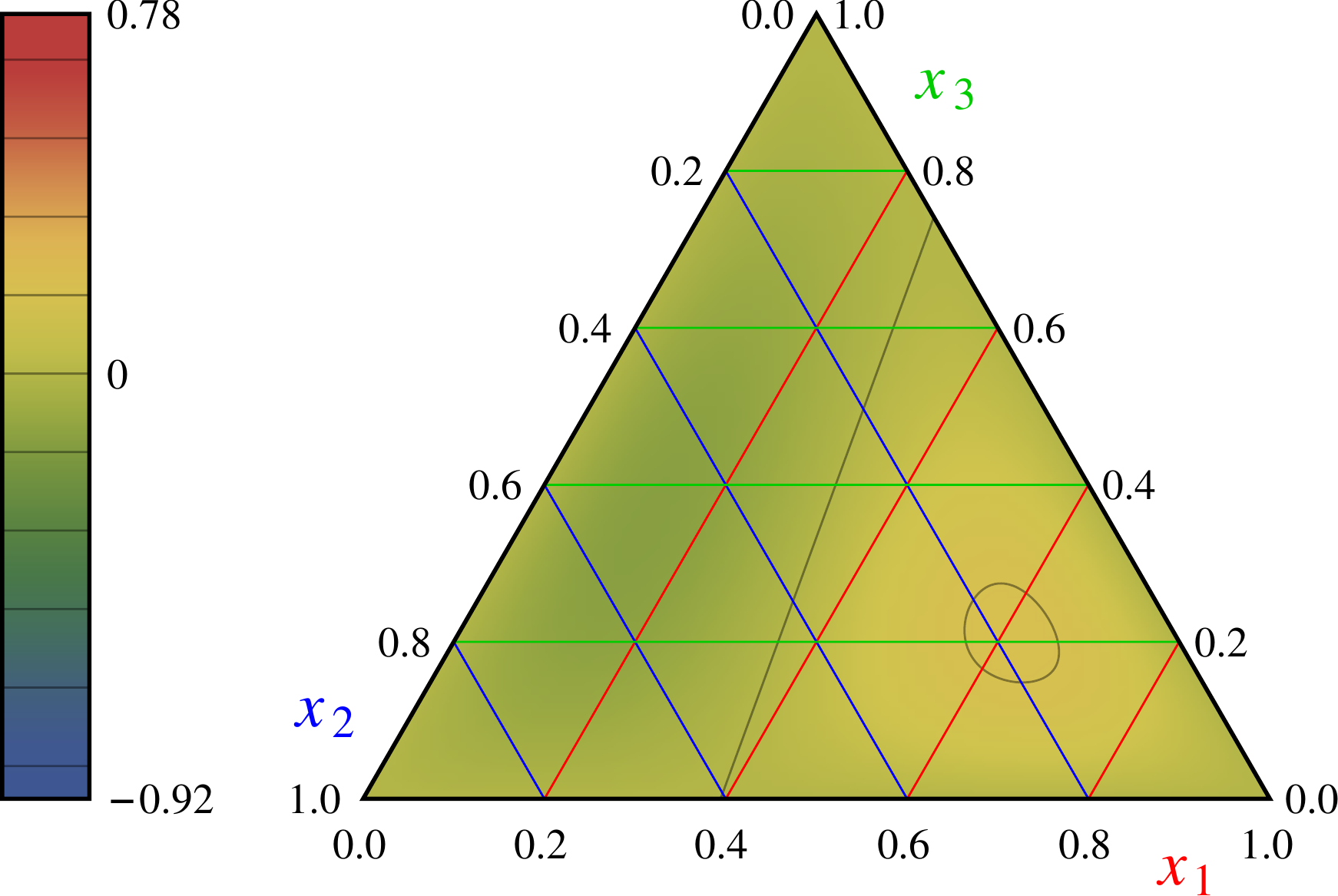}\raisebox{4.13cm}{$\mathllap{\mathclap{\boxed{\phi^N-\phi^{\star}}}\hspace{0.31\textwidth}}$}\hspace{0.079\textwidth}%
\includegraphics[width=0.46\textwidth]{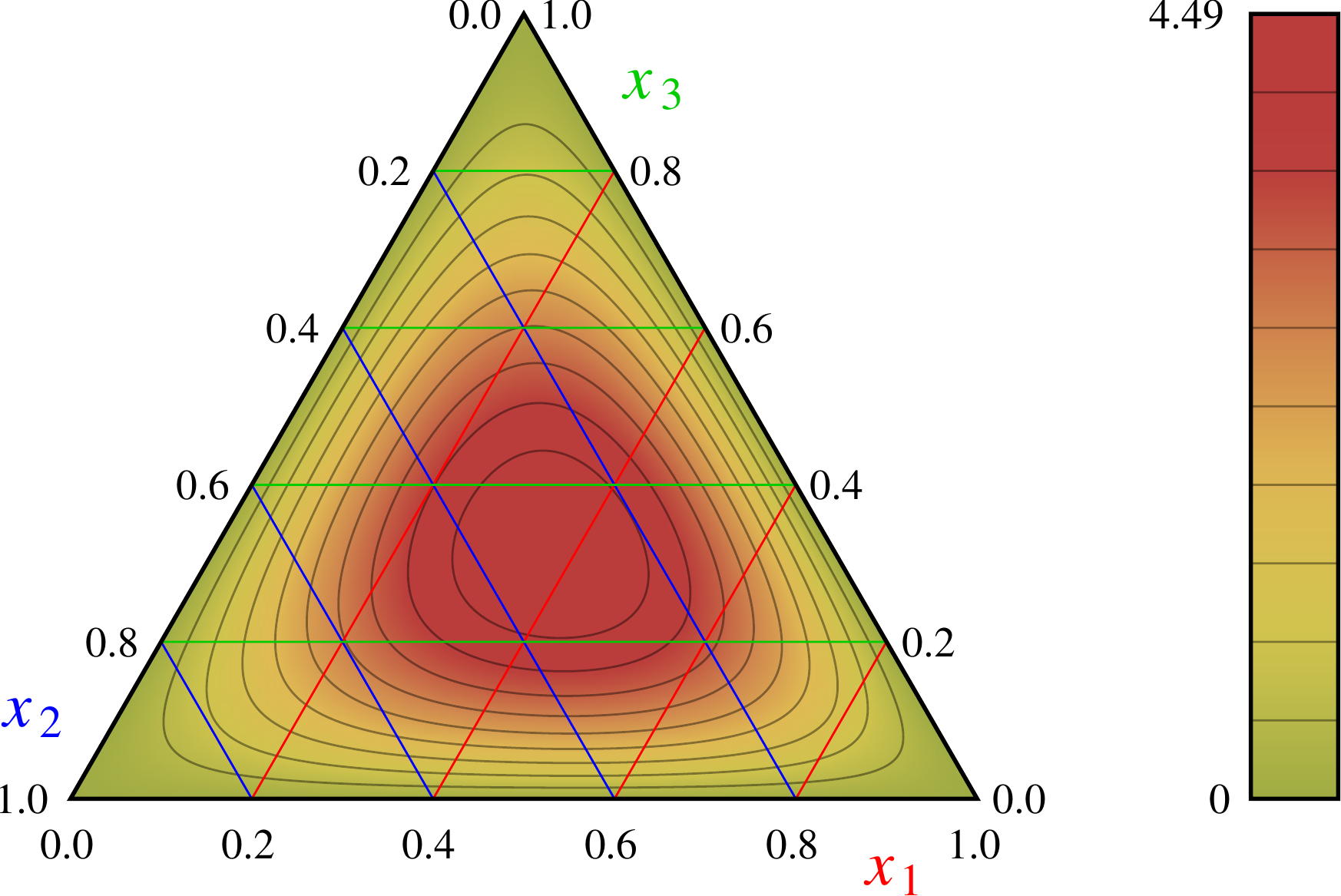}\raisebox{4.13cm}{$\mathllap{\mathclap{\boxed{\phi^{\star}}}\hspace{0.15\textwidth}}$}\\[\baselineskip]%
\includegraphics[width=0.46\textwidth]{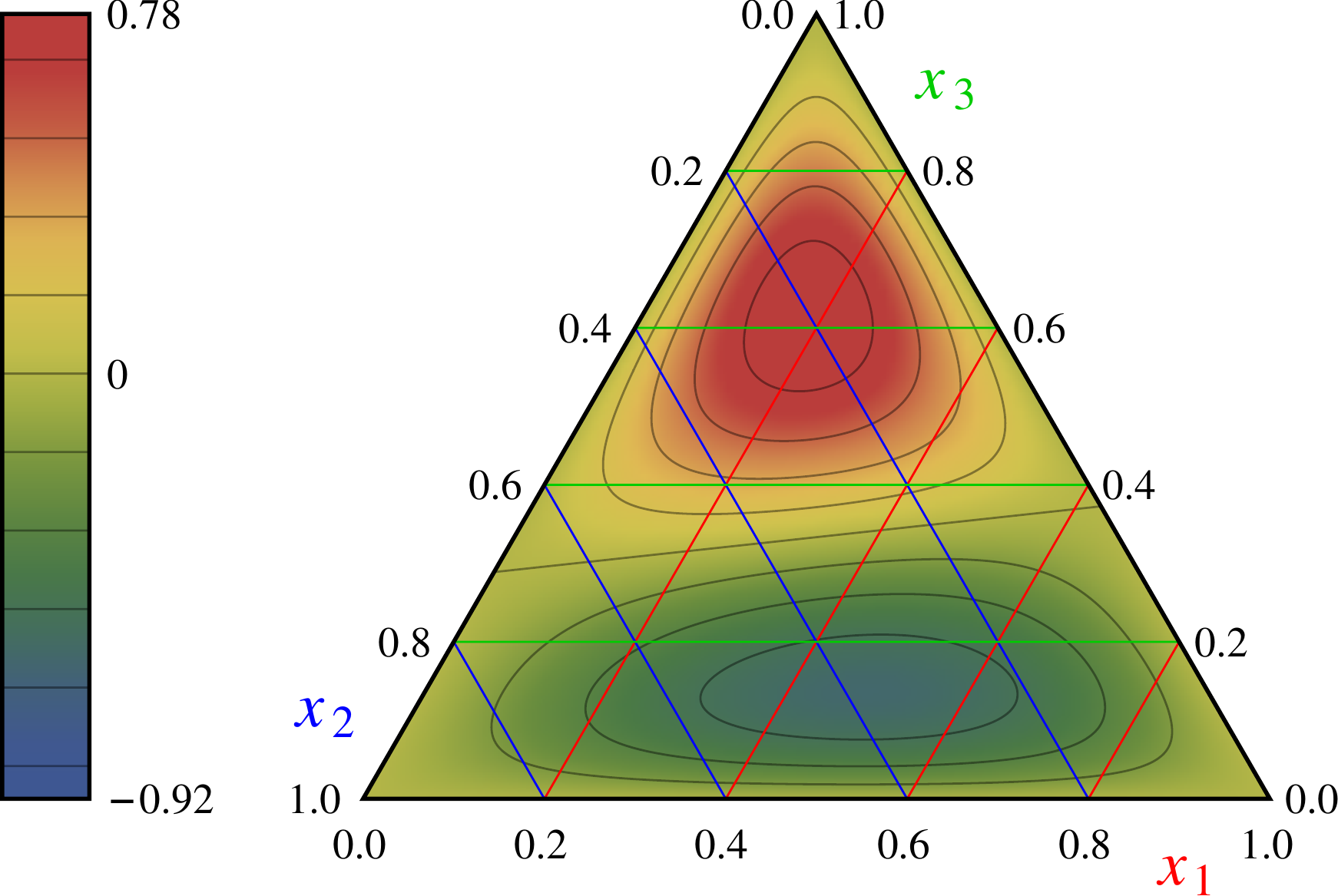}\raisebox{4.13cm}{$\mathllap{\mathclap{\boxed{\phi^\Sigma-\phi^{\star}}}\hspace{0.31\textwidth}}$}\hspace{0.079\textwidth}%
\includegraphics[width=0.46\textwidth]{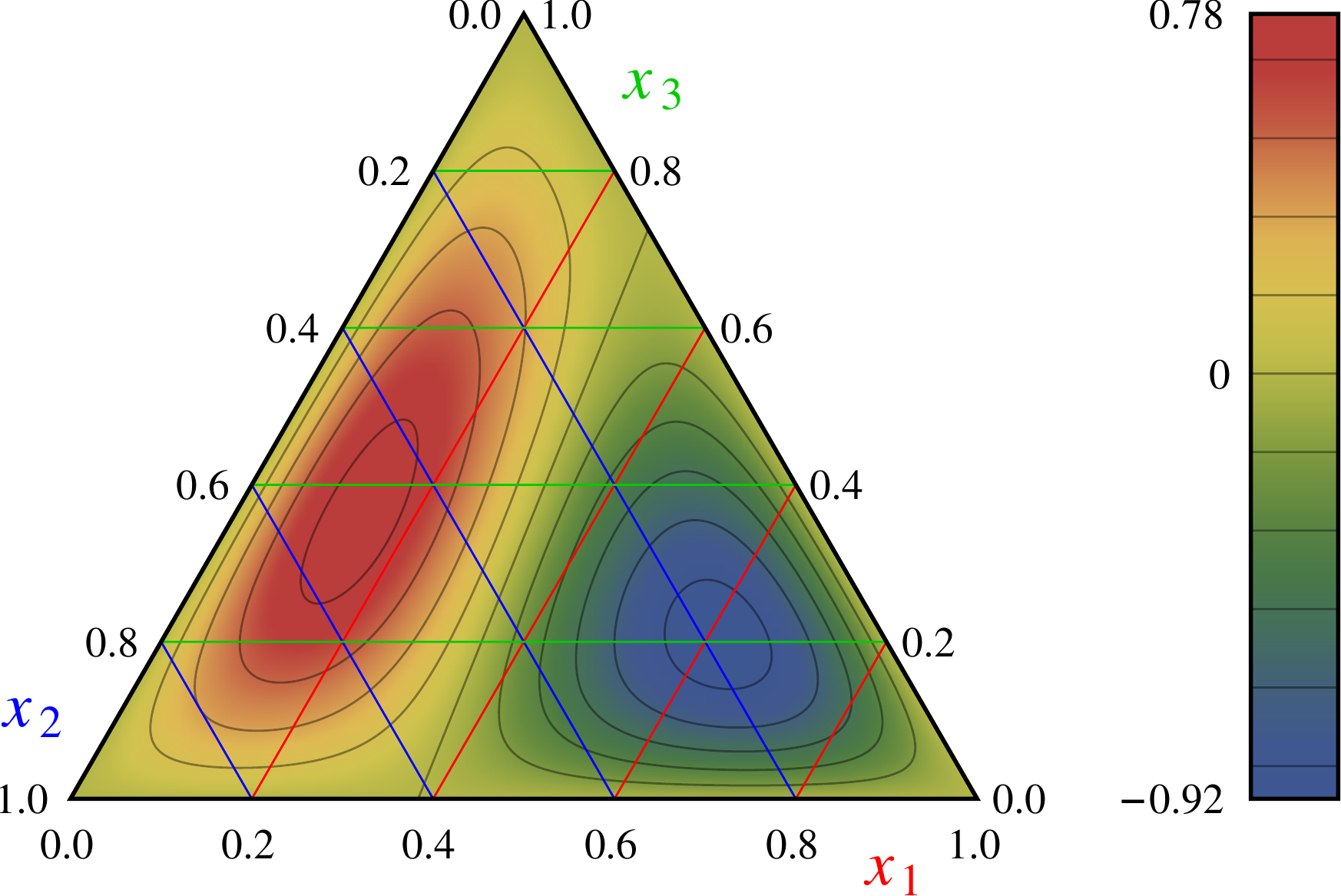}\raisebox{4.13cm}{$\mathllap{\mathclap{\boxed{\varpi^\Sigma-\phi^{\star}}}\hspace{0.15\textwidth}}$}\\[\baselineskip]%
\includegraphics[width=0.46\textwidth]{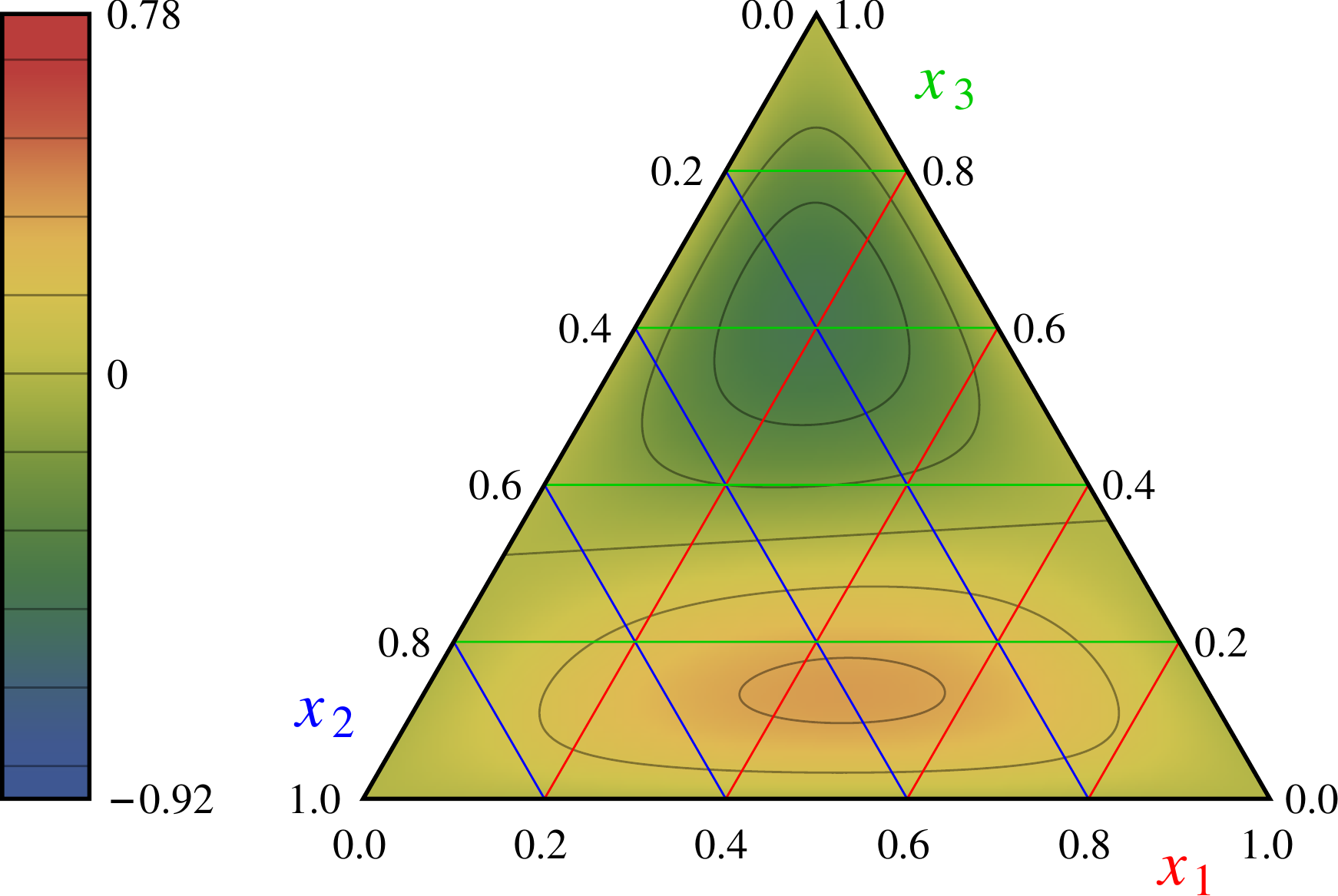}\raisebox{4.13cm}{$\mathllap{\mathclap{\boxed{\phi^\Xi-\phi^{\star}}}\hspace{0.31\textwidth}}$}\hspace{0.079\textwidth}%
\includegraphics[width=0.46\textwidth]{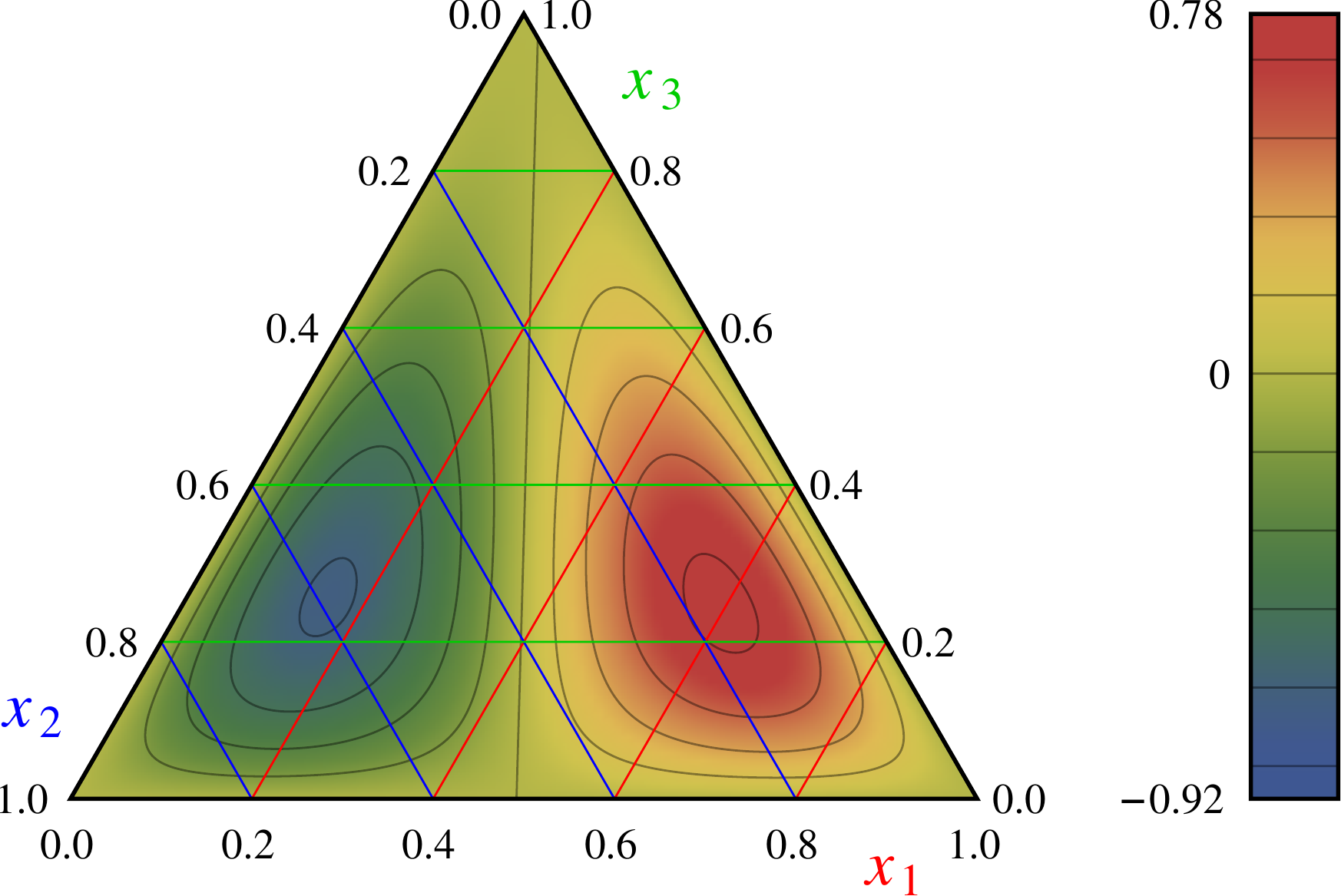}\raisebox{4.13cm}{$\mathllap{\mathclap{\boxed{\varpi^\Xi-\phi^{\star}}}\hspace{0.15\textwidth}}$}\\[\baselineskip]%
\includegraphics[width=0.46\textwidth]{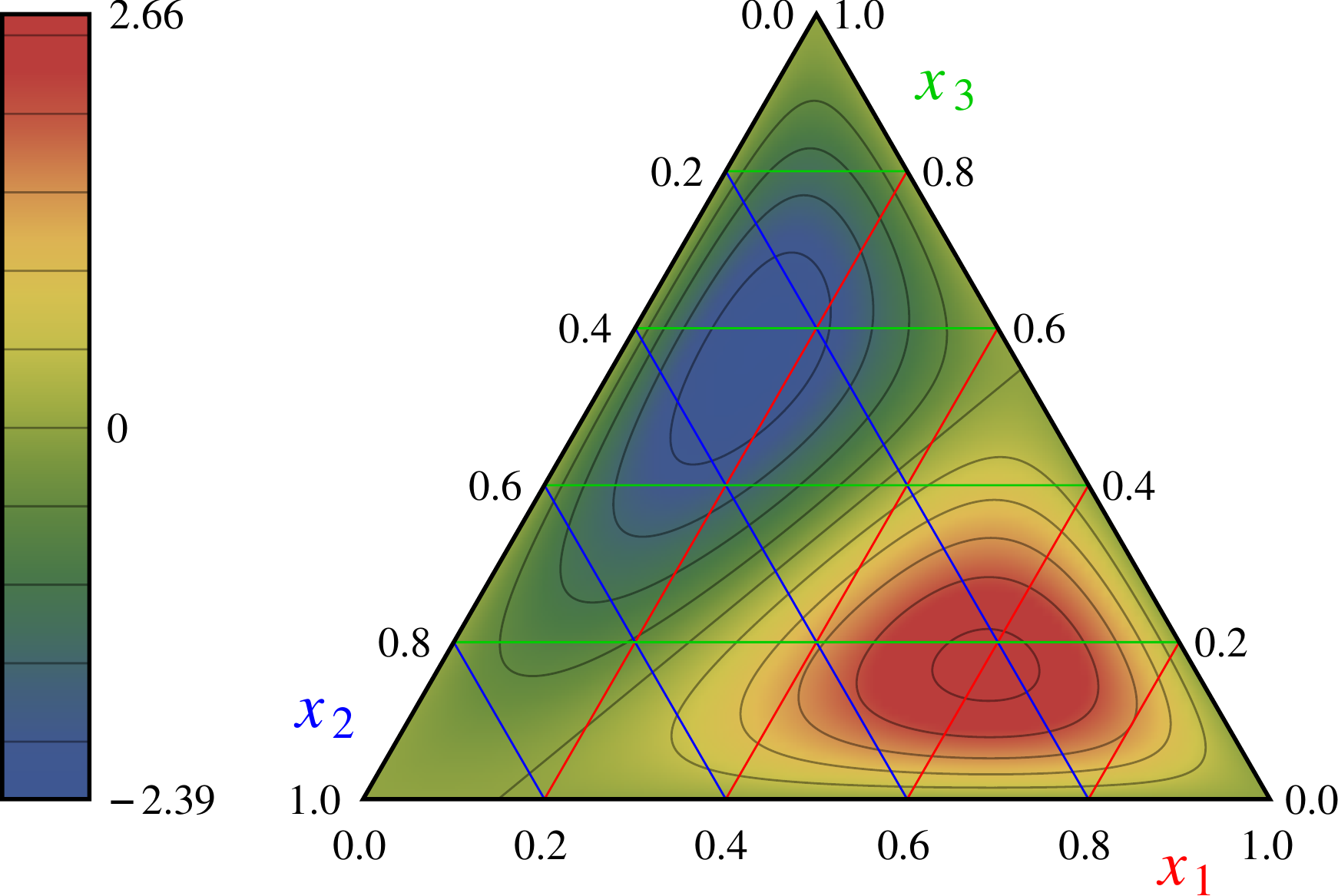}\raisebox{4.13cm}{$\mathllap{\mathclap{\boxed{\phi^\Lambda-\phi^{\star}}}\hspace{0.31\textwidth}}$}\hspace{0.079\textwidth}%
\includegraphics[width=0.46\textwidth]{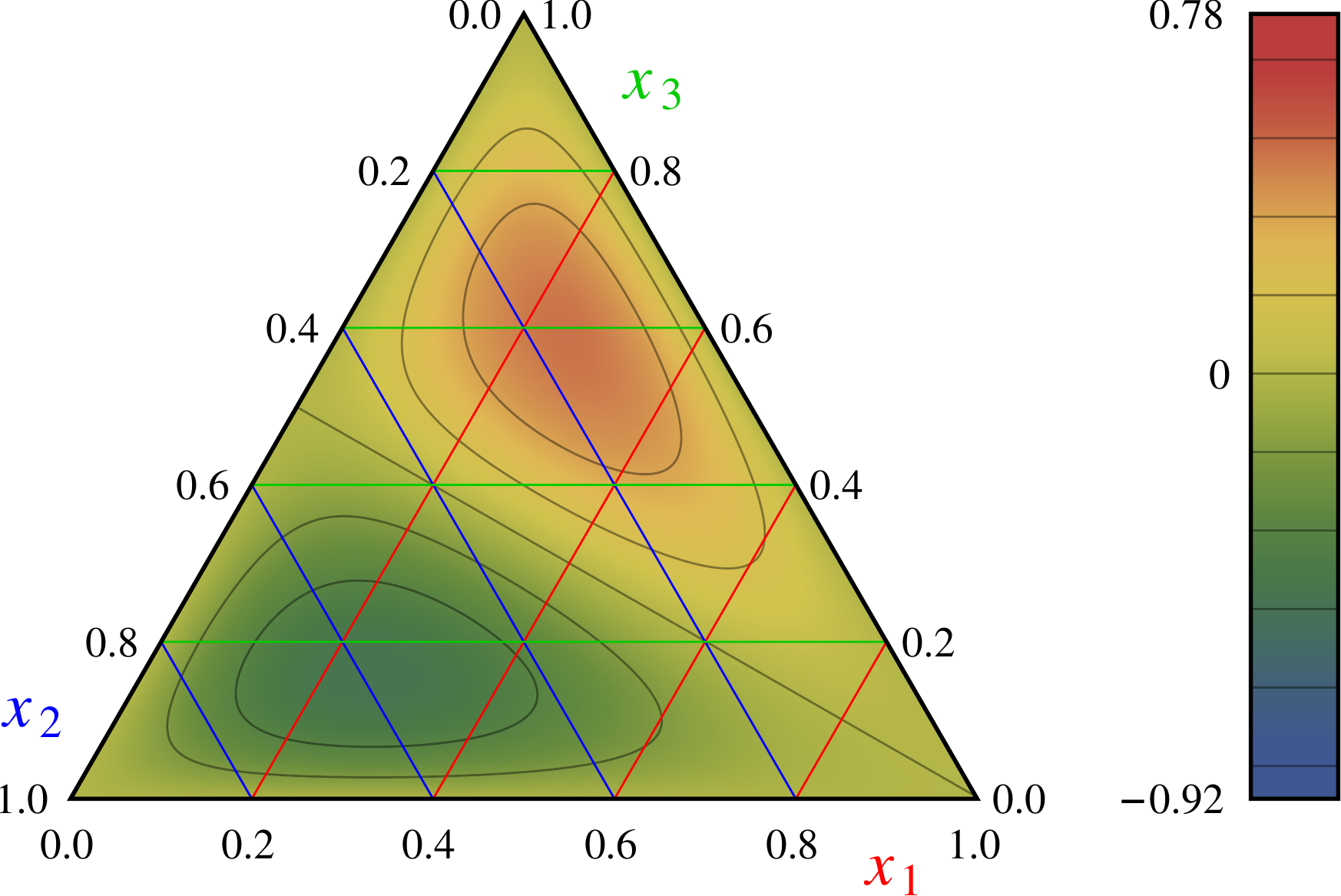}\raisebox{4.13cm}{$\mathllap{\mathclap{\boxed{\varpi^\Lambda-\phi^{\star}}}\hspace{0.15\textwidth}}$}%
\caption{\label{figure_barycentric_phipi}Barycentric plots ($x_1+x_2+x_3=1$) visualizing the \SU3 breaking in the shape functions~\eqref{eq_shape_functions}. The top right figure displays the momentum distribution for the flavor symmetric case, while the others show the deviations from it at the physical point.}%
\end{figure}%
\begin{figure}[p]%
\definecolor{mygreen}{rgb}{0,0.8,0}%
\centering
\includegraphics[width=0.46\textwidth]{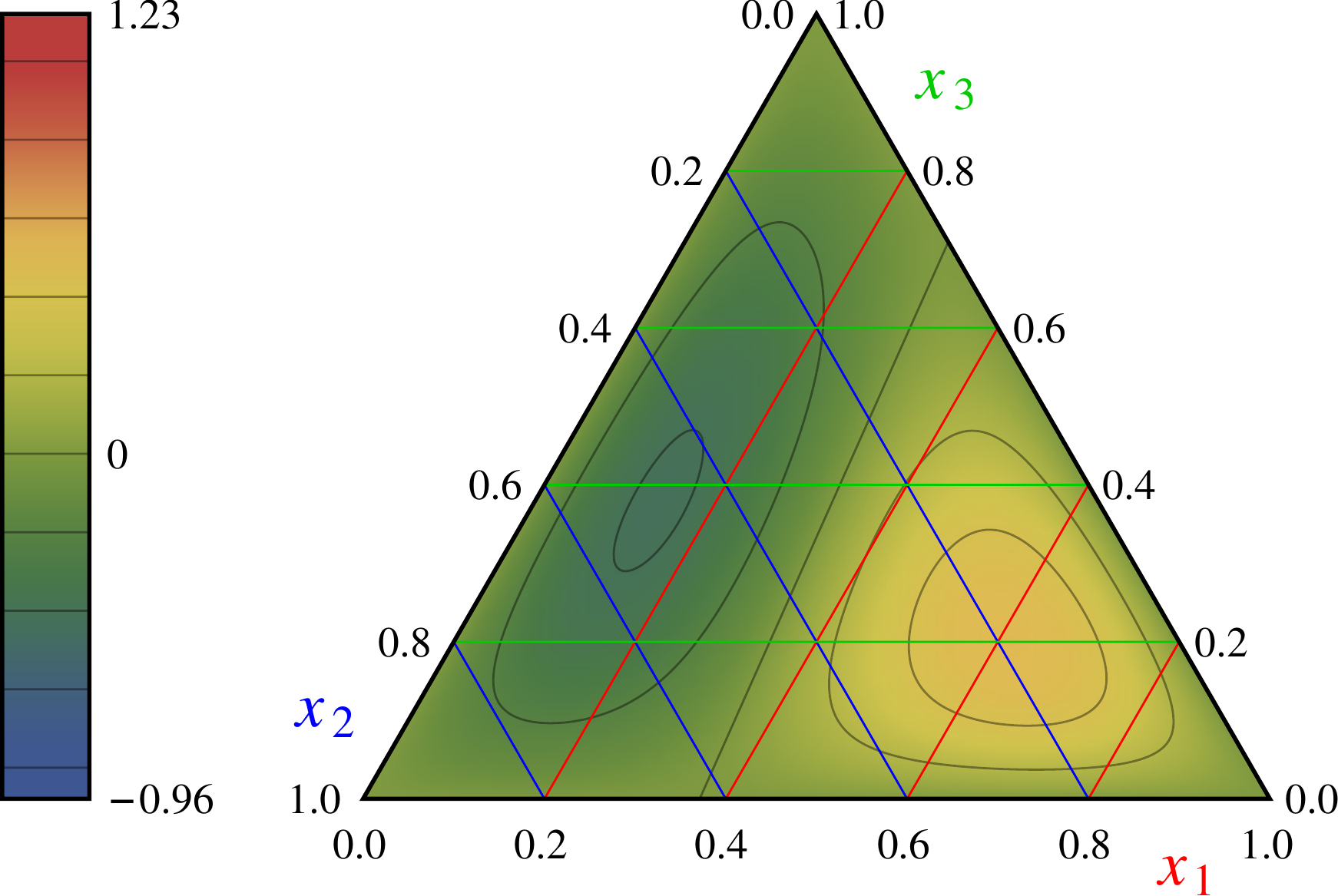}\raisebox{4.035cm}{$\mathllap{\mathclap{\boxed{\tfrac{[V-A]^{\mathrlap{N}}}{f^N}-\phi^{\text{as}}}}\hspace{0.32\textwidth}}$}\raisebox{4.13cm}{$\mathllap{\boxed{\color{red}u^\gooduparrow \color{blue}u^\gooddownarrow \color{mygreen}d^\gooduparrow}}$}\hspace{0.079\textwidth}%
\includegraphics[width=0.46\textwidth]{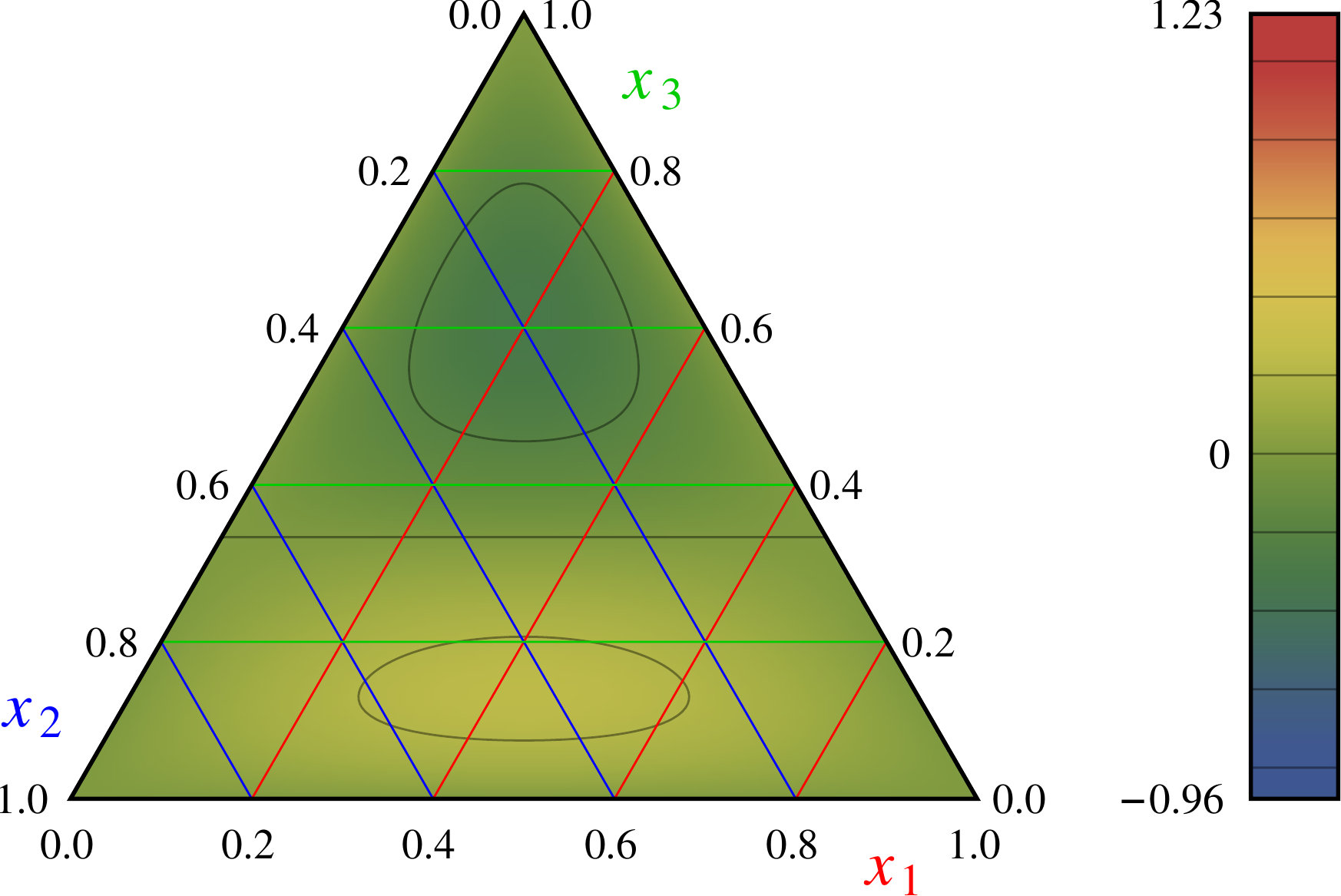}\raisebox{4.07cm}{$\mathllap{\mathclap{\boxed{\tfrac{T^N}{f^N}-\phi^{\text{as}}}}\hspace{0.14\textwidth}}$}\raisebox{4.13cm}{$\mathllap{\mathrlap{\boxed{\color{red}u^\gooduparrow \color{blue}u^\gooduparrow \color{mygreen}d^\gooddownarrow}}\hspace{0.46\textwidth}}$}\\[\baselineskip]%
\includegraphics[width=0.46\textwidth]{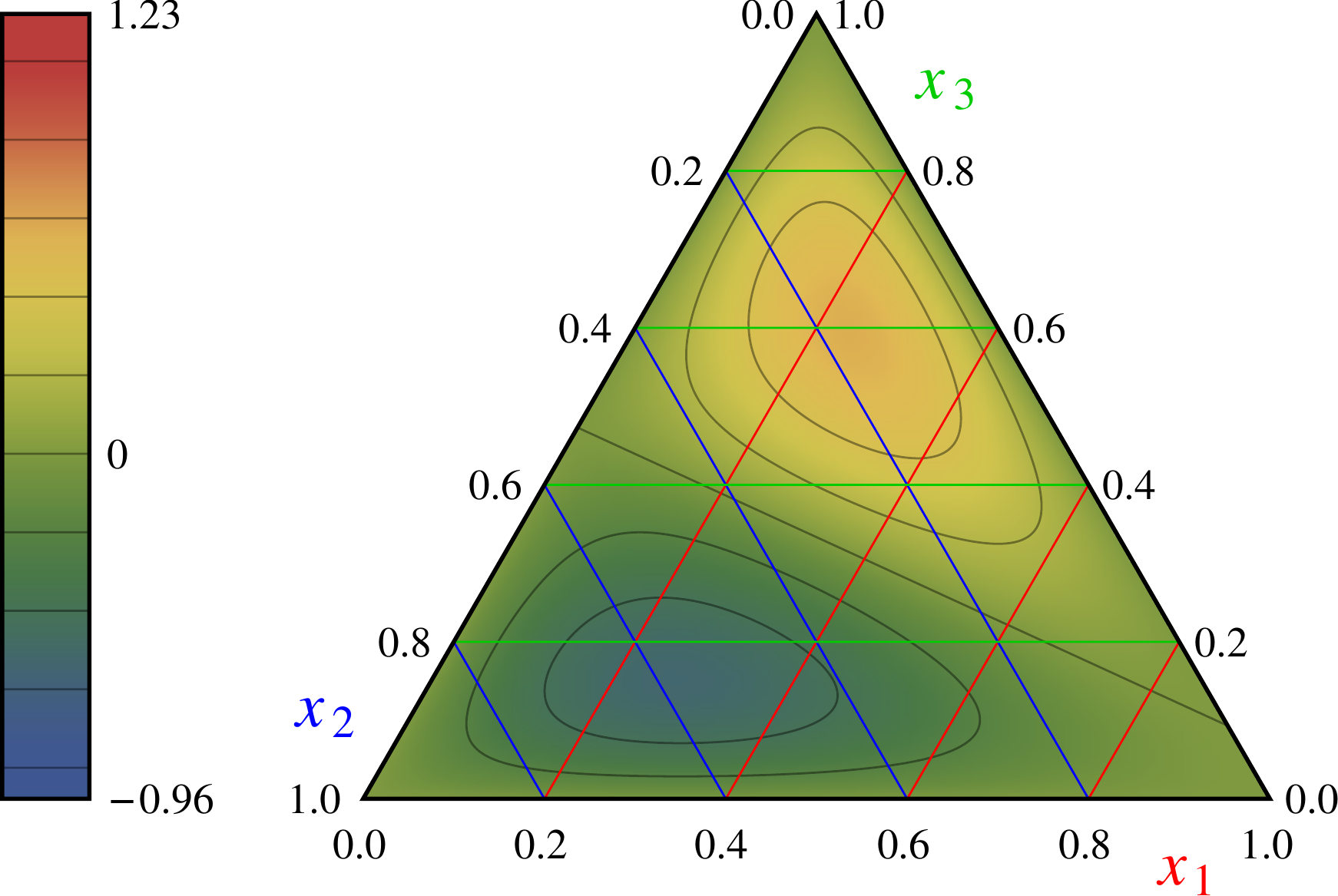}\raisebox{4.035cm}{$\mathllap{\mathclap{\boxed{\tfrac{[V-A]^{\mathrlap{\Sigma}}}{f^\Sigma}-\phi^{\text{as}}}}\hspace{0.32\textwidth}}$}\raisebox{4.13cm}{$\mathllap{\boxed{\color{red}d^\gooduparrow \color{blue}d^\gooddownarrow \color{mygreen}s^\gooduparrow}}$}\hspace{0.079\textwidth}%
\includegraphics[width=0.46\textwidth]{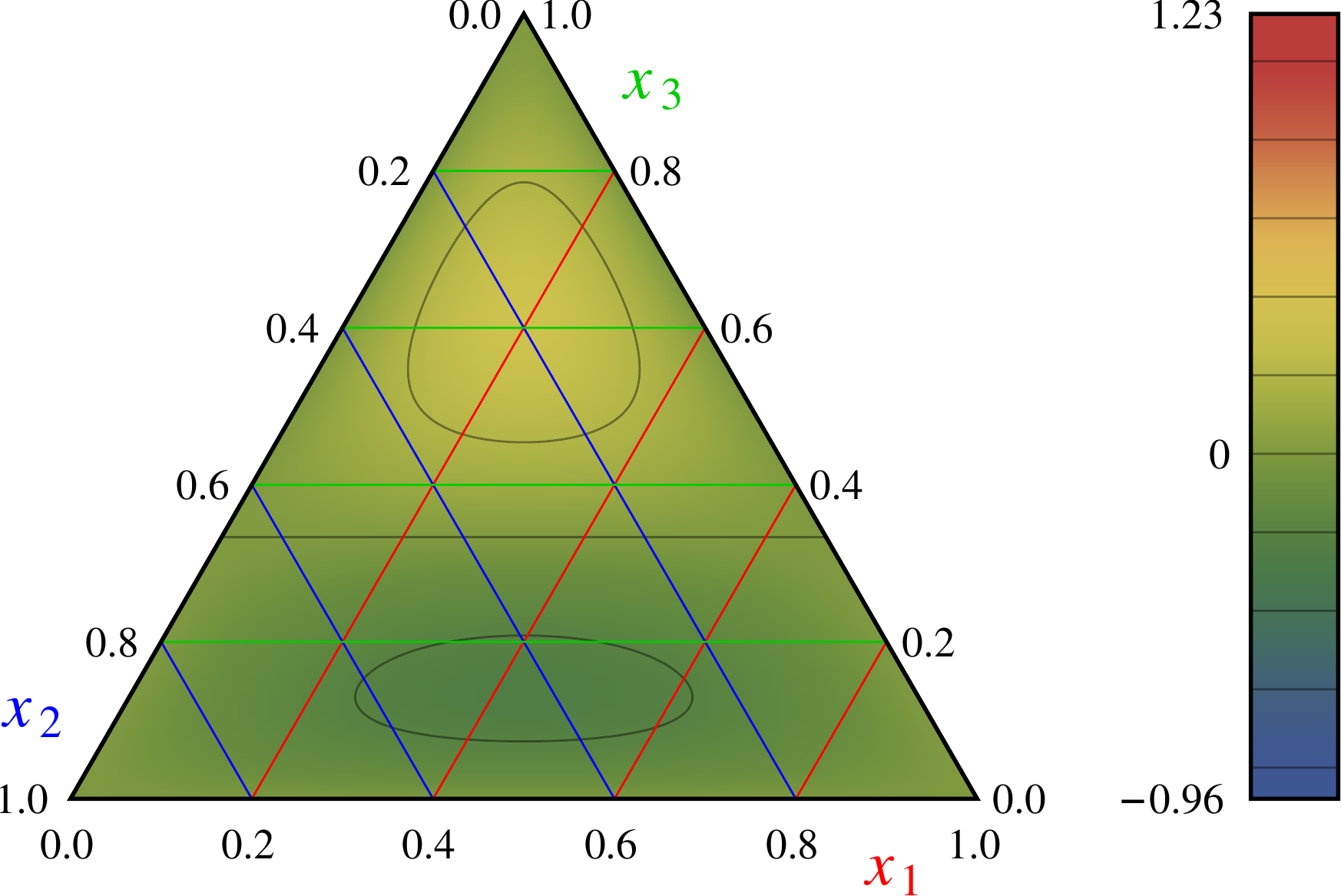}\raisebox{4.07cm}{$\mathllap{\mathclap{\boxed{\tfrac{T^\Sigma}{f_T^\Sigma}-\phi^{\text{as}}}}\hspace{0.14\textwidth}}$}\raisebox{4.13cm}{$\mathllap{\mathrlap{\boxed{\color{red}d^\gooduparrow \color{blue}d^\gooduparrow \color{mygreen}s^\gooddownarrow}}\hspace{0.46\textwidth}}$}\\[\baselineskip]%
\includegraphics[width=0.46\textwidth]{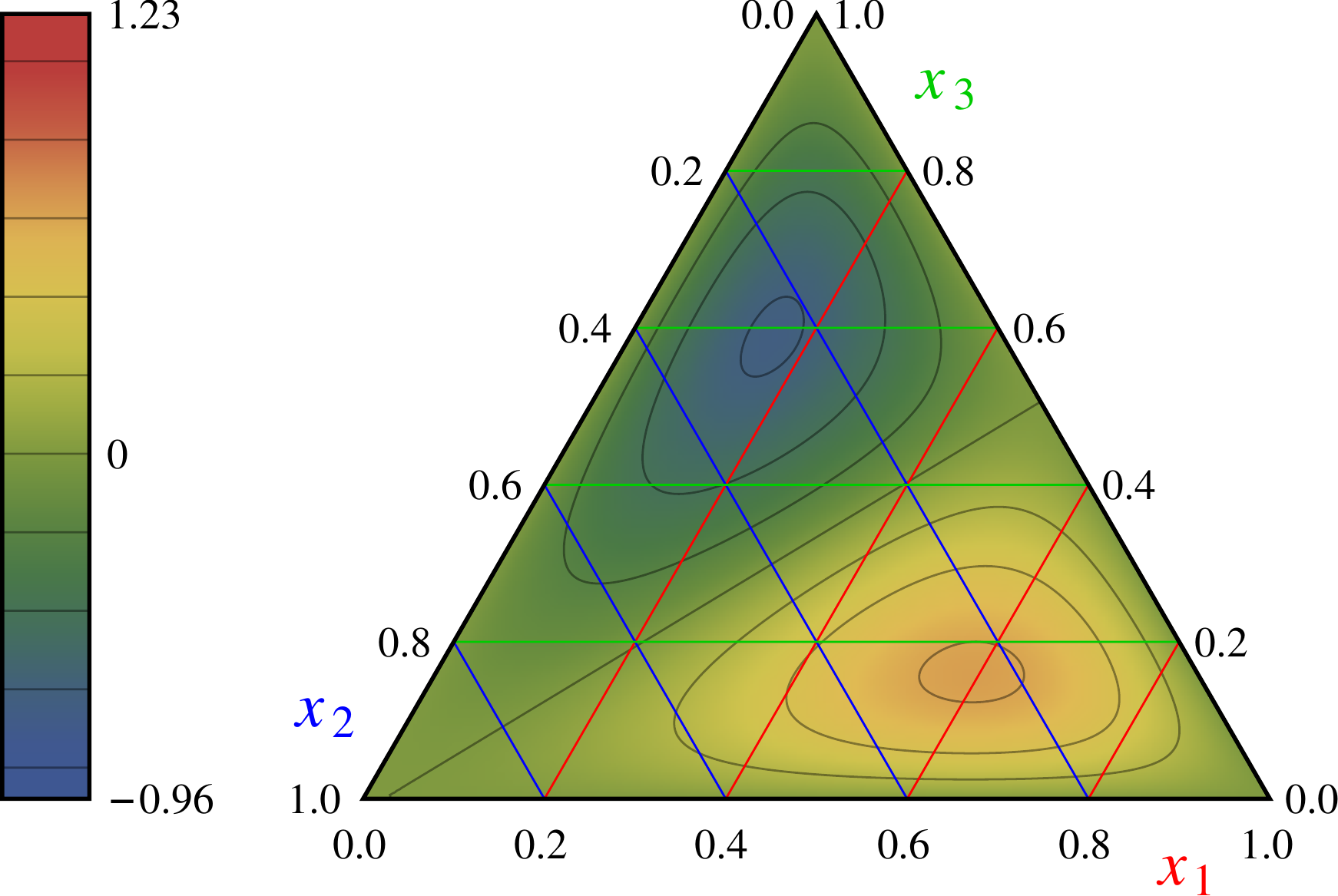}\raisebox{4.035cm}{$\mathllap{\mathclap{\boxed{\tfrac{[V-A]^{\mathrlap{\Xi}}}{f^\Xi}-\phi^{\text{as}}}}\hspace{0.32\textwidth}}$}\raisebox{4.13cm}{$\mathllap{\boxed{\color{red}s^\gooduparrow \color{blue}s^\gooddownarrow \color{mygreen}u^\gooduparrow}}$}\hspace{0.079\textwidth}%
\includegraphics[width=0.46\textwidth]{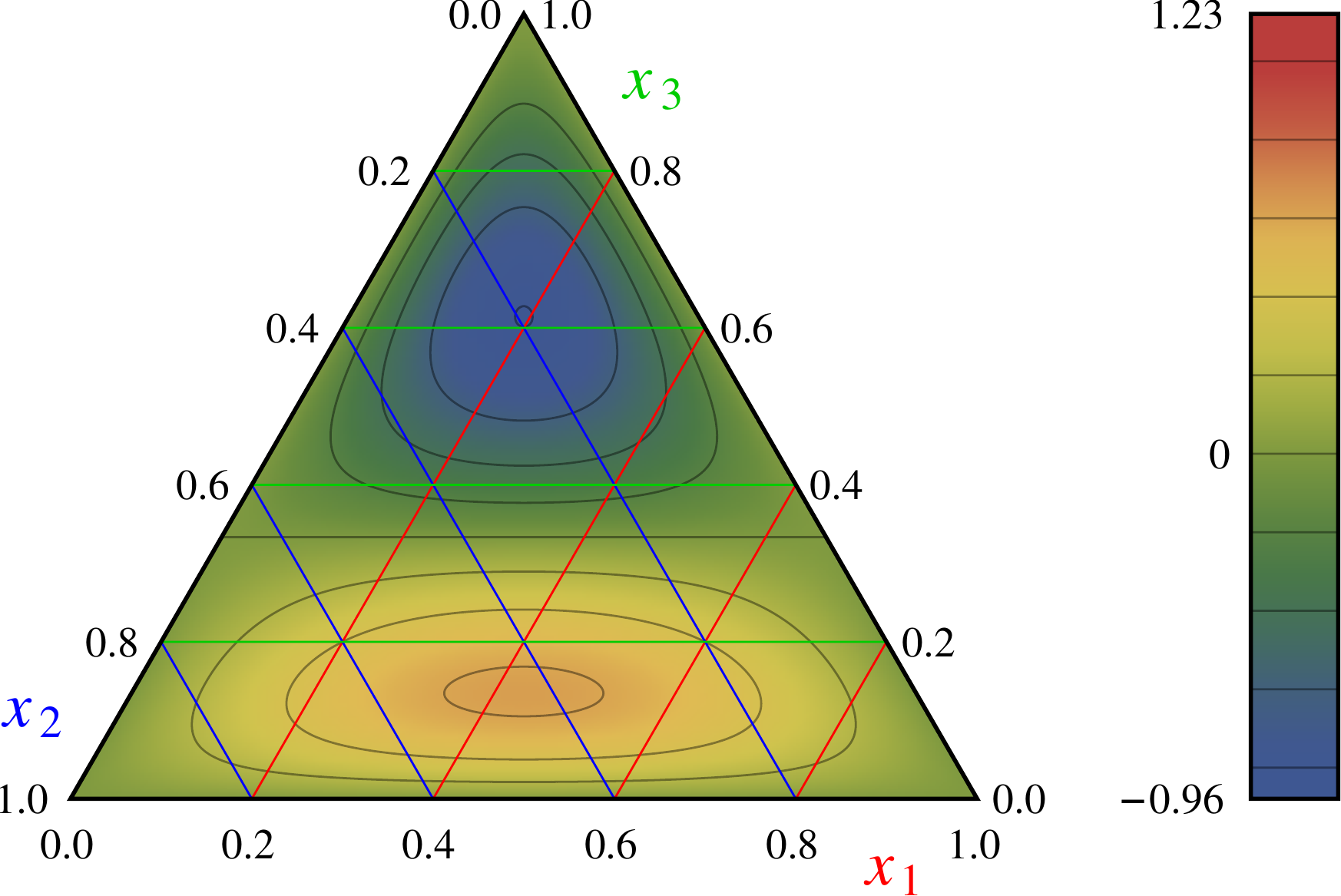}\raisebox{4.07cm}{$\mathllap{\mathclap{\boxed{\tfrac{T^\Xi}{f_T^\Xi}-\phi^{\text{as}}}}\hspace{0.14\textwidth}}$}\raisebox{4.13cm}{$\mathllap{\mathrlap{\boxed{\color{red}s^\gooduparrow \color{blue}s^\gooduparrow \color{mygreen}u^\gooddownarrow}}\hspace{0.46\textwidth}}$}\\[\baselineskip]%
\includegraphics[width=0.46\textwidth]{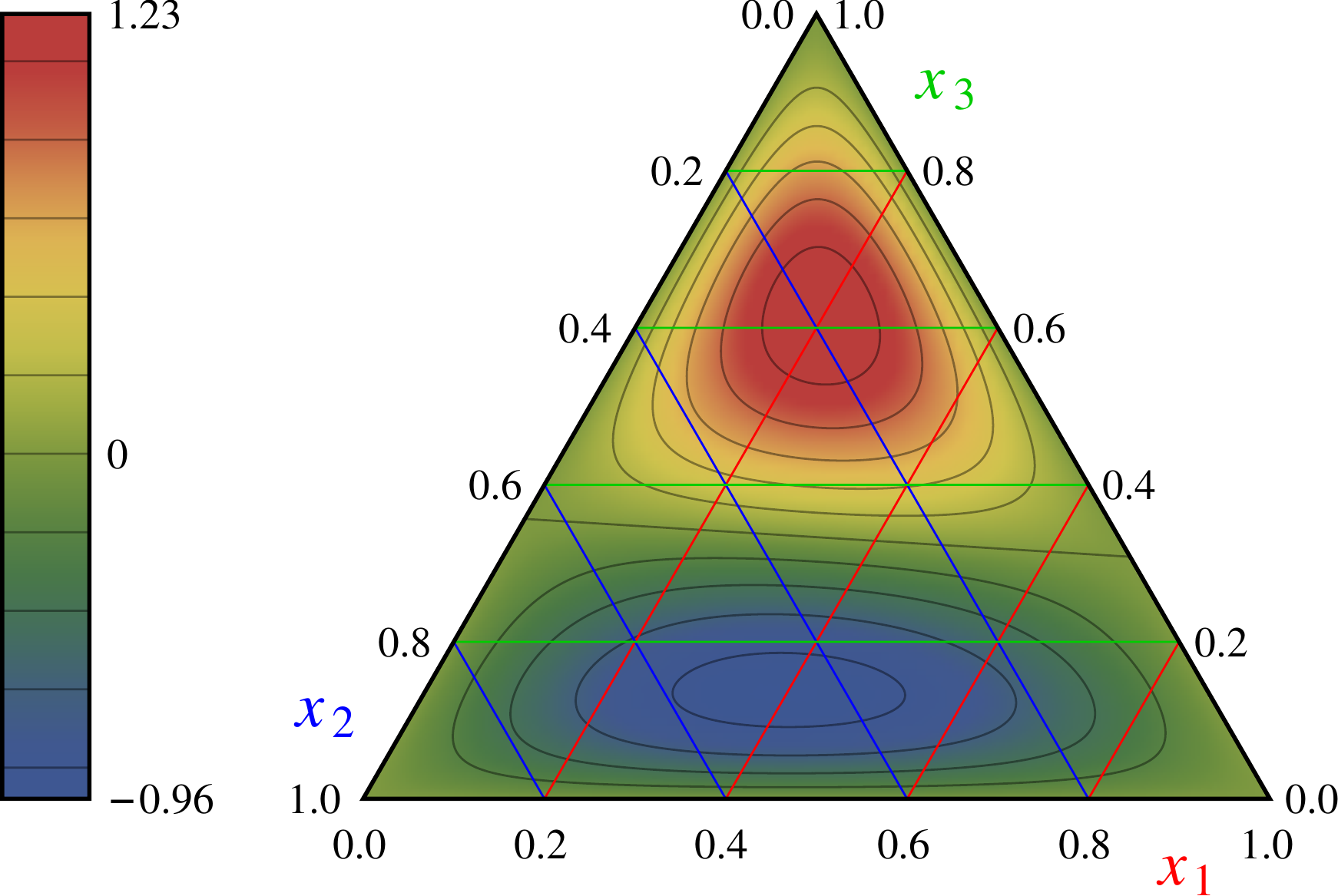}\raisebox{4.035cm}{$\mathllap{\mathclap{\boxed{\tfrac{[V-A]^{\mathrlap{\Lambda}}}{{\scriptscriptstyle\sqrt{\frac32}}f^\Lambda}-\phi^{\text{as}}}}\hspace{0.32\textwidth}}$}\raisebox{4.13cm}{$\mathllap{\boxed{\color{red}u^\gooduparrow \color{blue}d^\gooddownarrow \color{mygreen}s^\gooduparrow}}$}\hspace{0.079\textwidth}%
\includegraphics[width=0.46\textwidth]{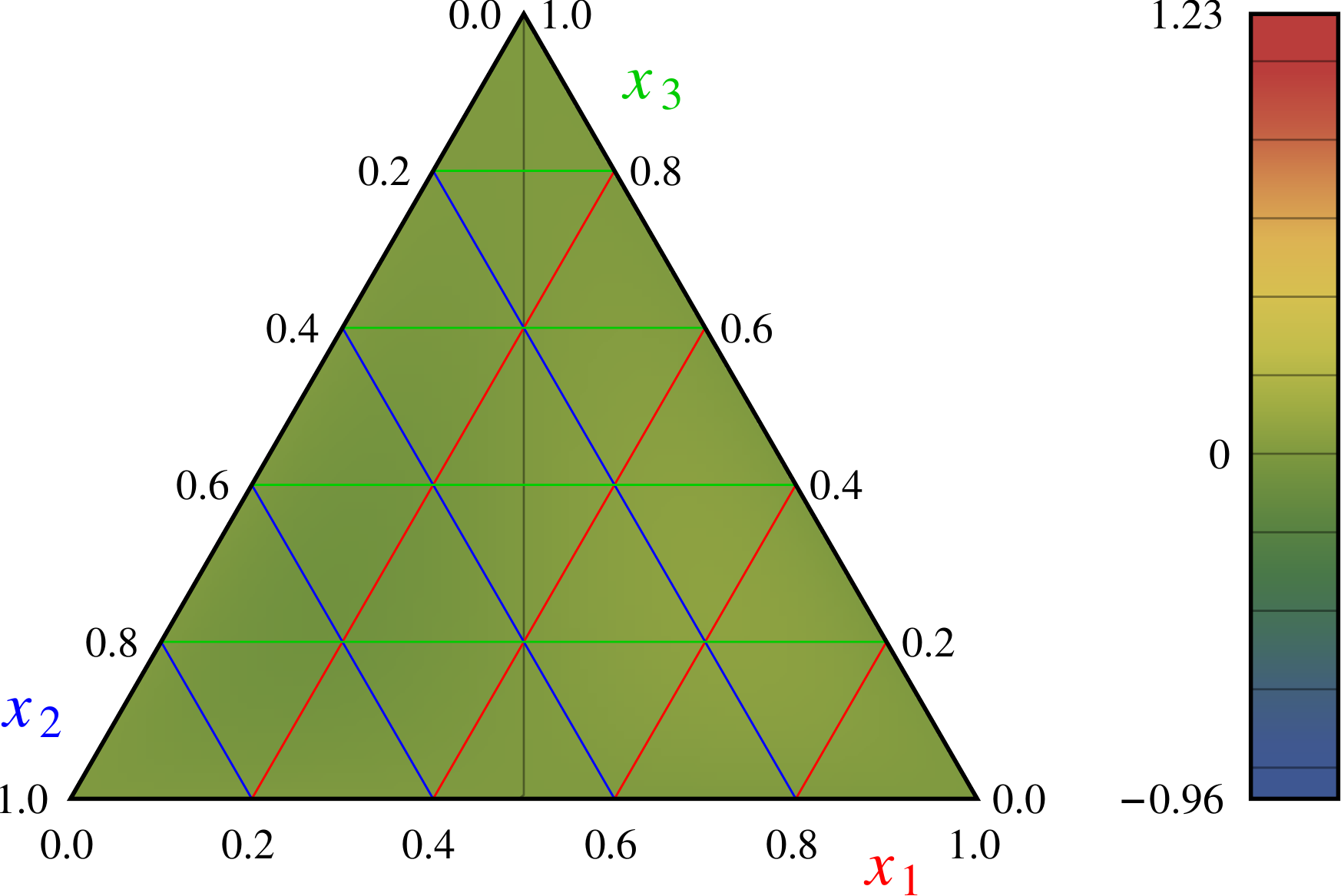}\raisebox{4.07cm}{$\mathllap{\mathclap{\boxed{\tfrac{T^\Lambda}{{\scriptscriptstyle\sqrt{\frac16}} f^\Lambda}}}\hspace{0.14\textwidth}}$}\raisebox{4.13cm}{$\mathllap{\mathrlap{\boxed{\color{red}u^\gooduparrow \color{blue}d^\gooduparrow \color{mygreen}s^\gooddownarrow}}\hspace{0.46\textwidth}}$}%
\caption{\label{figure_barycentric_VAT}Barycentric plots ($x_1+x_2+x_3=1$) showing the deviations of the DAs $[V-A]^B$ and $T^B$ from the asymptotic shape $\phi^{\text{as}}\equiv 120 x_1 x_2 x_3$. $T^\Lambda$ vanishes in the asymptotic limit, see eq.~\eqref{eq_da_parametrization_pi_lambda}. In this representation the coordinates $x_i$ directly correspond to quarks of definite flavor and helicity.}%
\end{figure}%
\afterpage{\clearpage}%
The \SU3 breaking in the shape of the leading twist DAs can be represented in many ways. Consider, e.g., normalized combinations of symmetric and antisymmetric DAs%
\begin{align}\label{eq_shape_functions}%
  \phi^B &= \frac{\Phi_+^B+\Phi_-^B}{f^B} \,,
&
  \varpi^{B\neq\Lambda} &= \frac{\Pi^B+\Phi_-^B}{f_T^B} \,,
&
  \varpi^\Lambda &= \frac{\Phi_+^\Lambda+\Pi^\Lambda}{f^\Lambda} \,,
\end{align}%
all of which are equal both in the asymptotic limit, $\phi^{\text{as}}\equiv\phi^{B,\text{as}}=\varpi^{B,\text{as}}=120 x_1 x_2 x_3$, and in the limit of exact flavor symmetry, $\phi^\star\equiv \phi^{B\star}=\varpi^{B\star}$. Due to isospin symmetry $\varpi^N=\phi^N$. Hence, there are seven independent functions that can be used to visualize the deviations from the DA $\phi^\star$ in the \SU3 flavor symmetry limit. These seven functions, $\phi^N-\phi^*$, etc., are shown in figure~\ref{figure_barycentric_phipi} together with $\phi^\star$ itself, which is almost (but not exactly) symmetric in $x_1$, $x_2$, $x_3$ due to small (but nonvanishing) values of $\varphi_{11}^\star$ and $\varphi_{10}^\star$ (cf.\ figures~\ref{figure_phi11} and~\ref{figure_phi10}).\par%
In phenomenological applications it is more convenient to consider the standard representation of DAs in terms of $[V-A]^B$ and $T^B$. In this way also the physical interpretation is more straightforward as every momentum fraction can be attributed to a quark of definite helicity and flavor. $[V-A]^B$ and $T^B$ do not coincide, however, at the flavor symmetric point, so that for these DAs it is more natural to show the deviations from the asymptotic shape $\phi^{\text{as}}$ rather than from $\phi^*$, see figure~\ref{figure_barycentric_VAT}. The plots in the left and in the right column show normalized DAs $[V-A]^B$ and $T^B$ after the subtraction of the asymptotic DA. Note that the amplitudes $T^{B\neq\Lambda}$ are symmetric under the interchange of $x_1$ and $x_2$ by construction. The approximate symmetry of $[V-A]^N$ under the  exchange of $x_2$ and $x_3$ is, in contrast, nontrivial. It is due to the approximate equality of the two nucleon shape parameters $\varphi_{10}^N\approx \varphi_{11}^N$ mentioned above. In the nucleon Fock state $u^\gooduparrow u^\gooddownarrow d^\gooduparrow$ this is equivalent to a symmetric distribution of momentum between the second and third quark. In agreement with earlier studies~\cite{Chernyak:1984bm,Chernyak:1987nu,Braun:2014wpa}, we observe that the ``leading'' $u^\gooduparrow$ quark, which has the same helicity as the nucleon, carries a larger momentum fraction. In the $u^\gooduparrow u^\gooduparrow d^\gooddownarrow$ nucleon state, which is described by $T^N$, the peak of the distribution is shifted towards the two $u$ quarks in a symmetric manner. $T^N$, however, is not an independent DA. Taking into account the isospin relation~\eqref{eq_isospin_relation_nucleon}, the spin-flavor structure of the nucleon light-cone wave function~\eqref{eq_wavefunction_not_lambda} can be presented, schematically, as $[V-A]^N u^\gooduparrow (u^\gooddownarrow d^\gooduparrow - d^\gooddownarrow u^\gooduparrow)$. In this picture our result for $[V-A]^N$ corresponds to a shift of the momentum distribution towards the $u^\gooduparrow$ quark, which carries the nucleon helicity, and the symmetry under $x_2\leftrightarrow  x_3$ may be interpreted as an indication for the remaining valence quarks forming a dynamical scalar ``diquark'', which is assumed in many models.\par%
For the $\Sigma$ baryon state $d^\gooduparrow d^\gooddownarrow s^\gooduparrow$ one sees that the maximum of the distribution is shifted from $d^\gooddownarrow$ towards $s^\gooduparrow$, whereas in the $d^\gooduparrow d^\gooduparrow s^\gooddownarrow$ state the $s$ quark gathers additional momentum from both $d$ quarks equally. The overall size of the deviations from the asymptotic distribution is, however, quite small, smilar to the nucleon case. For the $\Xi$ baryon the deviations are slightly larger. In the $s^\gooduparrow s^\gooddownarrow u^\gooduparrow$ state, the distribution is tilted towards the $s^\gooduparrow$ quark and leaves less momentum for the $u^\gooduparrow$ quark. $T^\Xi$ is clearly dominated by the two $s$ quarks. In summary, for the isospin-nonsinglet baryons one can identify two competing patterns: First, the strange quarks carry, in general, a larger fraction of the momentum. Second, in the $\lvert\uparrow\downarrow\uparrow\rangle$ state the first quark is favored over the second, while in the $\lvert\uparrow\uparrow\downarrow\rangle$ state the first two quarks behave identically. These rules do not apply to the $\Lambda$ baryon due to its reversed symmetry properties, see eq.~\eqref{eq_symmetry_VAT}: In the $u^\gooduparrow d^\gooddownarrow s^\gooduparrow$ state the maximum of the distribution is shifted towards the $s$ quark. $T^\Lambda$ is a special case, since it does not contain the leading asymptotic part due to the antisymmetry under exchange of $x_1$ and $x_2$. Hence, for the $\Lambda$ baryon, the Fock state $u^\gooduparrow d^\gooduparrow s^\gooddownarrow$ is expected to be highly suppressed.\par%
In order to quantify this picture, we consider normalized first moments of $[V-A]^B$ and $T^B$%
\begin{align}\label{eq_not_momentum_fractions}%
  \langle x_i \rangle^{B} &= \frac{1}{\varphi_{00,(1)}^B}\int \! [dx] \; x_i  [V-A]^B \,, & \langle x_i \rangle^{B\neq\Lambda}_T &= \frac{1}{\pi_{00,(1)}^B} \int \! [dx] \; x_i  T^B \,,
\end{align}%
which are sometimes referred to as momentum fractions in the literature. Note that this name is imprecise since the averaging is done with the DA and not a wave function squared, and, in particular, for $T^\Lambda$, which has no asymptotic part, the interpretation as momentum fractions breaks down completely. The $\langle x_i\rangle $ can be calculated in terms of the shape parameters as follows:%
\begin{table}[t]%
\centering%
\caption{\label{table_not_momentum_fractions}Normalized first moments of the DAs $[V-A]^B$ and $T^{B\neq\Lambda}$ in the $\MSbar$ scheme at a scale $\mu^2=\unit{4}{\squaren{\giga\electronvolt}}$, obtained via eq.~\eqref{eq_first_moments_explicit}.}%
\begin{tabular}{l<{\quad}>{\quad}cc<{\quad}>{\quad}cc<{\quad}>{\quad}cc<{\quad}>{\quad}cc}%
  \toprule
  $B$ & \multicolumn{2}{c}{$N$} & \multicolumn{2}{c}{$\Sigma$} & \multicolumn{2}{c}{$\Xi$} & \multicolumn{2}{c}{\quad$\Lambda$}\\
  \midrule
  $\langle x_1 \rangle^B$ & $u^\gooduparrow$   & 0.358 & $d^\gooduparrow$   & 0.331 & $s^\gooduparrow$   & 0.361 & $u^\gooduparrow$   & 0.310\\
  $\langle x_2 \rangle^B$ & $u^\gooddownarrow$ & 0.319 & $d^\gooddownarrow$ & 0.310 & $s^\gooddownarrow$ & 0.333 & $d^\gooddownarrow$ & 0.304\\
  $\langle x_3 \rangle^B$ & $d^\gooduparrow$   & 0.323 & $s^\gooduparrow$   & 0.359 & $u^\gooduparrow$   & 0.306 & $s^\gooduparrow$   & 0.386\\
  \midrule
  $\langle x_1 \rangle^B_T$ & $u^\gooduparrow$   & 0.340 & $d^\gooduparrow$   & 0.326 & $s^\gooduparrow$   & 0.352 &  \multicolumn{2}{c}{\quad---}\\
  $\langle x_2 \rangle^B_T$ & $u^\gooduparrow$   & 0.340 & $d^\gooduparrow$   & 0.326 & $s^\gooduparrow$   & 0.352 &  \multicolumn{2}{c}{\quad---}\\
  $\langle x_3 \rangle^B_T$ & $d^\gooddownarrow$ & 0.319 & $s^\gooddownarrow$ & 0.348 & $u^\gooddownarrow$ & 0.296 &  \multicolumn{2}{c}{\quad---}\\
  \bottomrule
\end{tabular}%
\end{table}%
\begin{subequations}\label{eq_first_moments_explicit}%
\begin{align}%
 \langle x_1 \rangle^{B\neq\Lambda} &= \frac13 + \frac13\widehat\varphi_{11}^{\,B}  + \widehat\varphi_{10}^{\,B} \,, &
 \langle x_2 \rangle^{B\neq\Lambda} &= \frac13 - \frac23\widehat\varphi_{11}^{\,B} \,, &
 \langle x_3 \rangle^{B\neq\Lambda} &= \frac13 + \frac13\widehat\varphi_{11}^{\,B} - \widehat\varphi_{10}^{\,B} \,,
\\*
 \langle x_1 \rangle^{B\neq\Lambda}_T &= \frac13 + \frac13\widehat\pi_{11}^{\,B} \,, &
 \langle x_2 \rangle^{B\neq\Lambda}_T &= \frac13 + \frac13\widehat\pi_{11}^{\,B} \,, &
 \langle x_3 \rangle^{B\neq\Lambda}_T &= \frac13 - \frac23\widehat\pi_{11}^{\,B} \,,
\\*
 \langle x_1 \rangle^\Lambda &= \frac13 + \frac13\widehat\varphi_{11}^{\,\Lambda} - \frac13\widehat\varphi_{10}^{\,\Lambda} \,, &
 \langle x_2 \rangle^\Lambda &= \frac13 - \frac23\widehat\varphi_{11}^{\,\Lambda} \,, &
 \langle x_3 \rangle^\Lambda &= \frac13 + \frac13\widehat\varphi_{11}^{\,\Lambda} + \frac13\widehat\varphi_{10}^{\,\Lambda} \,,
\end{align}%
\end{subequations}%
where%
\begin{align}%
 \widehat \varphi_{nk}^{\,B} &= \frac{\varphi_{nk}^B}{\varphi_{00,(1)}^B}\,, & \widehat \pi_{11}^{\,B\neq\Lambda} &= \frac{\pi_{11}^B}{\pi_{00,(1)}^B} \,.
\end{align}%
The results are summarized in table~\ref{table_not_momentum_fractions}. They support the qualitative picture suggested by the discussion of figure~\ref{figure_barycentric_VAT}.\par%
Finally, we consider the higher twist matrix elements that are related to the normalization of the $P$-wave light-cone wave functions and also appear as low energy constants in effective theories for generic GUT models~\cite{Claudson:1981gh}. We obtain, for the nucleon, $\lambda_2^N \approx -2\lambda_1^N$, which is well known, see, e.g., refs.~\cite{Leinweber:2004it,Braun:2008ur,Aoki:2008ku}. The same relation also holds for the $\Sigma$ and $\Xi$ hyperons but not for the $\Lambda$ baryon. Instead, we find $\lambda_2^\Lambda\approx-2\lambda_T^\Lambda$, i.e., the matrix element in eq.~\eqref{eq_lambdaLambda2} is zero within the error bars. The likely interpretation (similar to the familiar relations for isospin-nonsinglet baryons) is that the corresponding matrix elements vanish in the nonrelativistic quark model limit.%
\section{Conclusions and outlook\label{sect_conclusion}}%
In this work we have performed the first $N_f=2+1$ lattice QCD analysis of the normalization constants and (leading twist) first moments of the octet baryon distribution amplitudes with pion masses down to $\unit{222}{\mega\electronvolt}$. The results are scheme- and scale-dependent and, thus, have to be renormalized. To this end we first carried out a nonperturbative renormalization in a $\RI$ scheme, followed by a conversion to the $\MSbar$ scheme applying continuum perturbation theory at one-loop accuracy. We extrapolated our results to the physical point using three-flavor BChPT formulas derived in ref.~\cite{Wein:2015oqa}.\par%
We find significant \SU3 flavor breaking effects for the leading twist normalization constants%
\begin{align}%
 \frac{f^\Sigma}{f^N} &= 1.41(4)\,, & \frac{f_T^\Sigma}{f^N} &= 1.36(4)\,,  & \frac{f^\Xi}{f^N} &= 1.50(4)\,, & \frac{f_T^\Xi}{f^N} &= 1.52(4)\,, &  \frac{f^\Lambda}{f^N} &= 1.22(4)\,,
\end{align}%
and somewhat smaller symmetry breaking for the higher twist couplings%
\begin{align}%
 \frac{\lambda_1^\Sigma}{\lambda_1^N} &= 0.93(2)\,, & \frac{\lambda_1^\Xi}{\lambda_1^N} &= 0.98(2)\,,  &  \frac{\lambda_1^\Lambda}{\lambda_1^N} &= 0.81(2)\,, & \frac{\lambda_T^\Lambda}{\lambda_1^N} &= 1.05(3) \,,
\end{align}%
where the number in parentheses gives a combined statistical and chiral extrapolation error, while the uncertainty from the renormalization procedure is negligible for these ratios. It is likely that these ratios are less sensitive to discretization effects than the couplings themselves.\par%
Deviations from the asymptotic DAs are quantified by the values of shape parameters. They are small for all baryons in the octet, in agreement with the findings of ref.~\cite{Braun:2014wpa} for the nucleon, and much smaller than results of old QCD sum rule calculations~\cite{Chernyak:1987nu}. The \SU3 breaking in the shape parameters is, however, very large, see table~\ref{table_extrapolatedUC}. For the isospin-nonsinglet baryons one can identify two competing patterns: First, the strange quarks carry, in general, a larger fraction of the momentum. Second, in the $f^\gooduparrow g^\gooddownarrow h^\gooduparrow$ state (using our flavor conventions~\eqref{eq_baryon_flavors}) the first quark is favored over the second, while in the $f^\gooduparrow g^\gooduparrow h^\gooddownarrow$ state the first two quarks behave identically. These rules do not apply to the $\Lambda$ baryon due to its reversed symmetry properties, see eq.~\eqref{eq_symmetry_VAT}. The interplay of these two patterns leads to the rather elaborate structure shown in figure~\ref{figure_barycentric_VAT}.\par%
To first order in the \SU3 symmetry breaking parameter we have derived the following relation between the DAs of the $\Sigma$ and $\Xi$ hyperons:%
\begin{align}%
       \Phi_+^\Sigma(x_1,x_2,x_3) - \Pi^\Sigma(x_1,x_2,x_3)  &= \Pi^\Xi(x_1,x_2,x_3) - \Phi_+^\Xi(x_1,x_2,x_3)\,.
\end{align}%
This relation has the same theory status as the renowned Gell-Mann--Okubo relation for the masses, and is satisfied with similarly high accuracy $\sim 1\%$ in our data.\par%
The analysis presented here, using a trajectory with fixed mean quark mass, constitutes the first half of the twofold strategy pursued by the CLS effort. It will be complemented by a second set of lattices at fixed physical strange quark mass as indicated by the red line in figure~\ref{figure_ensembles}. The extrapolation to the physical point along this second path can be described using chiral perturbation theory with only two flavors, while any other path requires a full \SU3 treatment. The combination of these two methods will provide one with  an additional tool to estimate systematic errors. Its full implementation lies beyond the scope of this work, where we have focused on the development of the necessary formalism to describe patterns of \SU3 breaking at the wave function level. Future studies will have access to a rich landscape of CLS ensembles along both trajectories, including ensembles at (nearly) physical quark masses and various lattice spacings down to $a\approx\unit{0.04}{\femto\meter}$, thus allowing for a reliable continuum extrapolation.%
\acknowledgments
This work has been supported by the Deutsche Forschungsgemeinschaft (SFB/TRR-55), the Studienstiftung des deutschen Volkes, and the European Union under the Grant Agreement IRG 256594. The computations were done on the QPACE systems of the SFB/TRR-55, the Regensburg HPC-cluster iDataCool and computers of various institutions which we acknowledge below. The configurations were generated as part of the joint CLS effort~\cite{Bruno:2014jqa}. We thank Benjamin Gl\"a{\ss}le and Simon Heybrock for software development.\par%
The authors gratefully acknowledge the Gauss Centre for Supercomputing (GCS) for providing computing time for a GCS Large-Scale Project on the GCS share of the supercomputer JUQUEEN~\cite{juqueen} at J\"ulich Supercomputing Centre (JSC) as well as the GCS supercomputer SuperMUC at Leibniz Supercomputing Centre (LRZ, www.lrz.de). GCS is the alliance of the three national supercomputing centres HLRS (Universit\"at Stuttgart), JSC (Forschungszentrum J\"ulich), and LRZ (Bayerische Akademie der Wissenschaften), funded by the German Federal Ministry of Education and Research (BMBF) and the German State Ministries for Research of Baden-W{\"u}rttemberg (MWK), Bayern (StMWFK) and Nordrhein-Westfalen (MIWF). The authors also gratefully acknowledge computer time provided by PRACE (Partnership for Advanced Computing in Europe) as part of the project ContQCD.%
\newpage%
\appendix%
\section{The baryon octet\label{app_octet}}%
Starting with the standard representation for the quark triplet%
\begin{align}%
   u &= \begin{pmatrix}1\\0\\0\end{pmatrix}\,, &  d &= \begin{pmatrix}0\\1\\0\end{pmatrix}\,, & s &= \begin{pmatrix}0\\0\\1\end{pmatrix}\,,
\end{align}%
we define lowering operators $T_-$, $ U_-$ and $ V_-$ for the isospin, $U$-spin, and $V$-spin, respectively, in this (fundamental) representation as%
\begin{align}%
   T_- &= \begin{pmatrix}0&0&0\\1&0&0\\0&0&0\end{pmatrix}\,,
&  U_- &= \begin{pmatrix}0&0&0\\0&0&0\\0&1&0\end{pmatrix}\,,
&  V_- &= \begin{pmatrix}0&0&0\\0&0&0\\1&0&0\end{pmatrix}\,,
\end{align}%
so that%
\begin{align}%
  T_- u &= d \,, & U_- d &= s \,, & V_- u &= s \,.
\end{align}%
The baryon octet is usually presented as~\cite{Georgi:1985kw}%
\begin{align}%
 \begin{pmatrix}
\frac{\Lambda}{\sqrt{6}}+\frac{\Sigma^0}{\sqrt{2}} & \Sigma^+ & p
\\
\Sigma^-&\frac{\Lambda}{\sqrt{6}}-\frac{\Sigma^0}{\sqrt{2}}&n \\\Xi^-&\Xi^0&-2\frac{\Lambda}{\sqrt{6}} \end{pmatrix}
=
 p\, \mathcal{K}_p + n\, \mathcal{K}_n + \Sigma^0\,\mathcal{K}_{\Sigma^0} + \ldots \,,
\end{align}%
where $\mathcal{K}_B$ are matrices in flavor space, e.g.,%
\begin{align}%
 \mathcal{K}_p &= \begin{pmatrix}0&0&1\\0&0&0\\0&0&0\end{pmatrix}\,, &
 \mathcal{K}_\Lambda &= \frac{1}{\sqrt6}\begin{pmatrix}1&0&0\\0&1&0\\0&0&-2\end{pmatrix}\,, &&\text{etc.}
\end{align}%
We further \emph{define} the action of the lowering operators $T_-$, $ U_-$ and $ V_-$ on the octet by the usual expressions for the adjoint representation as%
\begin{align}%
  \hat T_-  \mathcal{K}_B &= [T_-, \mathcal{K}_B]\,, &   \hat U_-  \mathcal{K}_B &= [U_-, \mathcal{K}_B]\,, & \hat V_-  \mathcal{K}_B &= [V_-, \mathcal{K}_B]\,,
\end{align}%
without any additional phase factors.\par%
\begin{figure}[t]%
\centering%
\includegraphics{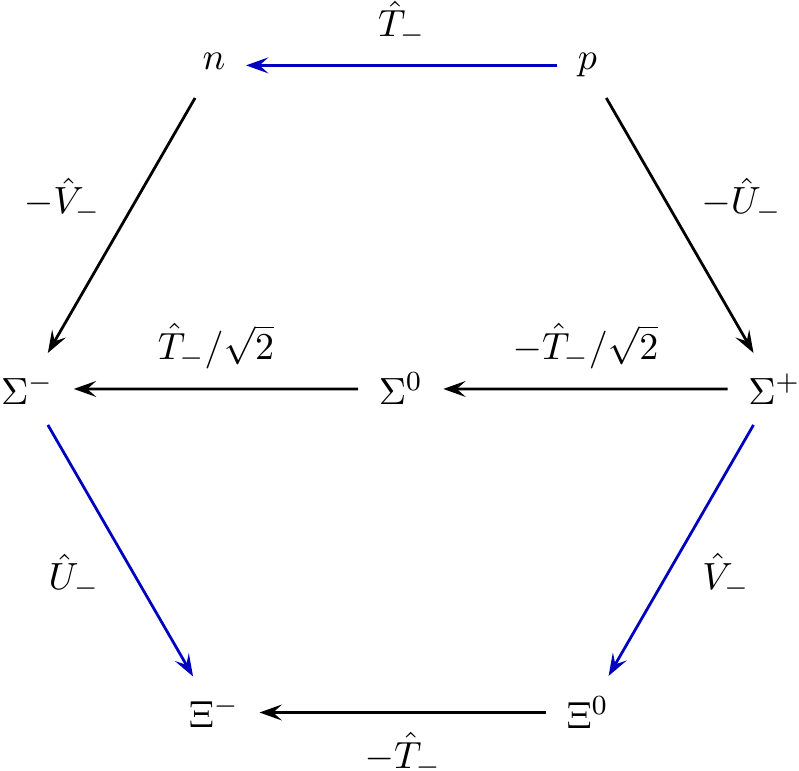}%
\caption{\label{fig_octet_phases}Illustration of our phase conventions. The $\Lambda$ baryon is not shown since one needs a linear combination for its construction, cf.\ eq.~\eqref{eq_construction_Lambda}. Blue arrows indicate the cases where one has to apply a Fierz transformation (see ref.~\cite{Braun:2000kw}) to relate the distribution amplitudes at the symmetric point. An explicit calculation shows that this always yields an additional minus sign that has to be taken into account in order to reproduce eq.~\eqref{eq_isospin_relations} and eq.~\eqref{eq_su3_relations}.}%
\end{figure}%
The above choices specify our phase conventions. Starting from the proton state, the complete octet can be constructed by applying the  following transformations as illustrated in figure~\ref{fig_octet_phases}:%
\begin{subequations}\label{eq_construction_octet}%
\begin{align}%
  \hat T_- | p \rangle &= | n \rangle  \,, \\*
  -\hat U_- | p \rangle&= | \Sigma^+ \rangle \,, \\*
 \tfrac{1}{\sqrt{2}}\hat T_-\hat U_- | p \rangle&= | \Sigma^0 \rangle  \,, \\*
 \tfrac12\hat T_-\hat T_-\hat U_- | p \rangle&= | \Sigma^- \rangle  \,, \\*
  -\hat V_-\hat U_- | p \rangle &= | \Xi^0 \rangle \,, \\*
  \hat T_-\hat V_-\hat U_- | p \rangle&= | \Xi^- \rangle \,, \\*
  \tfrac{-1}{\sqrt{6}} \bigl( \hat V_- + \hat U_- \hat T_- \bigr) | p \rangle&= | \Lambda \rangle \,. \label{eq_construction_Lambda}
\end{align}%
\end{subequations}%
Starting from the mixed-symmetric and mixed-antisymmetric flavor wave functions for the proton defined as%
\begin{align}%
|\text{MS},p\rangle &= \frac{1}{\sqrt{6}}(2 \fl{uud}-\fl{udu}-\fl{duu})\,, &  |\text{MA},p\rangle &= \frac{1}{\sqrt{2}}(\fl{udu}-\fl{duu})\,,
\end{align}%
the wave functions of the octet can now be constructed by applying the transformations in~\eqref{eq_construction_octet}, see tables~\ref{table_flavor_octet_1} and~\ref{table_flavor_octet_2}.\par%
Together with the choice of flavor ordering (cf.\ eq.~\eqref{eq_baryon_flavors})%
\begin{subequations}%
\begin{align}%
p&\mathrel{\hat=}uud\,, & n&\mathrel{\hat=}ddu\,, & \Sigma^+&\mathrel{\hat=}uus\,,
& \Sigma^0&\mathrel{\hat=}uds\,, \\
\Sigma^-&\mathrel{\hat=}dds\,, & \Xi^0&\mathrel{\hat=}ssu\,, & \Xi^-&\mathrel{\hat=}ssd\,, & \Lambda&\mathrel{\hat=}uds\,,
\end{align}%
\end{subequations}%
our conventions also fix the relative signs of the baryon DAs. As shown in ref.~\cite{Wein:2015oqa} this choice corresponds to%
\begin{subequations}\label{eq_isospin_relations}%
\begin{align}%
 \text{DA}^N &\equiv \text{DA}^p = -\text{DA}^n \,, \\
 \text{DA}^\Sigma &\equiv \text{DA}^{\Sigma^-}= \sqrt{2} \text{DA}^{\Sigma^0} = -\text{DA}^{\Sigma^+} \,, \\
 \text{DA}^\Xi &\equiv \text{DA}^{\Xi^0} = -\text{DA}^{\Xi^-} \,,
\end{align}%
\end{subequations}%
in the limit of exact isospin symmetry. This also fixes the relative phases at the flavor symmetric point in eq.~\eqref{eq_su3_relations}. All phases are now unambiguously determined up to a single unphysical global phase, which is commonly fixed by the condition that $f^N$ has to be positive.%
\begin{table}[t]%
\centering%
\caption{\label{table_flavor_singlet} Totally antisymmetric (A) flavor wave functions.}%
\begin{tabular}{lp{0.65\textwidth}}%
  \toprule
  $B$\hspace*{0.4cm} & $|\text{A},B\rangle=\sum_{f,g,h} F_{s}^{B,fgh} |fgh\rangle$\\
  \midrule
  $\Lambda$ & $(\fl{dus}-\fl{uds}+\fl{usd}-\fl{dsu}+\fl{sdu}-\fl{sud})/\sqrt{6}$\\
  \bottomrule
\end{tabular}%
\caption{\label{table_flavor_octet_1} Mixed-symmetric (MS) flavor wave functions.}%
\begin{tabular}{lp{0.65\textwidth}}%
  \toprule
  $B$\hspace*{0.4cm} & $|\text{MS},B\rangle=\sum_{f,g,h} F_{o1}^{B,fgh} |fgh\rangle$\\
  \midrule
  $N$        & $(2\fl{uud}-\fl{udu}-\fl{duu})/\sqrt{6}$\\
  $\Sigma$   & $(2\fl{dds}-\fl{dsd}-\fl{sdd})/\sqrt{6}$\\
  $\Xi$      & $(2\fl{ssu}-\fl{sus}-\fl{uss})/\sqrt{6}$\\
  $\Lambda$  & $(\fl{dsu}-\fl{usd}+\fl{sdu}-\fl{sud})/2$\\
  \bottomrule
\end{tabular}%
\caption{\label{table_flavor_octet_2} Mixed-antisymmetric (MA) flavor wave functions.}%
\begin{tabular}{lp{0.65\textwidth}}%
  \toprule
  $B$\hspace*{0.4cm} & $|\text{MA},B\rangle=\sum_{f,g,h} F_{o2}^{B,fgh} |fgh\rangle$\\
  \midrule
  $N$        & $(\fl{udu}-\fl{duu})/\sqrt{2}$\\
  $\Sigma$   & $(\fl{dsd}-\fl{sdd})/\sqrt{2}$\\
  $\Xi$      & $(\fl{sus}-\fl{uss})/\sqrt{2}$\\
  $\Lambda$  & $(2\fl{dus}-2\fl{uds}+\fl{dsu}-\fl{usd}+\fl{sud}-\fl{sdu})/\sqrt{12}$\\
  \bottomrule
\end{tabular}%
\caption{\label{table_flavor_decuplet} Totally symmetric (S) flavor wave functions.}%
\begin{tabular}{lp{0.65\textwidth}}%
  \toprule
  $B$\hspace*{0.4cm} & $|\text{S},B\rangle=\sum_{f,g,h} F_{d}^{B,fgh} |fgh\rangle$\\
  \midrule
  $N$        & $(\fl{uud}+\fl{udu}+\fl{duu})/\sqrt{3}$\\
  $\Sigma$   & $(\fl{dds}+\fl{dsd}+\fl{sdd})/\sqrt{3}$\\
  $\Xi$      & $(\fl{ssu}+\fl{sus}+\fl{uss})/\sqrt{3}$\\
  \bottomrule
\end{tabular}%
\end{table}%
\section{\label{app_operator_relations}Operator relations}%
\subsection{\label{app_operator_relations_lattice}Relation to \texorpdfstring{\H4}{spinorial H4} operators}%
In the following we will relate the operators defined in~\eqref{eq_operators_zeroth_moments} and~\eqref{eq_operators_first_moments} to those of ref.~\cite{Kaltenbrunner:2008pb}. It is implied that within the generic operators appearing on the right hand side of the equations, the quark flavors $f$, $g$ and $h$ are chosen such that they agree with the convention for the baryon $B$, see~\eqref{eq_baryon_flavors}. For the operators without derivatives we have%
\begin{align}%
  \mathcal{O}^{B,000}_{\mathcal{T},\mathfrak{A}} &=
  4\begin{pmatrix*}[l]-\mathcal{O}_9^{(6)}\\+\mathcal{O}_9^{(1)}\\-\mathcal{O}_9^{(12)}\\+\mathcal{O}_9^{(7)}\end{pmatrix*}\,,&
  \mathcal{O}^{B,000}_{\mathcal{T},\mathfrak{B}} &=
  4\begin{pmatrix*}[l]-\mathcal{O}_9^{(4)}\\+\mathcal{O}_9^{(3)}\\-\mathcal{O}_9^{(10)}\\+\mathcal{O}_9^{(9)}\end{pmatrix*}\,,&
  \mathcal{O}^{B,000}_{\mathcal{T},\mathfrak{C}} &=
  4\sqrt{2}\begin{pmatrix*}[l]+\mathcal{O}_9^{(2)}\\-\mathcal{O}_9^{(5)}\\+\mathcal{O}_9^{(8)}\\-\mathcal{O}_9^{(11)}\end{pmatrix*}\,,
\end{align}%
where the operators for the structure $\mathcal{V}+\mathcal{A}$ (or $\mathcal{V}-\mathcal{A}$) can be obtained by replacing $\mathcal{O}_9$ by $\mathcal{O}_7$ (or $\mathcal{O}_8$). For the operators with one derivative it is additionally implied that on the right hand side the position of the derivative is set as mandated by the superscripts $lmn$:%
\begin{align}%
  \mathcal{O}^{B,lmn}_{\mathcal{T},\mathfrak{A}} &=
  4\sqrt{2}\begin{pmatrix*}[l]+\mathcal{O}_{D7}^{(1)}\\-\mathcal{O}_{D7}^{(2)}\\-\mathcal{O}_{D7}^{(7)}\\+\mathcal{O}_{D7}^{(8)}\end{pmatrix*}\,,&
  \mathcal{O}^{B,lmn}_{\mathcal{T},\mathfrak{B}} &=
  4\sqrt{2}\begin{pmatrix*}[l]+\mathcal{O}_{D7}^{(3)}\\-\mathcal{O}_{D7}^{(4)}\\-\mathcal{O}_{D7}^{(9)}\\+\mathcal{O}_{D7}^{(10)}\end{pmatrix*}\,,&
  \mathcal{O}^{B,lmn}_{\mathcal{T},\mathfrak{C}} &=
  4\begin{pmatrix*}[l]+\mathcal{O}_{D7}^{(6)}\\+\mathcal{O}_{D7}^{(5)}\\-\mathcal{O}_{D7}^{(12)}\\-\mathcal{O}_{D7}^{(11)}\end{pmatrix*}\,,
\end{align}%
where the operators for the structure $\mathcal{V}+\mathcal{A}$ (or $\mathcal{V}-\mathcal{A}$) can be obtained by replacing $\mathcal{O}_{D7}$ by $\mathcal{O}_{D5}$ (or $\mathcal{O}_{D6}$).\par%
Similarly, the operators which are relevant for higher twist normalization constants (see eq.~\eqref{eq_correlators_higher_twist}) can be expressed in terms of $\mathcal{O}_{1-5}$. In the chiral odd sector we have%
\begin{align}%
    \mathcal{V}^{B,000} &=
  -2\sqrt2\begin{pmatrix*}[l]\mathcal{O}_3^{(1)}+\mathcal{O}_4^{(1)}\\\mathcal{O}_3^{(2)}+\mathcal{O}_4^{(2)}\\\mathcal{O}_3^{(3)}+\mathcal{O}_4^{(3)}\\\mathcal{O}_3^{(4)}+\mathcal{O}_4^{(4)}\end{pmatrix*}\,,&
    \mathcal{A}^{B,000} &=
  -2\sqrt2\begin{pmatrix*}[l]\mathcal{O}_3^{(1)}-\mathcal{O}_4^{(1)}\\\mathcal{O}_3^{(2)}-\mathcal{O}_4^{(2)}\\\mathcal{O}_3^{(3)}-\mathcal{O}_4^{(3)}\\\mathcal{O}_3^{(4)}-\mathcal{O}_4^{(4)}\end{pmatrix*} \,,
\end{align}%
relevant for $\lambda_1^B$, and%
\begin{align}%
    (\mathcal{S}-\mathcal{P})^{B,000} &=
  -2\sqrt2\begin{pmatrix*}[l]\mathcal{O}_5^{(1)}\\\mathcal{O}_5^{(2)}\\\mathcal{O}_5^{(3)}\\\mathcal{O}_5^{(4)}\end{pmatrix*}\,,
\end{align}%
relevant for $\lambda_T^\Lambda$. In the chiral even sector ($\lambda_2^B$) we obtain:%
\begin{align}%
    (\mathcal{S}+\mathcal{P})^{B,000} &=
  2\sqrt{\frac23}\begin{pmatrix*}[l]2\mathcal{O}_1^{(1)}+\mathcal{O}_2^{(1)}\\2\mathcal{O}_1^{(2)}+\mathcal{O}_2^{(2)}\\2\mathcal{O}_1^{(3)}+\mathcal{O}_2^{(3)}\\2\mathcal{O}_1^{(4)}+\mathcal{O}_2^{(4)}\end{pmatrix*}\,,&
    \mathcal{T}^{B,000} &=
  4\sqrt6\begin{pmatrix*}[l]\mathcal{O}_2^{(1)}\\\mathcal{O}_2^{(2)}\\\mathcal{O}_2^{(3)}\\\mathcal{O}_2^{(4)}\end{pmatrix*}\ .
\end{align}%
\subsection{\label{app_operator_relations_renormalization}Operator bases for renormalization}%
For the purpose of renormalization it is convenient to employ operator multiplets that transform irreducibly not only with respect to the spinorial hypercubic group \H4 but also with respect to the group $\mathcal S_3$ of permutations of the three quark flavors. The latter group has three inequivalent irreducible representations, which we label by the names of the corresponding ground state particle multiplets in a flavor symmetric world. Therefore, the one-dimensional trivial representation is labeled by $\mathscr D$ in the main text, the one-dimensional totally antisymmetric representation by $\mathscr S$ and the two-dimensional representation by $\mathscr O$.\par%
We construct multiplets with the desired transformation properties from the multiplets defined in ref.~\cite{Kaltenbrunner:2008pb}. For operators without derivatives in the representation $\tau^{\underbar{12}}_1$ of \H4 we have one doublet of operator multiplets transforming according to the two-dimensional representation of $\mathcal S_3$,%
\begin{subequations}%
\begin{align}%
 &\begin{cases*}
  \frac{1}{\sqrt{6}}(\mathcal{O}_7+\mathcal{O}_8-2\mathcal{O}_9)\\
  \frac{1}{\sqrt{2}}(\mathcal{O}_7-\mathcal{O}_8)
 \end{cases*}\,,
\intertext{(with the first multiplet being mixed-symmetric and the second one being mixed-anti\-symmetric) and one operator multiplet transforming trivially
under $\mathcal S_3$:}%
 &\tfrac{1}{\sqrt{3}}(\mathcal{O}_7+\mathcal{O}_8+\mathcal{O}_9) \,.
\end{align}%
\end{subequations}%
For operators without derivatives in the \H4 representation $\tau^{\underbar{4}}_1$ we have one multiplet that is totally antisymmetric under flavor permutations,%
\begin{subequations}%
\begin{align}%
 &\tfrac{1}{\sqrt{3}}(\mathcal{O}_3-\mathcal{O}_4-\mathcal{O}_5) \,,
\intertext{and two doublets of operator multiplets transforming according to the two-dimensional representation of $\mathcal S_3$:}%
 &\begin{cases*}
  \frac{1}{\sqrt{2}}(\mathcal{O}_3+\mathcal{O}_4)\\
  \frac{1}{\sqrt{6}}(-\mathcal{O}_3+\mathcal{O}_4-2\mathcal{O}_5)
 \end{cases*}\,,
\\
 &\begin{cases*}
  \mathcal{O}_2\\
  \frac{1}{\sqrt{3}}(2\mathcal{O}_1+\mathcal{O}_2)
 \end{cases*}\,.
\end{align}%
\end{subequations}%
For operators with one derivative in the \H4 representation $\tau^{\underbar{12}}_2$ we have one multiplet that is totally antisymmetric under $\mathcal S_3$,%
\begin{subequations}%
\begin{align}%
 &\tfrac{1}{\sqrt6}\bigl[(\mathcal{O}_{g5}-\mathcal{O}_{h5})+(\mathcal{O}_{h6}-\mathcal{O}_{f6})+(\mathcal{O}_{f7}-\mathcal{O}_{g7})\bigr] \,,
\intertext{four doublets of operator multiplets corresponding to the two-dimensional representation of $\mathcal S_3$,}%
 &\begin{cases*}
  \frac{1}{3\sqrt2}\bigl[(\mathcal{O}_{f5}+\mathcal{O}_{g5}+\mathcal{O}_{h5})+(\mathcal{O}_{f6}+\mathcal{O}_{g6}+\mathcal{O}_{h6})-2(\mathcal{O}_{f7}+\mathcal{O}_{g7}+\mathcal{O}_{h7})\bigr]\\
  \frac{1}{\sqrt6}\bigl[(\mathcal{O}_{f5}+\mathcal{O}_{g5}+\mathcal{O}_{h5})-(\mathcal{O}_{f6}+\mathcal{O}_{g6}+\mathcal{O}_{h6})\bigr]
 \end{cases*}\,,
\\
 &\begin{cases*}
  \frac{1}{6}\bigl[(-2\mathcal{O}_{f5}+\mathcal{O}_{g5}+\mathcal{O}_{h5})+(\mathcal{O}_{f6}-2\mathcal{O}_{g6}+\mathcal{O}_{h6})-2(\mathcal{O}_{f7}+\mathcal{O}_{g7}-2\mathcal{O}_{h7})\bigr]\\
  \frac{1}{2\sqrt3}\bigl[(-2\mathcal{O}_{f5}+\mathcal{O}_{g5}+\mathcal{O}_{h5})-(\mathcal{O}_{f6}-2\mathcal{O}_{g6}+\mathcal{O}_{h6})\bigr]
 \end{cases*}\,,
\\
 &\begin{cases*}
  \frac{1}{2}\bigl[(\mathcal{O}_{g5}-\mathcal{O}_{h5})-(\mathcal{O}_{h6}-\mathcal{O}_{f6})\bigr]\\
  \frac{1}{2\sqrt3}\bigl[(\mathcal{O}_{h5}-\mathcal{O}_{g5})+(\mathcal{O}_{f6}-\mathcal{O}_{h6})-2(\mathcal{O}_{g7}-\mathcal{O}_{f7})\bigr]
 \end{cases*}\,,
\\
 &\begin{cases*}
  \frac{1}{\sqrt6}(\mathcal{O}_{f8}+\mathcal{O}_{g8}-2\mathcal{O}_{h8})\\
  \frac{1}{\sqrt2}(\mathcal{O}_{f8}-\mathcal{O}_{g8})
 \end{cases*}\,,
\intertext{and three operator multiplets transforming trivially under flavor permutations:}%
 &\tfrac{1}{3}\bigl[(\mathcal{O}_{f5}+\mathcal{O}_{g5}+\mathcal{O}_{h5})+(\mathcal{O}_{f6}+\mathcal{O}_{g6}+\mathcal{O}_{h6})+(\mathcal{O}_{f7}+\mathcal{O}_{g7}+\mathcal{O}_{h7})\bigr] \,,
\\
 &\tfrac{1}{3\sqrt2}\bigl[(-2\mathcal{O}_{f5}+\mathcal{O}_{g5}+\mathcal{O}_{h5})+(\mathcal{O}_{f6}-2\mathcal{O}_{g6}+\mathcal{O}_{h6})+(\mathcal{O}_{f7}+\mathcal{O}_{g7}-2\mathcal{O}_{h7})\bigr] \,,
\\
 &\tfrac{1}{\sqrt3}(\mathcal{O}_{f8}+\mathcal{O}_{g8}+\mathcal{O}_{h8}) \ .
\end{align}%
\end{subequations}%
\section{\label{appendix_renormalization}Renormalization procedure}%
Every local three-quark operator can be represented as a linear combination of the operators%
\begin{align}%
\Psi^{fgh}_{\alpha \beta \gamma}(\bar{l},\bar{m},\bar{n};x) &= \epsilon^{ijk}
( D_{\bar l} f(x))^i_\alpha
( D_{\bar m} g(x))^j_\beta
( D_{\bar n} h(x))^k_\gamma
\end{align}%
with the same multi-index notation as above. Aiming at a mass-independent renormalization scheme we assign the same mass to all flavors and eventually consider the chiral limit where this mass is sent to zero. In order to conveniently display the behavior of the operators under permutations of the three quarks we write the above operators in the form%
\begin{align}%
\Psi^{f_1 f_2 f_3}_{\alpha_1 \alpha_2 \alpha_3}(\bar{l}_1,\bar{l}_2,\bar{l}_3;x)
\end{align}%
or, in an abbreviated notation, as $\Psi^f_\alpha (\bar{l};x)$. Then we have%
\begin{align}\label{eq_symm}%
\Psi^{f_\pi}_{\alpha_\pi} (\bar{l}_\pi;x) &= \Psi^f_\alpha (\bar{l};x)
\end{align}%
for all permutations $\pi$ in the symmetric group $\mathcal S_3$ of three elements, where%
\begin{align}%
\Psi^{f_\pi}_{\alpha_\pi} (\bar{l}_\pi;x) &=
\Psi^{f_{\pi(1)} f_{\pi(2)} f_{\pi(3)}}_{\alpha_{\pi(1)} \alpha_{\pi(2)} \alpha_{\pi(3)}}(\bar{l}_{\pi(1)},\bar{l}_{\pi(2)},\bar{l}_{\pi(3)};x) \,.
\end{align}%
From these ``elementary'' operators we construct the operators of interest with the help of flavor coefficients $F$ and spinor coefficients $S$ according to%
\begin{align}\label{eq_genop}%
F^f S^{\bar{l}}_\alpha \Psi^f_\alpha (\bar{l};x) \,,
\end{align}%
where a sum over all (multi-)indices which appear twice is implied.\par%
Under \SU3 the quark fields transform according to the fundamental representation~$3$ and for our three-quark operators we have the decomposition $3\otimes 3 \otimes 3 = 1 \oplus 8 \oplus 8 \oplus 10$. The flavor-singlet (flavor-decuplet) representation corresponds to the totally antisymmetric (totally symmetric or trivial) representation of $\mathcal S_3$. The two flavor octets, called mixed-symmetric (MS) and mixed-antisymmetric (MA), form a basis for the two-dimensional representation of $\mathcal S_3$. More explicitly, we have the singlet flavor structure $F^{B,f_1 f_2 f_3}_s$ with%
\begin{align}%
F^{B,f_\pi}_s &= \operatorname{sgn}(\pi) F^{B,f}_s \,,
\intertext{decuplet flavor structures $F^{B,f_1 f_2 f_3}_d$ with}%
F^{B,f_\pi}_d &= F^{B,f}_d \,,
\end{align}%
and the octet flavor structures $F^{B,f_1 f_2 f_3}_{ot}$, where $t=1$ corresponds to MS and $t=2$ corresponds to MA.\par%
The spinor structures should be chosen to yield a flavor-spinor structure that is totally symmetric under simultaneous permutations of the flavor, spinor and derivative indices $f_a$, $\alpha_a$ and $\bar{l}_a$ ($a=1,2,3$). Starting from the operator multiplets given in ref.~\cite{Kaltenbrunner:2008pb}, which transform irreducibly under the spinorial hypercubic group \H4, we construct multiplets of spinor structures%
\begin{align}%
S^{(m,i),\bar{l}}_{s,\alpha}  \,,\
S^{(m,i),\bar{l}}_{d,\alpha}  \,,\
S^{(m,i),\bar{l}}_{ot,\alpha} \,,
\end{align}%
which transform under $\mathcal S_3$ identically to their flavor counterparts:%
\begin{align}%
S^{(m,i),\bar{l}_\pi}_{s,\alpha_\pi} &= \operatorname{sgn}(\pi) S^{(m,i),\bar{l}}_{s,\alpha} \,,\ \text{etc.}
\end{align}%
Here $m$ labels the different \H4 multiplets and $i$ labels the different members of the multiplets. Then%
\begin{align}%
\smashoperator{\sum_{t=1}^2} F^{B,f}_{ot} S^{(m,i),\bar{l}}_{ot,\alpha}
\end{align}%
is indeed totally symmetric under simultaneous permutations of the flavor, spinor and derivative indices. An analogous statement holds for singlet  and decuplet. The corresponding operators $S^{\bar{l}}_\alpha \Psi^f_\alpha (\bar{l};x)$ with generic flavors are listed in appendix~\ref{app_operator_relations_renormalization}.\par%
In the case of the octet baryons we find%
\begin{align}%
F^{B,f}_{ot} S^{(m,i),\bar{l}}_{ot,\alpha} \Psi^f_\alpha (\bar{l};x) &= \frac{1}{2} \smashoperator[r]{\sum_{t^\prime=1}^2} F^{B,f}_{ot^\prime} S^{(m,i),\bar{l}}_{ot^\prime,\alpha}
  \Psi^f_\alpha (\bar{l};x)
\end{align}%
for $t \in \{ 1,2 \}$. Therefore, we can always work with the MA flavor structure ($t=2$) and assume that the flavor-spinor structure factorizes  into a flavor structure and a spinor structure as in~\eqref{eq_genop}. For the singlet and decuplet baryons this factorization is trivially  satisfied.\par%
Let us now describe our renormalization procedure, which is similar to the well-known $\RI$ scheme. In particular, we compute the quark field renormalization factor $Z_q$ from the quark propagator as in ref.~\cite{Gockeler:2010yr}.\par%
Under renormalization, multiplets of operators transforming according to the same representation of \H4 and having  the same or lower dimension will in general mix. Since mixing with operators of lower dimension is particularly problematic we have chosen the operators such that this type of mixing is absent. Moreover, there is no mixing between operators transforming according to inequivalent representations of $\mathcal S_3$.\par
\begin{figure}[t]%
\centering%
\includegraphics[width=.3\textwidth]{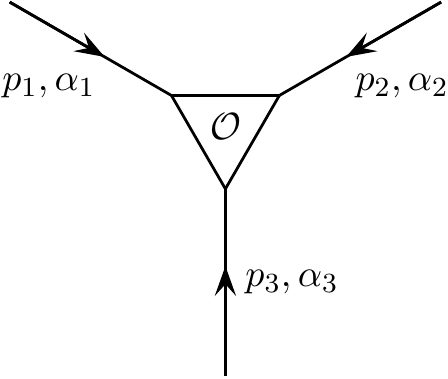}%
\caption{\label{fig_4point}Pictorial representation of the four-point function.}%
\end{figure}

%
For an operator of the form~\eqref{eq_genop} we consider (in a fixed gauge) the vertex function%
\begin{align}%
\Lambda (p_1,p_2,p_3)^{f_1 f_2 f_3}_{\alpha_1 \alpha_2 \alpha_3}
&\equiv \Lambda (p)^f_\alpha
= \smashoperator{\sum_{\pi \in \mathcal S_3}} F^{f_\pi} S^{\bar{l}}_\beta H^\beta_{\alpha_\pi} ({\bar{l}};p_\pi)
= \smashoperator{\sum_{\pi \in \mathcal S_3}} F^{f_\pi} S^{\bar{l}_\pi}_{\beta_\pi} H^{\beta_\pi}_{\alpha_\pi} (\bar{l}_\pi;p_\pi) \,.
\end{align}%
Here $H^\beta_\alpha (\bar{l};p) \equiv H^{\beta_1 \beta_2 \beta_3}_{\alpha_1 \alpha_2 \alpha_3} (\bar{l}_1,\bar{l}_2,\bar{l}_3;p_1,p_2,p_3)$ denotes the ``flavorless'' amputated four-point function with open spinor indices $\alpha_1,\alpha_2,\alpha_3$ ($\beta_1,\beta_2,\beta_3$) at the external quark lines (at the operator), pictorially represented in figure~\ref{fig_4point}. More explicitly, we have%
\begin{align}%
\begin{split}
H^{\beta_1 \beta_2 \beta_3}_{\alpha_1 \alpha_2 \alpha_3} (\bar{l}_1, \bar{l}_2, \bar{l}_3;p_1, p_2, p_3)
&= \smashoperator{\sum_{x_1,x_2,x_3}} e^{i (p_1 \cdot x_1 + p_2 \cdot x_2 + p_3 \cdot x_3)}
 \epsilon^{i_1 i_2 i_3} \epsilon^{j_1 j_2 j_3}  \\
& \quad \times \left \langle G^{i_1 j_1}_{\beta^{\phantom{\prime}}_1 \alpha^\prime_1} (\bar{l}_1;0,x_1) \,
              G^{i_2 j_2}_{\beta^{\phantom{\prime}}_2 \alpha^\prime_2} (\bar{l}_2;0,x_2) \,
              G^{i_3 j_3}_{\beta^{\phantom{\prime}}_3 \alpha^\prime_3} (\bar{l}_3;0,x_3) \right \rangle \\
& \quad \times G_2^{-1} (p_1)_{\alpha^\prime_1 \alpha^{\phantom{\prime}}_1}
  G_2^{-1} (p_2)_{\alpha^\prime_2 \alpha^{\phantom{\prime}}_2}
  G_2^{-1} (p_3)_{\alpha^\prime_3 \alpha^{\phantom{\prime}}_3} \,.
\end{split}
\end{align}%
The momentum space propagator $G_2 (p)$ is defined from%
\begin{align}%
\sum_x e^{i p \cdot x}
\left \langle f^i_{\alpha} (0) \bar{g}^j_\beta (x)
\right \rangle &=
\sum_x e^{i p \cdot x}
\left \langle G^{ij}_{\alpha \beta} (0,x) \right \rangle \delta_{fg} =
G_2 (p)_{\alpha \beta} \, \delta_{fg} \, \delta_{ij} \,,
\end{align}%
where $G^{ij}_{\alpha \beta}(x,y)$ is the quark propagator on a given gauge field configuration and $\langle \cdots \rangle$ indicates the average over the gauge fields fixed to the Landau gauge. Propagators with covariant derivatives acting at $x$ are denoted by $G^{ij}_{\alpha \beta}(\bar{l};x,y)$. Since in the present context of renormalization all quark masses are equal, the propagators do not need a flavor label. The external momenta are chosen such that $p_1^2 = p_2^2 = p_3^2 = (p_1+p_2+p_3)^2 = (p_1+p_2)^2 = (p_1+p_3)^2 = \mu^2$ with the renormalization scale $\mu$.\par%
We write the mixing operator multiplets for a fixed flavor structure $F^{B,f}_{o2}$ in the form%
\begin{align}%
\mathscr O^{(i)}_m (x) &= F^{B,f}_{o2} S^{(m,i),\bar{l}}_{o2,\alpha} \Psi^f_\alpha (\bar{l};x)
\end{align}%
and, analogously, with $o2$ replaced by $o1$, $s$ or $d$. The corresponding vertex functions are given by%
\begin{align}%
\Lambda (\mathscr O^{(i)}_m | p)^f_\alpha &= \smashoperator{\sum_{\pi \in \mathcal S_3}} F^{B,f_\pi}_{o2} S^{(m,i),\bar{l}}_{o2,\beta} H^\beta_{\alpha_\pi} (\bar{l};p_\pi) \,.
\end{align}%
The renormalized vertex functions take the form%
\begin{align}%
\Lambda^{\text{R}} (\mathscr O^{(i)}_m | p)^f_\alpha &= \smashoperator{\sum_{\pi \in \mathcal S_3}} F^{B,f_\pi}_{o2}
  \Bigl[ S^{(m,i),\bar{l}}_{o2,\beta} H^\beta_{\alpha_\pi} (\bar{l};p_\pi)
  \Bigr]^{\text{R}} \,,
\end{align}%
where%
\begin{align}%
  \Bigl[ S^{(m,i),\bar{l}}_{o2,\beta} H^\beta_\alpha (\bar{l};p) \Bigr]^{\text{R}} &= \sum_{m^\prime} Z_q^{-3/2} Z_{m m^\prime}
  S^{(m^\prime,i),\bar{l}}_{o2,\beta} H^\beta_\alpha (\bar{l};p) \,.
\end{align}%
The renormalization and mixing coefficients $Z_{m m^\prime}$ are fixed by the renormalization condition%
\begin{align}%
\sum_i \Lambda^{\text{R}} ( \mathscr O^{(i)}_m | p)^f_\alpha
\Bigl( \Lambda^{\text{Born}} ( \mathscr O^{(i)}_{m^\prime} | p)^f_\alpha \Bigr)^*
&= \sum_i \Lambda^{\text{Born}} ( \mathscr O^{(i)}_m | p)^f_\alpha
\Bigl( \Lambda^{\text{Born}} ( \mathscr O^{(i)}_{m^\prime} | p)^f_\alpha \Bigr)^*
\,.
\end{align}%
Here and in the following the superscript ``Born'' indicates the corresponding tree level expression (Born term). This is taken with all lattice artefacts included. More explicitly, our renormalization condition can be written as%
\begin{align}%
\smashoperator{\sum_{m^{\prime\prime}}} Z_q^{-3/2} Z_{m m^{\prime\prime}} L_{m^{\prime\prime} m^\prime} &= R_{m m^\prime}
\end{align}%
with%
\begin{align}%
L_{m m^\prime} &= \sum_i \smashoperator{\sum_{t=1}^2} S^{(m,i),\bar{l}}_{ot,\beta}
  \Bigl( S^{(m^\prime,i),\bar{l}^\prime}_{ot,\beta^\prime} \Bigr)^* H^\beta_\alpha (\bar{l};p)
  \Bigl( H^{\beta^\prime}_\alpha (\bar{l}^\prime;p)^{\text{Born}} \Bigr)^*
\shortintertext{and}%
R_{m m^\prime} &= \sum_i \smashoperator{\sum_{t=1}^2} S^{(m,i),\bar{l}}_{ot,\beta}
  \Bigl( S^{(m^\prime,i),\bar{l}^\prime}_{ot,\beta^\prime} \Bigr)^*
               H^\beta_\alpha (\bar{l};p)^{\text{Born}}
  \Bigl( H^{\beta^\prime}_\alpha (\bar{l}^\prime;p)^{\text{Born}} \Bigr)^* \,.
\end{align}%
For singlet and decuplet one gets analogous equations where, of course, no sums over $t$ appear. So we have%
\begin{align}%
Z_{m m^\prime} &= Z_q^{3/2} \bigl( R L^{-1} \bigr)_{m m^\prime} \,.
\end{align}\par%
This procedure yields (matrices of) renormalization factors leading from the bare operators on the lattice to operators renormalized according to the MOM scheme introduced above. However, in the applications we need operators renormalized in the $\MSbar$ scheme. These are constructed with the help of (matrices of) conversion factors calculated in continuum perturbation theory, where we use the particular version of the $\MSbar$ scheme introduced in ref.~\cite{Kraenkl:2011qb}. Due to the complexity of higher-loop calculations we had to limit ourselves to one-loop accuracy. Also the anomalous dimensions of our operators are in general only known to one loop, with the exception of the operators without derivatives, for which the anomalous dimensions have been calculated to three loops~\cite{Gracey:2012gx}.\par
Let us now give a few more technical details of the computation of the renormalization matrices. The required propagators have been evaluated on gauge field ensembles which were generated on lattices of size $32^4$ for four different quark masses using periodic (antiperiodic) boundary conditions for the quark fields in spatial (temporal) direction and periodic boundary conditions for the gauge fields in all four directions. For the external momenta we have taken%
\begin{align}%
p_1 &= \frac{\mu}{2} (+1,+1,+1,+1) \,, &
p_2 &= \frac{\mu}{2} (-1,-1,-1,+1) \,, &
p_3 &= \frac{\mu}{2} (+1,-1,-1,-1) \,,
\end{align}%
employing twisted boundary conditions. The extrapolation to the chiral limit is performed linearly in the square of the pseudoscalar mass.\par%
The values to be used in the analysis of the physical matrix elements are determined by interpolating the (chirally extrapolated) numerical data linearly in $\mu^2$ to our target scale $\mu_0^2 = \unit{4}{\giga\electronvolt\squared}$. While the statistical errors are quite small, systematic uncertainties are far more important. In order to estimate their impact we proceed as follows. Let the value of the renormalization coefficient under study be $z_0$. We consider two additional procedures to determine this coefficient. In the first procedure we take the (interpolated) data at $\unit{10}{\giga\electronvolt\squared}$ and evolve them down to $\unit{4}{\giga\electronvolt\squared}$ with the help of the perturbative renormalization group using as many loops in the anomalous dimensions as are available. Let us call the resulting number $z_1$. In the second procedure we read off our result directly at $\unit{4}{\giga\electronvolt\squared}$, but use for the perturbative conversion to the $\MSbar$ scheme one loop order less than before. (This means in our case that the conversion matrix is set to unity.) This gives the value $z_2$. Then we perform the complete analysis with the three choices $z_0$, $z_1$, $z_2$ for the needed renormalization and mixing coefficients. Defining $\delta_i$ as the difference between the outcome of the analysis employing $z_i$ and the outcome of the analysis employing $z_0$, we estimate the systematic uncertainties due to the renormalization factors as $\sqrt{\delta_1^2 + (\delta_2/2)^2}$. Here we have multiplied $\delta_2$ by $1/2$, because going from one to two or more loops in the conversion factors is expected to yield a smaller change than going from zero loops to one loop.\par%
\providecommand{\href}[2]{#2}\begingroup\raggedright\endgroup


\begin{thebibliography}{10}

\bibitem{Lepage:1980fj}
G.~P. Lepage and S.~J. Brodsky, {\it {Exclusive Processes in Perturbative
  Quantum Chromodynamics}},  {\em Phys. Rev.} {\bf D22} (1980) 2157.

\bibitem{Efremov:1979qk}
A.~V. Efremov and A.~V. Radyushkin, {\it {Factorization and Asymptotical
  Behavior of Pion Form-Factor in QCD}},  {\em Phys. Lett.} {\bf B94} (1980)
  245.

\bibitem{Chernyak:1983ej}
V.~L. Chernyak and A.~R. Zhitnitsky, {\it {Asymptotic Behavior of Exclusive
  Processes in QCD}},  {\em Phys. Rept.} {\bf 112} (1984) 173.

\bibitem{Chernyak:1984bm}
V.~L. Chernyak and I.~R. Zhitnitsky, {\it {Nucleon wave function and nucleon
  form-factors in QCD}},  {\em Nucl. Phys.} {\bf B246} (1984) 52.

\bibitem{King:1986wi}
I.~D. King and C.~T. Sachrajda, {\it {Nucleon Wave Functions and QCD Sum
  Rules}},  {\em Nucl. Phys.} {\bf B279} (1987) 785.

\bibitem{Chernyak:1987nu}
V.~L. Chernyak, A.~A. Ogloblin, and I.~R. Zhitnitsky, {\it {Wave functions of
  octet baryons}},  {\em Z. Phys.} {\bf C42} (1989) 569.

\bibitem{Chernyak:1987nv}
V.~L. Chernyak, A.~A. Ogloblin, and I.~R. Zhitnitsky, {\it {Calculation of
  Exclusive Processes With Baryons}},  {\em Z. Phys.} {\bf C42} (1989) 583.

\bibitem{Stefanis:1992nw}
N.~G. Stefanis and M.~Bergmann, {\it {On the Nucleon distribution amplitude:
  the Heterotic solution}},  {\em Phys. Rev.} {\bf D47} (1993) 3685,
  [\href{http://arxiv.org/abs/hep-ph/9211250}{{\tt hep-ph/9211250}}].

\bibitem{Mikhailov:1986be}
S.~V. Mikhailov and A.~V. Radyushkin, {\it {Nonlocal Condensates and QCD Sum
  Rules for the Pion Wave Function}},  {\em JETP Lett.} {\bf 43} (1986) 712.

\bibitem{Radyushkin:1990te}
A.~V. Radyushkin, {\it {Hadronic form-factors: Perturbative QCD versus QCD sum
  rules}},  {\em Nucl. Phys.} {\bf A532} (1991) 141.

\bibitem{Bolz:1996sw}
J.~Bolz and P.~Kroll, {\it {Modeling the nucleon wave function from soft and
  hard processes}},  {\em Z. Phys.} {\bf A356} (1996) 327,
  [\href{http://arxiv.org/abs/hep-ph/9603289}{{\tt hep-ph/9603289}}].

\bibitem{Braun:2001tj}
V.~M. Braun, A.~Lenz, N.~Mahnke, and E.~Stein, {\it {Light cone sum rules for
  the nucleon form-factors}},  {\em Phys. Rev.} {\bf D65} (2002) 074011,
  [\href{http://arxiv.org/abs/hep-ph/0112085}{{\tt hep-ph/0112085}}].

\bibitem{Braun:2006hz}
V.~M. Braun, A.~Lenz, and M.~Wittmann, {\it {Nucleon Form Factors in QCD}},
  {\em Phys. Rev.} {\bf D73} (2006) 094019,
  [\href{http://arxiv.org/abs/hep-ph/0604050}{{\tt hep-ph/0604050}}].

\bibitem{Anikin:2013aka}
I.~V. Anikin, V.~M. Braun, and N.~Offen, {\it {Nucleon Form Factors and
  Distribution Amplitudes in QCD}},  {\em Phys. Rev.} {\bf D88} (2013) 114021,
  [\href{http://arxiv.org/abs/1310.1375}{{\tt arXiv:1310.1375}}].

\bibitem{Gockeler:2008xv}
{\bf QCDSF} Collaboration, M.~G{\"o}ckeler et~al., {\it {Nucleon distribution
  amplitudes from lattice QCD}},  {\em Phys. Rev. Lett.} {\bf 101} (2008)
  112002, [\href{http://arxiv.org/abs/0804.1877}{{\tt arXiv:0804.1877}}].

\bibitem{Braun:2008ur}
{\bf QCDSF} Collaboration, V.~M. Braun et~al., {\it {Nucleon distribution
  amplitudes and proton decay matrix elements on the lattice}},  {\em Phys.
  Rev.} {\bf D79} (2009) 034504, [\href{http://arxiv.org/abs/0811.2712}{{\tt
  arXiv:0811.2712}}].

\bibitem{Braun:2014wpa}
R.~W. Schiel, V.~M. Braun, S.~Collins, B.~Gl{\"a}{\ss}le, M.~G{\"o}ckeler,
  A.~Sch{\"a}fer, W.~S{\"o}ldner, A.~Sternbeck, and P.~Wein, {\it {Light-cone
  distribution amplitudes of the nucleon and negative parity nucleon resonances
  from lattice QCD}},  {\em Phys. Rev.} {\bf D89} (2014) 094511,
  [\href{http://arxiv.org/abs/1403.4189}{{\tt arXiv:1403.4189}}].

\bibitem{Braun:2009jy}
V.~M. Braun et~al., {\it {Electroproduction of the $N^*(1535)$ resonance at
  large momentum transfer}},  {\em Phys. Rev. Lett.} {\bf 103} (2009) 072001,
  [\href{http://arxiv.org/abs/0902.3087}{{\tt arXiv:0902.3087}}].

\bibitem{Feldmann:2011xf}
T.~Feldmann and M.~W.~Y. Yip, {\it {Form Factors for $\Lambda_b \to \Lambda$
  Transitions in SCET}},  {\em Phys. Rev.} {\bf D85} (2012) 014035,
  [\href{http://arxiv.org/abs/1111.1844}{{\tt arXiv:1111.1844}}]. [{\it{Erratum
  ibid.}} {\bf{D86}} (2012) 079901].

\bibitem{Wang:2015ndk}
Y.-M. Wang and Y.-L. Shen, {\it {Perturbative Corrections to $\Lambda_b \to
  \Lambda$ Form Factors from QCD Light-Cone Sum Rules}},
  \href{http://arxiv.org/abs/1511.09036}{{\tt arXiv:1511.09036}}.

\bibitem{Lach:1995we}
J.~Lach and P.~Zenczykowski, {\it {Hyperon radiative decays}},  {\em Int. J.
  Mod. Phys.} {\bf A10} (1995) 3817.

\bibitem{Aznauryan:2012ba}
I.~G. Aznauryan et~al., {\it {Studies of Nucleon Resonance Structure in
  Exclusive Meson Electroproduction}},  {\em Int. J. Mod. Phys.} {\bf E22}
  (2013) 1330015, [\href{http://arxiv.org/abs/1212.4891}{{\tt
  arXiv:1212.4891}}].

\bibitem{Liu:2010bh}
Y.-L. Liu and M.-Q. Huang, {\it {A light-cone QCD sum rule approach for the
  $\Xi$ baryon electromagnetic form factors and the semileptonic decay $\Xi_c
  \to \Xi e^+ \nu_e$}},  {\em J. Phys.} {\bf G37} (2010) 115010,
  [\href{http://arxiv.org/abs/1102.4245}{{\tt arXiv:1102.4245}}].

\bibitem{Aliev:2014nla}
T.~M. Aliev and M.~Savc{\i}, {\it {Octet negative parity to octet positive
  parity electromagnetic transitions in light cone QCD}},  {\em J. Phys.} {\bf
  G41} (2014) 075007, [\href{http://arxiv.org/abs/1403.0096}{{\tt
  arXiv:1403.0096}}].

\bibitem{Bruno:2014jqa}
M.~Bruno et~al., {\it {Simulation of QCD with $N_{f} = 2 + 1$ flavors of
  non-perturbatively improved Wilson fermions}},  {\em JHEP} {\bf 02} (2015)
  043, [\href{http://arxiv.org/abs/1411.3982}{{\tt arXiv:1411.3982}}].

\bibitem{Bietenholz:2011qq}
{\bf QCDSF/UKQCD} Collaboration, W.~Bietenholz et~al., {\it {Flavour blindness
  and patterns of flavour symmetry breaking in lattice simulations of up, down
  and strange quarks}},  {\em Phys. Rev.} {\bf D84} (2011) 054509,
  [\href{http://arxiv.org/abs/1102.5300}{{\tt arXiv:1102.5300}}].

\bibitem{Wein:2015oqa}
P.~Wein and A.~Sch{\"a}fer, {\it {Model-independent calculation of
  SU(3)${}_{f}$ violation in baryon octet light-cone distribution amplitudes}},
   {\em JHEP} {\bf 05} (2015) 073, [\href{http://arxiv.org/abs/1501.07218}{{\tt
  arXiv:1501.07218}}].

\bibitem{Borsanyi:2012zs}
{\bf BMW} Collaboration, S.~Bors{\'a}nyi et~al., {\it {High-precision scale
  setting in lattice QCD}},  {\em JHEP} {\bf 09} (2012) 010,
  [\href{http://arxiv.org/abs/1203.4469}{{\tt arXiv:1203.4469}}].

\bibitem{Kraenkl:2011qb}
S.~Kr{\"a}nkl and A.~Manashov, {\it {Two-loop renormalization of three-quark
  operators in QCD}},  {\em Phys. Lett.} {\bf B703} (2011) 519,
  [\href{http://arxiv.org/abs/1107.3718}{{\tt arXiv:1107.3718}}].

\bibitem{Braun:2000kw}
V. Braun, R.~J. Fries, N.~Mahnke, and E.~Stein, {\it {Higher twist
  distribution amplitudes of the nucleon in QCD}},  {\em Nucl. Phys.} {\bf
  B589} (2000) 381, [\href{http://arxiv.org/abs/hep-ph/0007279}{{\tt
  hep-ph/0007279}}]. [{\it{Erratum ibid.}} {\bf{B607}} (2001) 433].

\bibitem{Ji:2002xn}
X.~Ji, J.-P. Ma, and F.~Yuan, {\it {Three quark light cone amplitudes of the
  proton and quark orbital motion dependent observables}},  {\em Nucl. Phys.}
  {\bf B652} (2003) 383, [\href{http://arxiv.org/abs/hep-ph/0210430}{{\tt
  hep-ph/0210430}}].

\bibitem{Belitsky:2005qn}
A.~V. Belitsky and A.~V. Radyushkin, {\it {Unraveling hadron structure with
  generalized parton distributions}},  {\em Phys. Rept.} {\bf 418} (2005) 1,
  [\href{http://arxiv.org/abs/hep-ph/0504030}{{\tt hep-ph/0504030}}].

\bibitem{Braun:2008ia}
V.~M. Braun, A.~N. Manashov, and J.~Rohrwild, {\it {Baryon Operators of Higher
  Twist in QCD and Nucleon Distribution Amplitudes}},  {\em Nucl. Phys.} {\bf
  B807} (2009) 89, [\href{http://arxiv.org/abs/0806.2531}{{\tt
  arXiv:0806.2531}}].

\bibitem{Gracey:2012gx}
J.~A. Gracey, {\it {Three loop renormalization of 3-quark operators in QCD}},
  {\em JHEP} {\bf 09} (2012) 052, [\href{http://arxiv.org/abs/1208.5619}{{\tt
  arXiv:1208.5619}}].

\bibitem{Claudson:1981gh}
M.~Claudson, M.~B. Wise, and L.~J. Hall, {\it {Chiral Lagrangian for Deep Mine
  Physics}},  {\em Nucl. Phys.} {\bf B195} (1982) 297.

\bibitem{Belyaev:1982cd}
V.~M. Belyaev and B.~L. Ioffe, {\it {Determination of the baryon mass and
  baryon resonances from the quantum-chromodynamics sum rule. Strange
  baryons}},  {\em Sov. Phys. JETP} {\bf 57} (1983) 716.

\bibitem{Ioffe:1983ty}
B.~L. Ioffe and A.~V. Smilga, {\it {Hyperon magnetic moments in QCD}},  {\em
  Phys. Lett.} {\bf B133} (1983) 436.

\bibitem{Leinweber:2004it}
D.~B. Leinweber, W.~Melnitchouk, D.~G. Richards, A.~G. Williams, and J.~M.
  Zanotti, {\it {Baryon spectroscopy in lattice QCD}},  {\em Lect. Notes Phys.}
  {\bf 663} (2005) 71, [\href{http://arxiv.org/abs/nucl-th/0406032}{{\tt
  nucl-th/0406032}}].

\bibitem{Edwards:2004sx}
{\bf SciDAC/LHPC/UKQCD} Collaboration, R.~G. Edwards and B.~Jo{\'o}, {\it {The
  Chroma software system for lattice QCD}},  {\em Nucl. Phys. B Proc. Suppl.}
  {\bf 140} (2005) 832, [\href{http://arxiv.org/abs/hep-lat/0409003}{{\tt
  hep-lat/0409003}}].

\bibitem{Arts:2015jia}
P.~Arts et~al., {\it {QPACE 2 and Domain Decomposition on the Intel Xeon Phi}},
   {\em PoS} {\bf LATTICE 2014} (2015) 021,
  [\href{http://arxiv.org/abs/1502.04025}{{\tt arXiv:1502.04025}}].

\bibitem{Heybrock:2014iga}
S.~Heybrock, B.~Jo{\'o}, D.~D. Kalamkar, M.~Smelyanskiy, K.~Vaidyanathan,
  T.~Wettig, and P.~Dubey, {\it {Lattice QCD with Domain Decomposition on Intel
  Xeon Phi Co-Processors}},  \href{http://arxiv.org/abs/1412.2629}{{\tt
  arXiv:1412.2629}}.

\bibitem{Frommer:2013fsa}
A.~Frommer, K.~Kahl, S.~Krieg, B.~Leder, and M.~Rottmann, {\it {Adaptive
  Aggregation Based Domain Decomposition Multigrid for the Lattice Wilson Dirac
  Operator}},  {\em SIAM J. Sci. Comput.} {\bf 36} (2014) A1581,
  [\href{http://arxiv.org/abs/1303.1377}{{\tt arXiv:1303.1377}}].

\bibitem{Luscher:2012av}
M.~L{\"u}scher and S.~Schaefer, {\it {Lattice QCD with open boundary conditions
  and twisted-mass reweighting}},  {\em Comput. Phys. Commun.} {\bf 184} (2013)
  519, [\href{http://arxiv.org/abs/1206.2809}{{\tt arXiv:1206.2809}}].

\bibitem{Gusken:1989qx}
S.~G{\"u}sken, {\it {A Study of smearing techniques for hadron correlation
  functions}},  {\em Nucl. Phys. B Proc. Suppl.} {\bf 17} (1990) 361.

\bibitem{Falcioni:1984ei}
M.~Falcioni, M.~L. Paciello, G.~Parisi, and B.~Taglienti, {\it {Again on SU(3)
  glueball mass}},  {\em Nucl. Phys.} {\bf B251} (1985) 624.

\bibitem{Luscher:2011kk}
M.~L{\"u}scher and S.~Schaefer, {\it {Lattice QCD without topology barriers}},
  {\em JHEP} {\bf 07} (2011) 036, [\href{http://arxiv.org/abs/1105.4749}{{\tt
  arXiv:1105.4749}}].

\bibitem{Gockeler:2011ze}
{\bf QCDSF/UKQCD} Collaboration, M.~G{\"o}ckeler et~al., {\it {Baryon Axial
  Charges and Momentum Fractions with $N_f=2+1$ Dynamical Fermions}},  {\em
  PoS} {\bf LATTICE 2010} (2010) 163,
  [\href{http://arxiv.org/abs/1102.3407}{{\tt arXiv:1102.3407}}].

\bibitem{Cooke:2013qqa}
{\bf QCDSF/UKQCD} Collaboration, A.~N. Cooke et~al., {\it {SU(3) flavour
  breaking and baryon structure}},  {\em PoS} {\bf LATTICE 2013} (2014) 278,
  [\href{http://arxiv.org/abs/1311.4916}{{\tt arXiv:1311.4916}}].

\bibitem{Kaltenbrunner:2008pb}
T.~Kaltenbrunner, M.~G{\"o}ckeler, and A.~Sch{\"a}fer, {\it {Irreducible
  Multiplets of Three-Quark Operators on the Lattice: Controlling Mixing under
  Renormalization}},  {\em Eur. Phys. J.} {\bf C55} (2008) 387,
  [\href{http://arxiv.org/abs/0801.3932}{{\tt arXiv:0801.3932}}].

\bibitem{Bruns:2012eh}
P.~C. Bruns, L.~Greil, and A.~Sch{\"a}fer, {\it {Chiral extrapolation of baryon
  mass ratios}},  {\em Phys. Rev.} {\bf D87} (2013) 054021,
  [\href{http://arxiv.org/abs/1209.0980}{{\tt arXiv:1209.0980}}].

\bibitem{Jenkins:1991es}
E.~E. Jenkins and A.~V. Manohar, {\it {Chiral corrections to the baryon axial
  currents}},  {\em Phys. Lett.} {\bf B259} (1991) 353.

\bibitem{Close:1993mv}
F.~E. Close and R.~G. Roberts, {\it {Consistent analysis of the spin content of
  the nucleon}},  {\em Phys. Lett.} {\bf B316} (1993) 165,
  [\href{http://arxiv.org/abs/hep-ph/9306289}{{\tt hep-ph/9306289}}].

\bibitem{Borasoy:1998pe}
B.~Borasoy, {\it {Baryon axial currents}},  {\em Phys. Rev.} {\bf D59} (1999)
  054021, [\href{http://arxiv.org/abs/hep-ph/9811411}{{\tt hep-ph/9811411}}].

\bibitem{Zhu:2000zf}
S.-L. Zhu, S.~Puglia, and M.~J. Ramsey-Musolf, {\it {Recoil order chiral
  corrections to baryon octet axial currents}},  {\em Phys. Rev.} {\bf D63}
  (2001) 034002, [\href{http://arxiv.org/abs/hep-ph/0009159}{{\tt
  hep-ph/0009159}}].

\bibitem{Lin:2007ap}
H.-W. Lin and K.~Orginos, {\it {First Calculation of Hyperon Axial Couplings
  from Lattice QCD}},  {\em Phys. Rev.} {\bf D79} (2009) 034507,
  [\href{http://arxiv.org/abs/0712.1214}{{\tt arXiv:0712.1214}}].

\bibitem{WalkerLoud:2011ab}
A.~Walker-Loud, {\it {Evidence for non-analytic light quark mass dependence in
  the baryon spectrum}},  {\em Phys. Rev.} {\bf D86} (2012) 074509,
  [\href{http://arxiv.org/abs/1112.2658}{{\tt arXiv:1112.2658}}].

\bibitem{GellMann:1961ky}
M.~Gell-Mann, {\it {The Eightfold Way: A Theory of strong interaction
  symmetry}},  {\em CTSL-20, TID-12608} (1961).

\bibitem{Wein:2011ix}
P.~Wein, P.~C. Bruns, T.~R. Hemmert, and A.~Sch{\"a}fer, {\it {Chiral
  extrapolation of nucleon wave function normalization constants}},  {\em Eur.
  Phys. J.} {\bf A47} (2011) 149, [\href{http://arxiv.org/abs/1106.3440}{{\tt
  arXiv:1106.3440}}].

\bibitem{Schiel:2011av}
{\bf QCDSF} Collaboration, R.~W. Schiel et~al., {\it {An Update on Distribution
  Amplitudes of the Nucleon and its Parity Partner}},  {\em PoS} {\bf LATTICE
  2011} (2011) 175, [\href{http://arxiv.org/abs/1112.0473}{{\tt
  arXiv:1112.0473}}].

\bibitem{Aoki:2008ku}
{\bf RBC/UKQCD} Collaboration, Y.~Aoki et~al., {\it {Proton lifetime bounds
  from chirally symmetric lattice QCD}},  {\em Phys. Rev.} {\bf D78} (2008)
  054505, [\href{http://arxiv.org/abs/0806.1031}{{\tt arXiv:0806.1031}}].

\bibitem{juqueen}
M. Stephan and J. Docter, {\it {JUQUEEN: IBM Blue
  Gene/Q${}^{\text{\textregistered}}$ Supercomputer System at the J{\"u}lich
  Supercomputing Centre}},  {\em JLSRF} {\bf 1} (2015) A1.

\bibitem{Georgi:1985kw}
H.~Georgi, {\em {Weak interactions and modern particle theory}}.
\newblock Benjamin Cummings, Menlo Park USA, 1984.

\bibitem{Gockeler:2010yr}
{\bf QCDSF/UKQCD} Collaboration, M.~G{\"o}ckeler et~al., {\it {Perturbative and
  Nonperturbative Renormalization in Lattice QCD}},  {\em Phys. Rev.} {\bf D82}
  (2010) 114511, [\href{http://arxiv.org/abs/1003.5756}{{\tt
  arXiv:1003.5756}}]. [{\it{Erratum ibid.}} {\bf{D86}} (2012) 099903].

\end{thebibliography}
\end{document}